\documentclass{aa}  

\usepackage{graphicx}
\usepackage{txfonts}
\newcommand{\msun}{{\rm M}_{\odot}}
\newcommand{\rsun}{{\rm R}_{\odot}}

\begin{document}

   \title{The circumstellar envelope around the S-type AGB star W Aql}

   \subtitle{Effects of an eccentric binary orbit}

   \author{S. Ramstedt
          \inst{1}
          \and
          S. Mohamed\inst{2,3,4} \and W.~H.~T. Vlemmings\inst{5} \and T. Danilovich\inst{6} \and M. Brunner\inst{7} \and E. De Beck\inst{5}\and E.~M.~L. Humphreys\inst{8} \and M. Lindqvist\inst{5} \and M. Maercker\inst{5} \and H. Olofsson\inst{5} \and F. Kerschbaum\inst{7} \and G. Quintana-Lacaci\inst{9}}

   \institute{Department of Physics and Astronomy, Uppsala University,\\
              \email{sofia.ramstedt@physics.uu.se}
         \and
             South African Astronomical Observatory, PO box 9, 7935 Observatory, South Africa              
             \and
             Astronomy Department, University of Cape Town, University of Cape Town, 7701, Rondebosch, South Africa
             \and
             South Africa National Institute for Theoretical Physics, Private Bag X1, Matieland, 7602, South Africa
             \and
             Dept. of Earth and Space Sciences, Chalmers University of Technology, Onsala Space Observatory, 439 92 Onsala Sweden
             \and
             Instituut voor Sterrenkunde, KU Leuven, Celestijnenlaan 200D, 3001 Leuven, Belgium
             \and
             Dept. of Astrophysics, University of Vienna, T\"urkenschanzstr. 17, 1180 Vienna, Austria
             \and
             ESO, Karl-Schwarzschild-Str. 2, 85748 Garching bei M\"unchen, Germany
             \and
             Instituto de Ciencia de Materiales de Madrid, CSIC, c/ Sor Juana In\'es de la Cruz 3, 28049 Cantoblanco, Madrid, Spain
             }

   \date{}

% \abstract{}{}{}{}{} 
% 5 {} token are mandatory
 
  \abstract
   {Recent observations at subarcsecond resolution, now possible also at submillimeter wavelengths, have shown intricate circumstellar structures around asymptotic giant branch (AGB) stars, mostly attributed to binary interaction. The results presented here are part of a larger project aimed at investigating the effects of a binary companion on the morphology of circumstellar envelopes (CSEs) of AGB stars.}
   {AGB stars are characterized by intense stellar winds that build CSEs around the stars. Here, the CO($J$\,=\,3$\rightarrow$2) emission from the CSE of the binary S-type AGB star W Aql has been observed at subarcsecond resolution using ALMA. The aim of this paper is to investigate the wind properties of the AGB star and to analyse how the known companion has shaped the CSE.}
   {The average mass-loss rate during the creation of the detected CSE is estimated through modelling, using the ALMA brightness distribution and previously published single-dish measurements as observational constraints. The ALMA observations are presented and compared to the results from a 3D smoothed particle hydrodynamics (SPH) binary interaction model with the same properties as the W Aql system and with two different orbital eccentricities. Three-dimensional radiative transfer modelling is performed and the response of the interferometer is modelled and discussed.}
   {The estimated average mass-loss rate of W~Aql is $\dot{M}$\,=\,3.0$\times$10$^{-6}$\,M$_{\odot}$\,yr$^{-1}$ and agrees with previous results based on single-dish CO line emission observations. The size of the emitting region is consistent with photodissociation models. The inner 10\arcsec~of the CSE is asymmetric with arc-like structures at separations of  2-3\arcsec\ scattered across the denser sections. Further out, weaker
 spiral structures at greater separations  are found, but this is at the limit of the sensitivity and field of view of the ALMA observations.}
   {The CO($J$\,=\,3$\rightarrow$2) emission is dominated by a smooth component overlayed with two weak arc patterns with different separations. The larger pattern is predicted by the binary interaction model with separations of  $\sim$10\arcsec\  and therefore likely due to the known companion. It is consistent with a binary orbit with low eccentricity. The smaller separation pattern is asymmetric and coincides with the dust distribution, but the separation timescale (200 yrs) is not consistent with any known process of the system. The separation of the known companions of the system is large enough to not have a very strong effect on the circumstellar morphology. The density contrast across the envelope of a binary with an even larger separation will not be easily detectable, even with ALMA, unless the orbit is strongly asymmetric or the AGB star has a much larger mass-loss rate. }


   \maketitle
%
%________________________________________________________________

\section{Introduction}
Stars with a main-sequence initial mass between 0.8--8\,M$_{\odot}$ will evolve up the asymptotic giant branch (AGB) before nuclear burning is finally extinguished and the stars die \citep{habiolof03}. The gas and dust expelled from the stars by the intense AGB wind \citep[][]{wils00} is later lit up by the hard radiation from the exposed stellar cores and form planetary nebulae (PNe). Several findings suggest that not all stars in the full AGB mass range will form PNe. For example, the range of measured carbon isotopic ratios is much smaller for PNe than for AGB stars  \citep[e.g.][]{ramsolof14}. 

In  recent  decades there has been intense research to find out how and when the complex morphologies seen among the majority of PNe arise. Interacting winds \citep[e.g.][]{kwok02,stefetal13}, rotation \citep{dorfhofn96, garcetal16}, global magnetic fields \citep{garcetal05}, and binary (or large planet) interaction \citep{nordblac06,dema09,stafetal16} have been proposed as possible shaping agents for stars already on the AGB, and/or during the transition. Recently, the community has reached some consensus that interaction with a binary companion is required to explain the morphologies exhibited by most PNe \citep{dema14}, thereby suggesting that PNe formation is more common among binary stars \citep[e.g.][]{moedema06}.

It is likely -- and recent, high spatial resolution observations have also shown \citep[e.g.][]{maeretal12,maurhugg13,cernetal15,kimetal17} -- that the shaping of the circumstellar material will start on the (late) AGB. Circumstellar wind dynamics in binary systems with a mass-losing primary are therefore a very active field of research \citep[e.g.][]{kimetal15,toupetal15}. To provide observational constraints for recently developed 3D hydrodynamical models \citep[e.g.][]{mohapods12}, we have designed a project for ALMA to observe the CO gas distribution and kinematics around known binary AGB stars. Resolved images of the binary pair are available for the initial sample stars, R~Aqr, $o$~Cet (Mira), W~Aql, and $\pi^{1}$~Gru, and the sources cover extensive ranges in the important modelling parameters, i.e. binary separation, $a$\,$\sim$\,20--400\,AU, and AGB wind expansion velocity, $v_{\rm{e}}$\,$\approx$\,10--100\,km\,s$^{-1}$. Initial results have been published in \citet[][on Mira]{ramsetal14}, \citet[][also on W~Aql, and $\pi^{1}$~Gru]{ramsetal15}, and \citet[][on $\pi^{1}$~Gru]{doanetal17}. 

This paper  focuses on the S-type AGB star W~Aql. S-type stars are classified by the presence by ZrO bands and are thought to represent an intermediate evolutionary stage where the atmospheric C/O-ratio is close to 1. W~Aql has a known companion at a separation of $a$\,=\,0.46\arcsec \citep{ramsetal11} or $\approx$180\,AU, and the distance to the system is estimated to 395\,pc \citep{danietal14}. Several resolved observations of the system and the circumstellar material exist. The dust emission at 11.15 $\mu$m from the inner arcsecond around W~Aql was measured using the three-element Infrared Spatial Interferometer (ISI) in 2004 \citep{tateetal06}. The observations showed brighter emission on the east side of the source indicating dust excess very close to the star (0\farcs5$\sim$200\,AU at D=395\,pc). Assuming a typical outflow velocity of 20\,km\,s$^{-1}$, \citet{tateetal06} suggest that a dust shell has been expelled from the star ``within the last 35 years'' (from 2004). The dust distribution on a 1\arcmin\  scale around W~Aql was imaged by \citet{ramsetal11} using polarimetry to detect the dust scattered light. This revealed an asymmetric dust distribution around the star with more dust on the south-west side. The first resolved Hubble Space Telescope (HST\footnote{HST Proposal 10185, PI: Raghvendra Sahai}) image of the binary pair was also presented. The companion was later classified by \citet{danietal15} as a F8 to G0 main-sequence star. \citet{danietal15} established that the known companion has an effective temperature in the range  5900 - 6170 K and a mass of 1.04 - 1.09 M$_{\odot}$. They find a mass for the AGB star of 1.04 - 3 M$_{\odot}$. The thermal dust emission from W~Aql (on even larger scales) was imaged by Herschel/PACS as part of the  Mass-loss of Evolved StarS (MESS) key program \citep{groeetal11}. \citet{mayeetal13} investigated the large-scale dust distribution, and found that the south-west dust asymmetry also extends to larger scales. 

\citet{danietal14} modelled the molecular line emission from W Aql assuming a spherically symmetric circumstellar envelope (CSE), and derived the abundances of several chemically important molecules (including the first detection of NH$_{3}$ in an S-type star) and a mass-loss rate of $\dot{M}$\,=\,4.0$\times$10$^{-6}$\,M$_{\odot}$\,yr$^{-1}$. In this paper the 1D CO radiative transfer model is further constrained by the ALMA data. The data is presented in Sect.~\ref{obs}, and the 1D radiative transfer modelling to estimate the average mass-loss rate in Sect.~\ref{mdot}.  The 3D models used to analyse the wind shaping effects due to the companion are introduced in Sects~\ref{sph} and~\ref{lime}. In Sect.~\ref{res} the results from the observations and the models are given, and further discussed in Sect.~\ref{dis}. Finally, conclusions are drawn in Sect.~\ref{conc}.
 
%__________________________________________________________________

\section{Observations}
\label{obs}

\subsection{Observations}
\begin{figure*}
\center
\includegraphics[width=6cm]{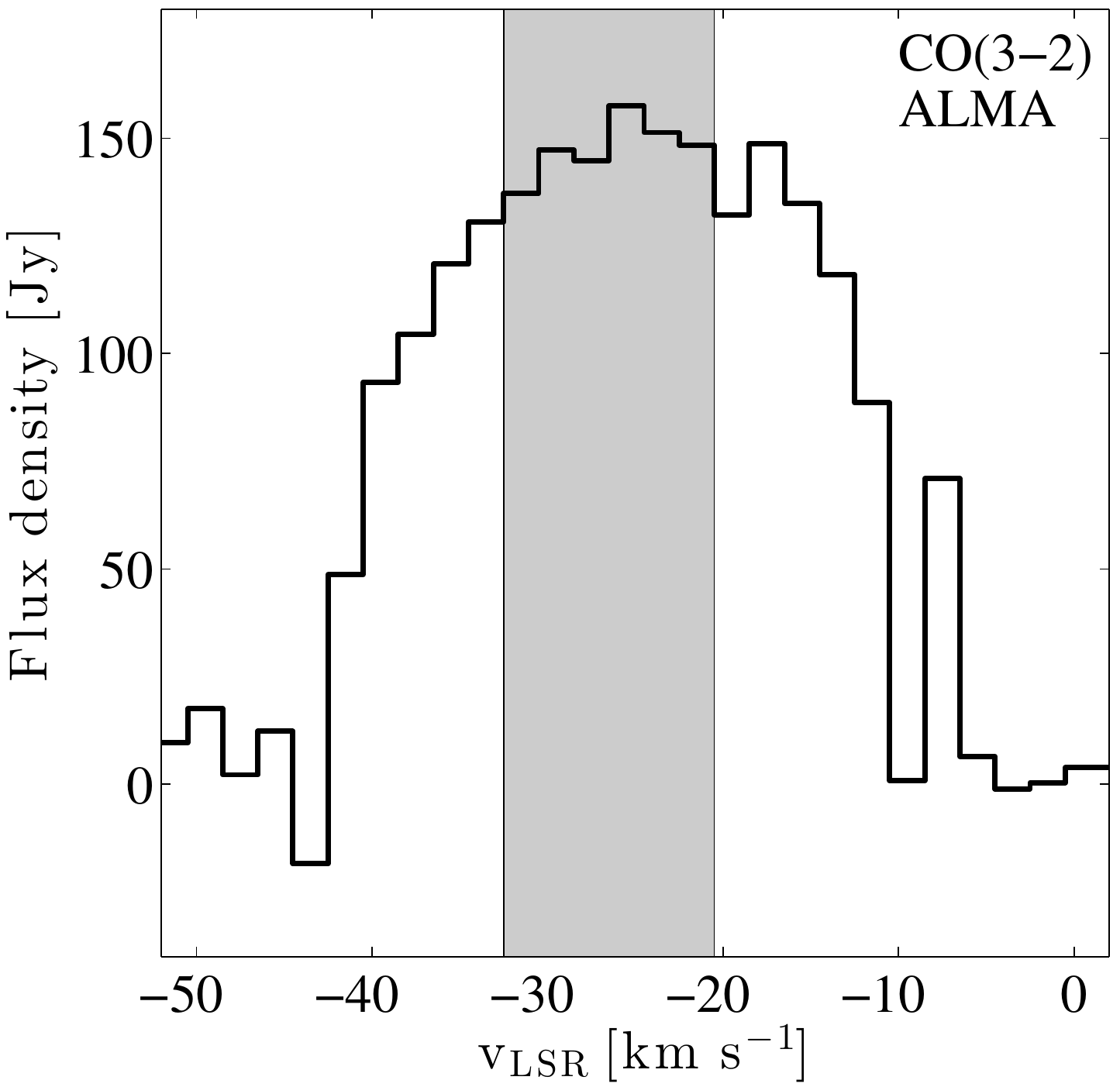}
\includegraphics[width=6cm]{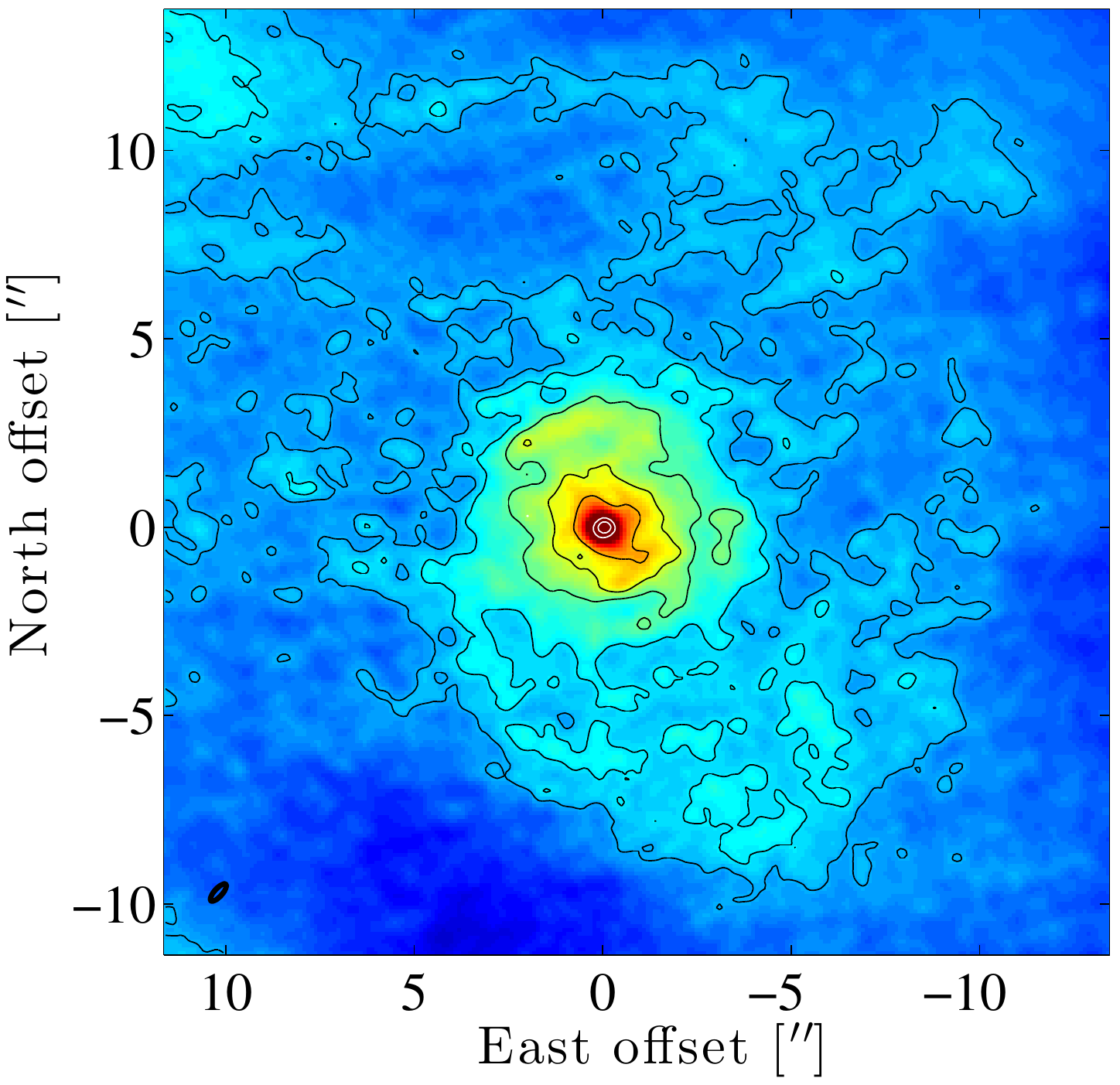}
\includegraphics[width=6cm]{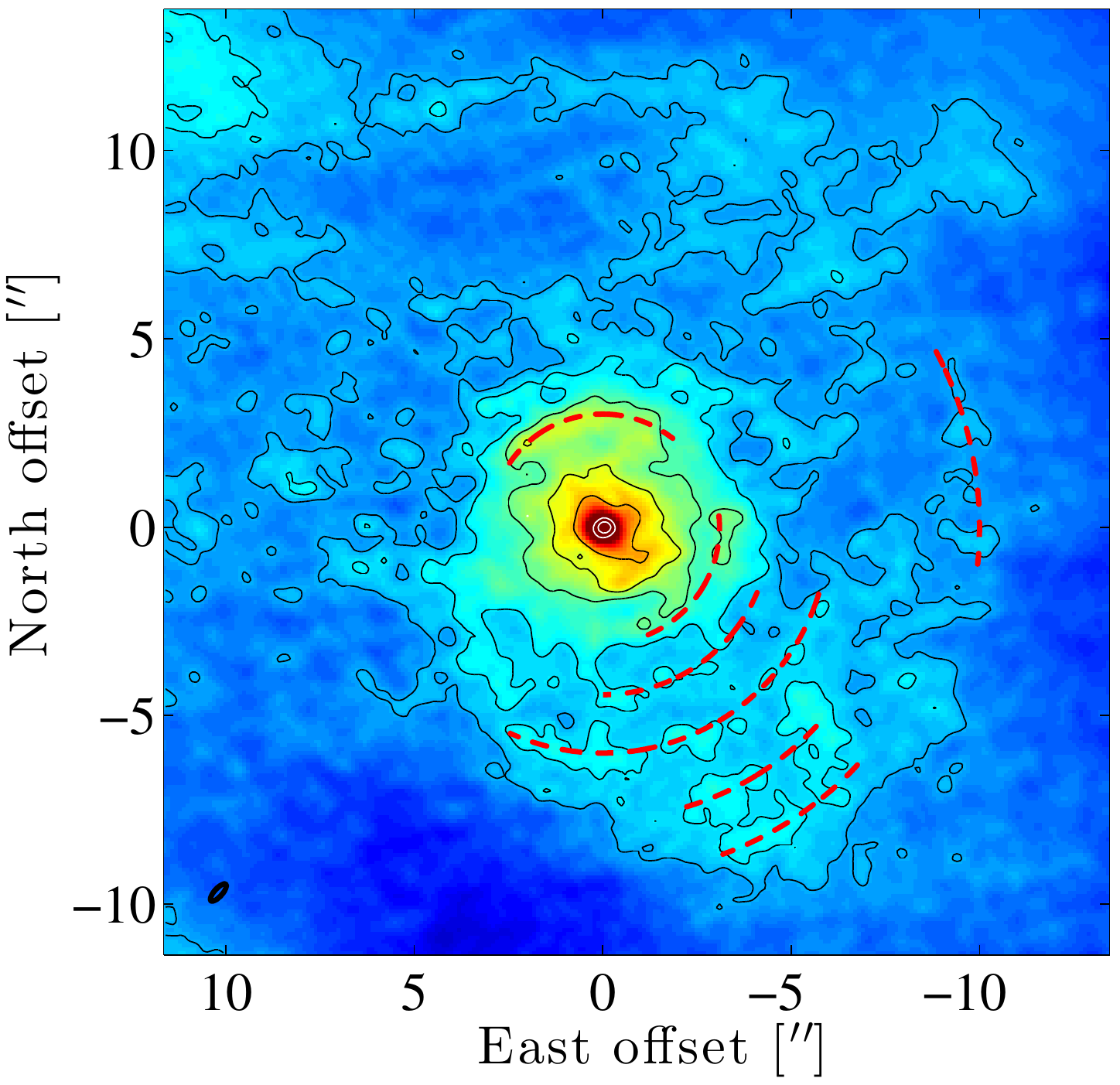}
\caption{{\it Left:}  ALMA CO($J$\,=\,3$\rightarrow$2) emission line at 2\,km\,s$^{-1}$ spectral resolution. The dip at $v_{\rm{LSR}}$=-10\,km\,s$^{-1}$ is due to a bad channel (see Appendix) and is not a real feature. The grey shaded area shows the averaged velocity range  to create the image to the right. Middle: Image average of the central channels of the line (grey shaded area in left plot). The image shows arc-like structures and increased emission on the south-west and west sides of the star. Black contours are drawn at 5, 10, 20, 30, and 40$\sigma$ (where $\sigma$ is measured over the emission-free channels). The white contours show the position of the AGB star from the continuum emission at 345\,GHz. The beam is drawn in the lower left corner. {\it Right:} Same as middle image with arcs outlined by red dashed lines.}
\label{lineandimage}
\end{figure*}

W Aql was observed with ten Atacama Compact Array (ACA) 7\,m antennas on  6 March 2014. It was also observed on  20 March 2014 with 33, and on  14 April 2014 with 34 main array 12\,m antennas. The main array observations were performed as a ten-pointing mosaic covering an area of approximately 30\arcsec$\times$30$\arcsec$. The same area was covered by the ACA in three mosaic pointings. Finally, it was observed with three 12\,m antennas of the Total Power (TP) array on the 12 June 2015. The correlator was set up with four spectral windows $\sim$2\,GHz in width  centred  on 331, 333, 343, and 345\,GHz to cover  the $^{12}$CO and $^{13}$CO $J$\,=\,3\,$\rightarrow$\,2 lines in the same setting. The total observing time was 23\,mins for the ACA, and 7.25\,mins and 7.3\,mins for the 12m array on the first and second run, respectively, and 67\,mins total time for the TP array. Baseline lengths range from 15 to 558\,m for the main array, and from 9 to 49\,m for the ACA. This results in a maximum recoverable scale of $\sim$7\arcsec~for the main array and  $\sim$12\arcsec~for the ACA. The restoring beam of the TP observation is 18\farcs84$\times$18\farcs84 at 345.2\,GHz. Calibration was carried out following standard procedure. The quasar J1911-2006 was used as complex gain calibrator; quasar J1924-2914 was used as bandpass calibrator; and Ceres, Titan, and quasar J1924-292 were used as flux calibrators. Total power observations are calibrated using standard single-dish observing procedures with regular pointing checks and focus calibration. The intensity scale is converted from Kelvin to Jansky using conversion factors of 46.723, 48.568, 47.036, and 48.196 across the four different spectral windows. The main array data was combined with the ACA data, which was weighted  by a factor of 0.25 to account for the lower sensitivity of the 7m antennas, and the combined data was imaged. The combined main array+ACA visibilities were finally combined with the TP visibility data using the CASA task SD2VIS\footnote{Developed at the Nordic ALMA ARC: www.oso.nordic-alma.se}. A continuum image was made using the emission free channels from all spectral windows, but the continuum was too weak to be used for self-calibration of the offset mosaic pointings. The position and measured continuum flux density at 338\,GHz is determined from image fitting and given in Table~\ref{cont}. Final imaging was done after subtracting the continuum and averaging to reach a spectral resolution of 2\,km\,s$^{-1}$. The beam of the final images has a full width at half maximum of 0\farcs47$\times$0\farcs41, and a position angle of 65.3$^{\circ}$. The rms noise level in the emission-free channels reaches 15\,mJy/beam. Previous $^{12}$CO $J$\,=\,3\,$\rightarrow$\,2 line emission observations with APEX \citep{ramsetal09} measured a peak intensity of $\sim$145\,Jy showing that the same amount of flux was recovered by the combined ALMA observations (Fig.~\ref{lineandimage}, left). The $^{12}$CO $J$\,=\,3\,$\rightarrow$\,2 line emission is centred at a local standard of rest (LSR) velocity of $v_{\rm{LSR}}$=-25.5\,km\,s$^{-1}$.

\begin{table}[htp]
\caption{Current position (J2000.0) and continuum flux density at 338\,GHz measured from the ALMA data.}
\begin{center}
\begin{tabular}{lccc}
\hline \hline
Source & RA & Dec & $S_{\nu}$ \\
\hline
W Aql & 19:15:23.379 & -07:02:50.38 & 29$\pm$3\,mJy \\
\hline
\end{tabular}
\end{center}
\label{cont}
\end{table}%

%__________________________________________________________________

\section{Analysis}

\begin{figure}
\includegraphics[width=\columnwidth]{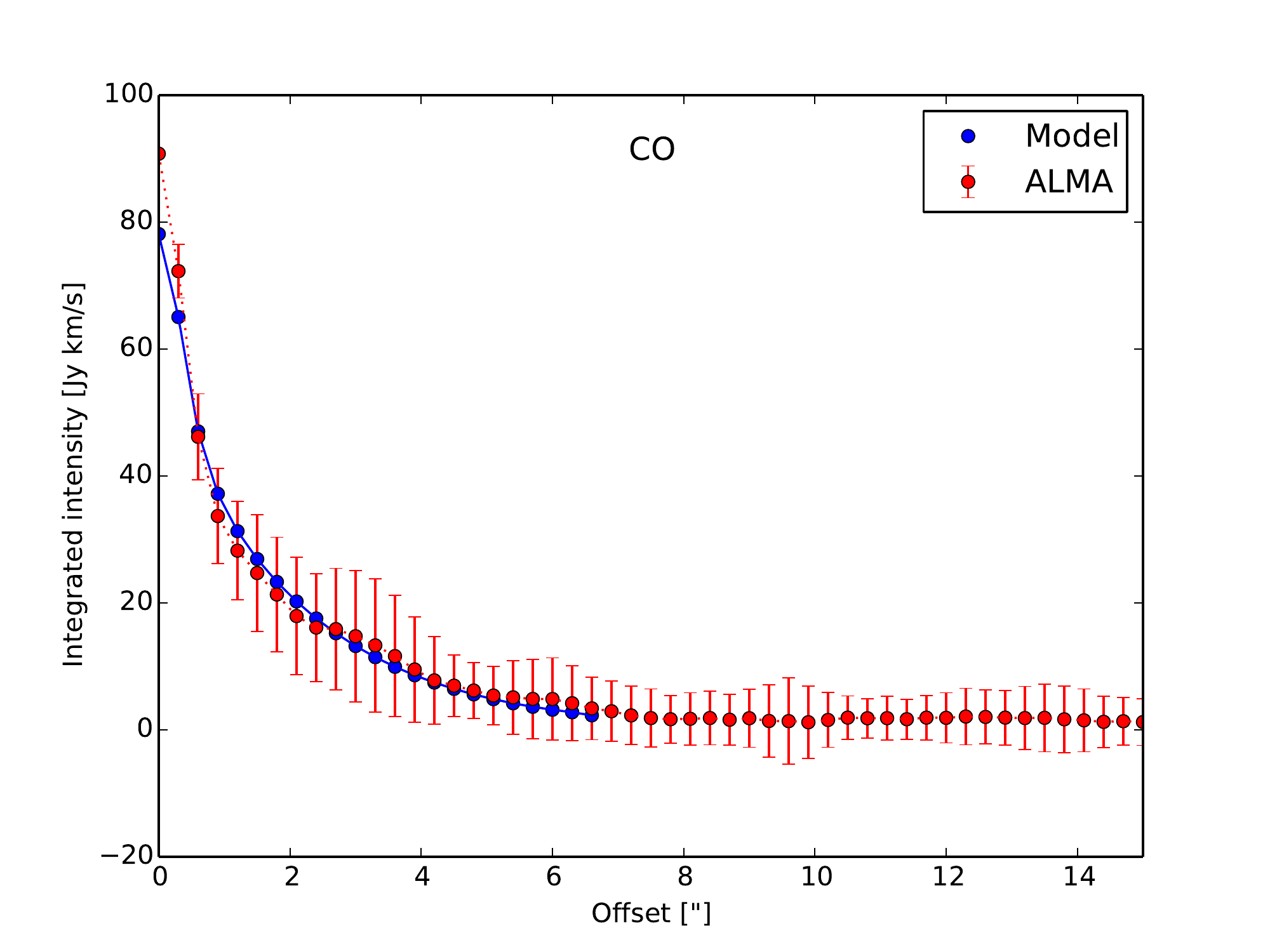}
\caption{Comparison between the direction-averaged CO($J$\,=\,3$\rightarrow$2) brightness distribution from the ALMA observations (red dots with error bars) and that derived from the radiative transfer model assuming a spherically symmetric CSE (blue dots). }
\label{bright}
\end{figure}

\subsection{Determining the average mass-loss rate}
\label{mdot}
\citet{danietal14} modelled all the available single-dish CO rotational lines of W~Aql (up to $J$\,=\,25$\rightarrow$24 in the vibrational ground state, $v$\,=\,0) using the non-LTE, non-local Monte Carlo radiative transfer code described in \citet{schoolof01}, and assuming a distance of 395\,pc. The radiative transfer code assumes a spherically symmetric CSE created by a constant mass-loss rate. For a further description on the assumptions made for the W~Aql model in particular and the code in general see \citet{danietal14} and \citet{schoolof01}, respectively.

The model by \citet{danietal14} is  further constrained here by the brightness distribution derived from the CO($J$\,=\,3$\rightarrow$2) map from ALMA. The CO envelope size was previously based on the photodissociation models by \citet{mamoetal88,stanetal95}. Now the brightness distribution provides direct constraints on the size of the CO envelope, which can be treated as a free parameter. The CO($J$\,=\,3$\rightarrow$2) brightness distribution is derived by smoothing the full image cube to a resolution of 0\farcs6 using imsmooth in CASA. Then the brightness distribution is measured by the integrated line flux (within a 0\farcs6 beam) at increasing distances from the centre (determined by the position of W~Aql A) in four directions (north, east, south, and west) and finally averaged over direction. The corresponding brightness distribution (using the same beam size and radii) is calculated from the radiative transfer modelling results and compared to the observed values until a fit is found (Fig.~\ref{bright}). The error in the measured line flux is assumed to be 10\% with the variation across the different directions taken into account. 

\subsection{Hydrodynamical modelling}
\label{sph}
W~Aql was modelled using the Evolved Stellar
Interactions with {\small GADGET} in 3D ($^{\rm{3D}}$ESI-Gadget code), which is based on a modified version of the collisionless and gasdynamical cosmological code {\small GADGET2} \citep{Spr05}. The fluid equations are solved using smoothed particle hydrodynamics (SPH), a Lagrangian method particularly suited to studying flows with arbitrary geometries. Modules include radiative cooling, stellar winds, and binary interactions \citep{Moh07,Moh12}.  

\begin{table}[htp]
\caption{Properties assumed in the hydrodynamic models. }
\begin{center}
\begin{tabular}{|l|l|}
\hline
Parameter & Assumed value \\
\hline
M$_{\rm AGB}$  & 2 $\msun$ \\
M$_{\rm companion}$ & 1 $\msun$ \\
R$_{\rm AGB}$  & 400 $\rsun$ \\
T$_{\rm AGB}$  & 3000 K\\
$\dot{M}_{\rm AGB}$ & 3$\times10^{-6}\, \msun$\,yr$^{-1}$ \\
$v_{\infty{\rm AGB}}$ & 16.5\,km\,s$^{-1}$ \\
$a$ & 180 AU \\
P$_{\rm orb}$ & 1395 years\\
$e$ & 0.2 and 0.6\\
\hline
\end{tabular}
\end{center}
\label{gadget}
\end{table}

The parameters assumed for the models are given in Table~\ref{gadget}; the stars are treated as point masses, and two different eccentricities for the binary orbit were assumed -- $e=0.2$ and $e=0.6$. The stellar wind from the AGB star was simulated by periodically injecting particles at a boundary given by the surface of the star. An acceleration parameter was added to the momentum equation to ensure that a smooth continuous wind with a terminal velocity of 16.5\,km\,s$^{-1}$ was produced. It was not possible to resolve the surface of the companion;   this was instead treated as an accretion boundary (the boundary radius is set to 1\,AU) where particles lose mass as they approach the companion and are removed from the simulation once they are 1\% of their original mass. The wind particles are injected into initial vacuum conditions, thus the outermost arc is denser than it would be otherwise. Each simulation was run for long enough to ensure that this region is outside  the observed ALMA field of view. After $\approx$ 11\,800 years of evolution, 6$\times 10^6$ particles had been injected. The initial mass of the particles is determined empirically so that it leads to the estimated mass-loss rate. They all have the same initial mass and the mass of the AGB star does not change as these particles are injected. Over the timescales considered, the amount of mass injected is insignificant (relative to the stellar mass). The output from these  high-resolution models was then gridded using the SPH cubic spline kernel to a 512$^3$ grid in order to be used as input for the post-processing radiative transfer routine. 

\subsection{Post-processing radiative transfer}
\label{lime}
The output from the hydrodynamical model, i.e. 512$^{3}$ grids of the H$_{2}$ number density, the temperature, and the velocity distributions, is used as input to calculate the expected CO line emission. The non-LTE, non-local 3D radiative transfer code LIME \citep[LIne Modelling Engine,][]{brinhoge10} is used to calculate the CO level populations and solve the radiative transfer equations. Grid points are distributed across the computational volume using Delauney triangulation and the probability for placing a grid point is weighted with the density. Each grid cell is the corresponding Voronoi cell where the local conditions (density, temperature, excitation, etc.) are constant across each cell. For the calculations presented here, 100000 grid points were used in a volume with a radius of 15000\,AU. A constant CO/H$_{2}$ fractional abundance of 6.0$\times$10$^{-4}$ is assumed for the S-type star. The molecular excitation is calculated including 41 rotational transitions in the vibrational ground state and first excited state. Higher energy levels are not expected to be significantly populated. Collisions with H$_{2}$ are included assuming an ortho-to-para ratio of 3. Collisional rate coefficients, as well as energy levels and radiative transition probabilities, are taken from \citet{schoetal05}\footnote{http://home.strw.leidenuniv.nl/$\sim$moldata/}. For the continuum thermal dust emission, amorphous silicate grains are assumed \citep{justtiel92}.

The output images from the radiative transfer model are produced over 100 channels using a spectral resolution of 1\,km\,s$^{-1}$, and a spatial resolution of 0\farcs2. The source distance is assumed to be 395\,pc. Finally, the uv-coverage of the combined ALMA arrays is taken into account by running this image through the ALMA simulator using the array configurations at the time of the W~Aql observations (in Cycle 1).

%__________________________________________________________________

\begin{figure*}
\center
\includegraphics[height=5.5cm,width=6cm]{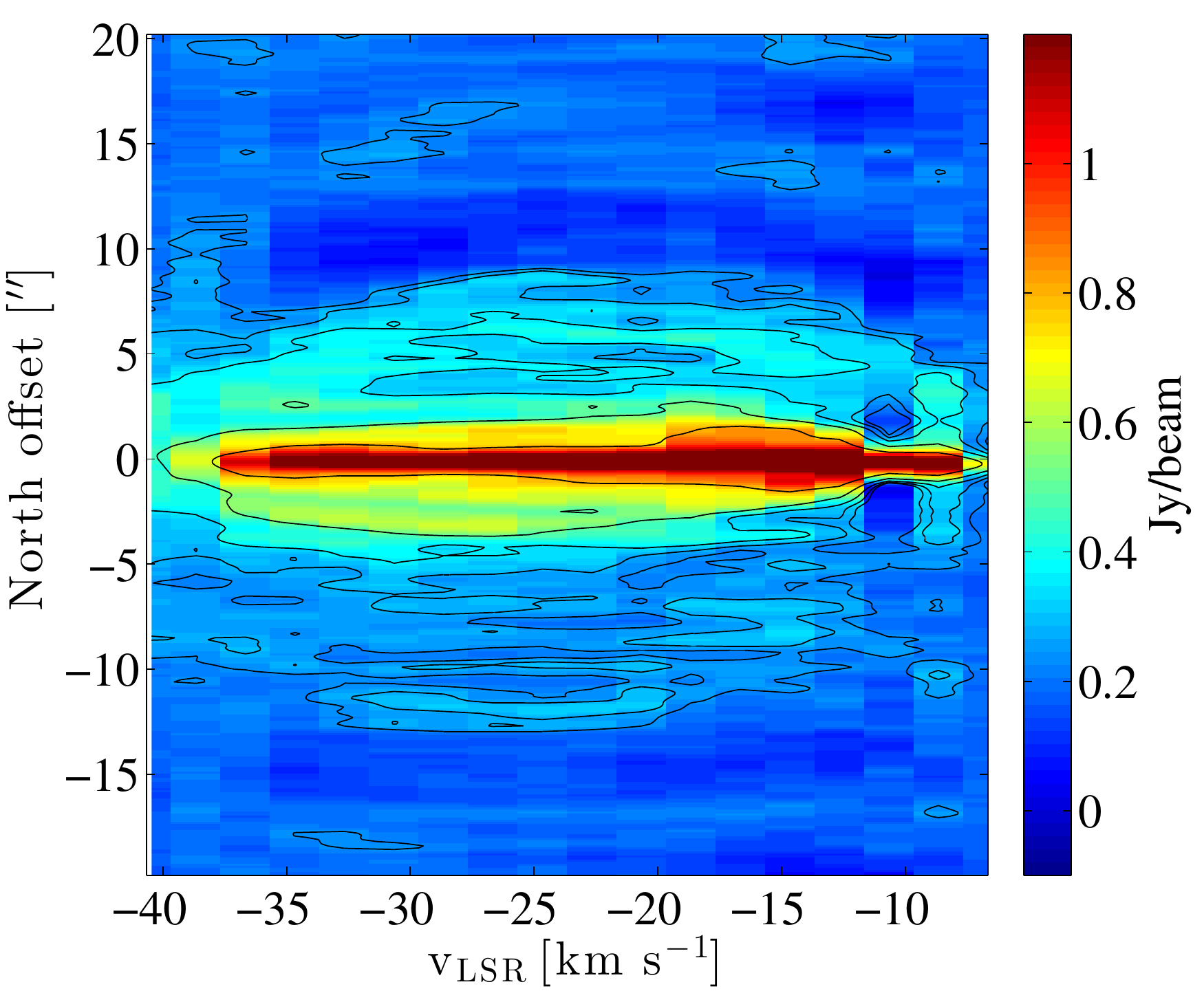}
\includegraphics[height=5.5cm,width=6cm]{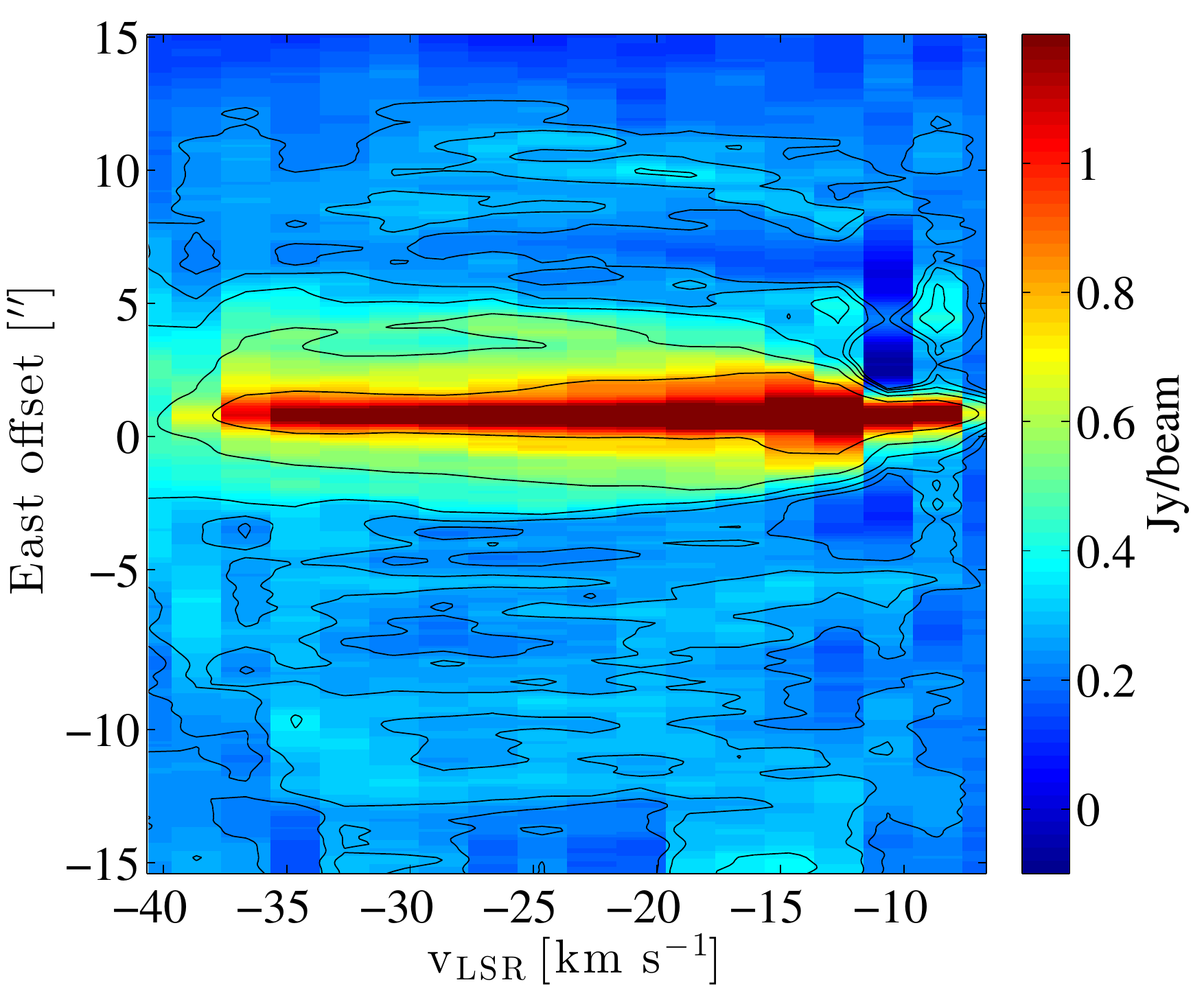}
\includegraphics[height=5.5cm,width=6cm]{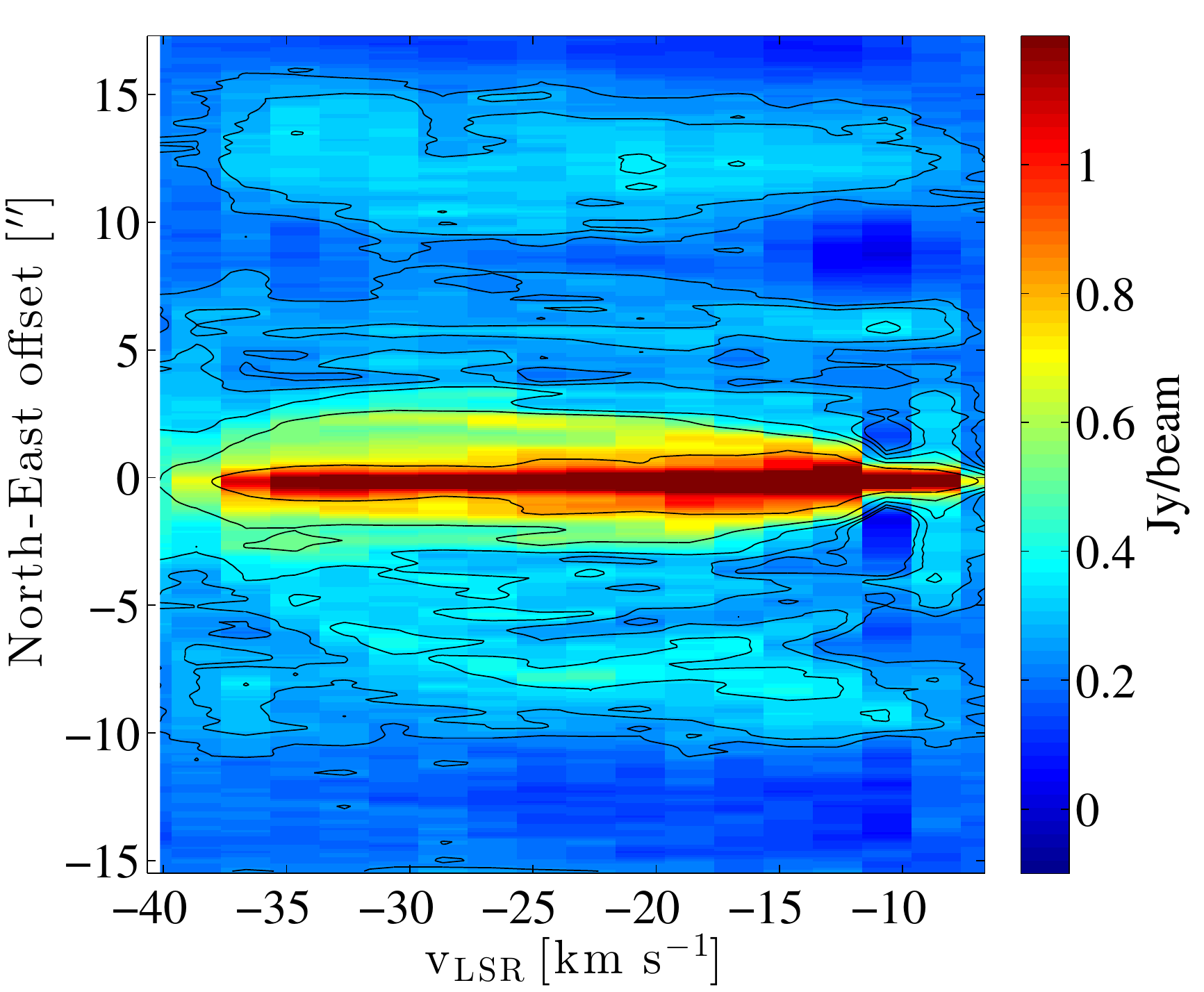}
\caption{{\it Left:} Position-velocity diagram across the north-south direction (PA=0$^{\circ}$). The contours are drawn at 3, 5, 10, 20, and 40$\sigma$ (where $\sigma$ is measured over the emission-free channels). {\it Middle:}  Position-velocity diagram across the east-west direction (PA=90$^{\circ}$) showing the brighter and more structured emission on the west side of the stars.  {\it Right:} Position-velocity diagram along the binary axis (PA=35$^{\circ}$) showing the structure within the south-west bright emission (Fig.~\ref{lineandimage}).  }
\label{pv90}
\end{figure*}

\section{Results}
\label{res}

\subsection{Wind properties}
The average mass-loss rate during the creation of the CSE probed by the observations is estimated to $\dot{M}$\,=\,3.0$\times$10$^{-6}$\,M$_{\odot}$\,yr$^{-1}$. The line widths are best fitted by a slowly accelerating velocity field which reaches the terminal expansion velocity of $v_{\rm{e}}$\,=\,16.5\,km\,s$^{-1}$ at r\,$\sim$\,2$\times$10$^{16}$\,cm, i.e. about  100 times the inner radius of the (modelled) CSE. This agrees well with previous estimates of the average mass-loss rate during the creation of the CSE around W~Aql. Also, the size of the CSE required to produce the CO(3-2) brightness distribution (Fig.~\ref{bright}) agrees with the results from the model by \citet{mamoetal88}.  

\subsection{Circumstellar morphology}
The channel maps of the CO($J$\,=\,3$\rightarrow$2) emission around W~Aql (see Appendix~\ref{apenchan}) shows a circumstellar CO emission distribution quite similar to that seen around  R~Scl and other stars  \citep[e.g.][]{maeretal16}. The emitting region starts out small and essentially circular at the blue edge of the line, reaches maximum size at the line centre where several prominent arc-like structures are seen, and then  shrinks again toward the red line edge. The small-scale structure seen closer to the line centre is more well-defined, while the emission appears rather smooth closer to the line edges. Unlike R~Scl, W~Aql is not surrounded by a spherically symmetric thin gas shell. Figure~\ref{lineandimage} (middle, right) shows the image averaged over the central channels of the line (Fig.~\ref{lineandimage}, left). The strength of the brightness distributions in  the four different directions over which it was measured vary only marginally  (Sect.~\ref{mdot}), indicating that most of the line emission comes from a smooth extended component. In Fig.~\ref{lineandimage} (middle, right), however, the emission appears slightly brighter and less smooth on the west side and in particular on the  south-west side of the source. There are prominent arc-like structures close to the star (within $\pm$3\arcsec), while these structures become less well-defined further out. In Fig.~\ref{lineandimage}, right, the arc-like structures  are marked since the contrast between the arcs and the background is weak. The full imaged area reaches about 20\arcsec~from the position of the continuum source, but structures beyond $\sim$12\arcsec~are less reliable, due to the rapidly declining image quality when moving away from the image centre.

Figure~\ref{pv90} shows position--velocity (PV) diagrams generated over a narrow slit  (9 pixels)  across the north-south and east-west directions at a position angle (PA) of 0$^\circ$ and 90$^\circ$. The right diagram is generated along the apparent binary axis as it appears in the resolved HST image \citep[at PA=35$^\circ$,][]{ramsetal11}. In the PV diagrams, the asymmetry is even more pronounced, and the arc-like structures on the west and south-west sides (marked in Fig.~\ref{lineandimage}) appear more clearly. The contrast is weak, but there is a hint at an arc periodicity of about 3-4\arcsec\  on both the east and   west sides in the middle PV diagram at PA=90$^\circ$. In the central-channel image in Fig.~\ref{lineandimage}, middle and right, the arc-like structures appear clearest along the north-east--south-west diagonal, or along  the binary axis, which is also shown in the PV-diagram at PA=35$^\circ$  (Fig.~\ref{pv90}, far right) where the arcs are separated by 2-3\arcsec. However, it should be noted that the orientation of the orbit is not known and the direction of the binary axis in the HST image is  only as it appears projected onto the plane of the sky. 

\subsection{Binary interaction model results}
\begin{figure*}
\center
\includegraphics[height=4.2cm]{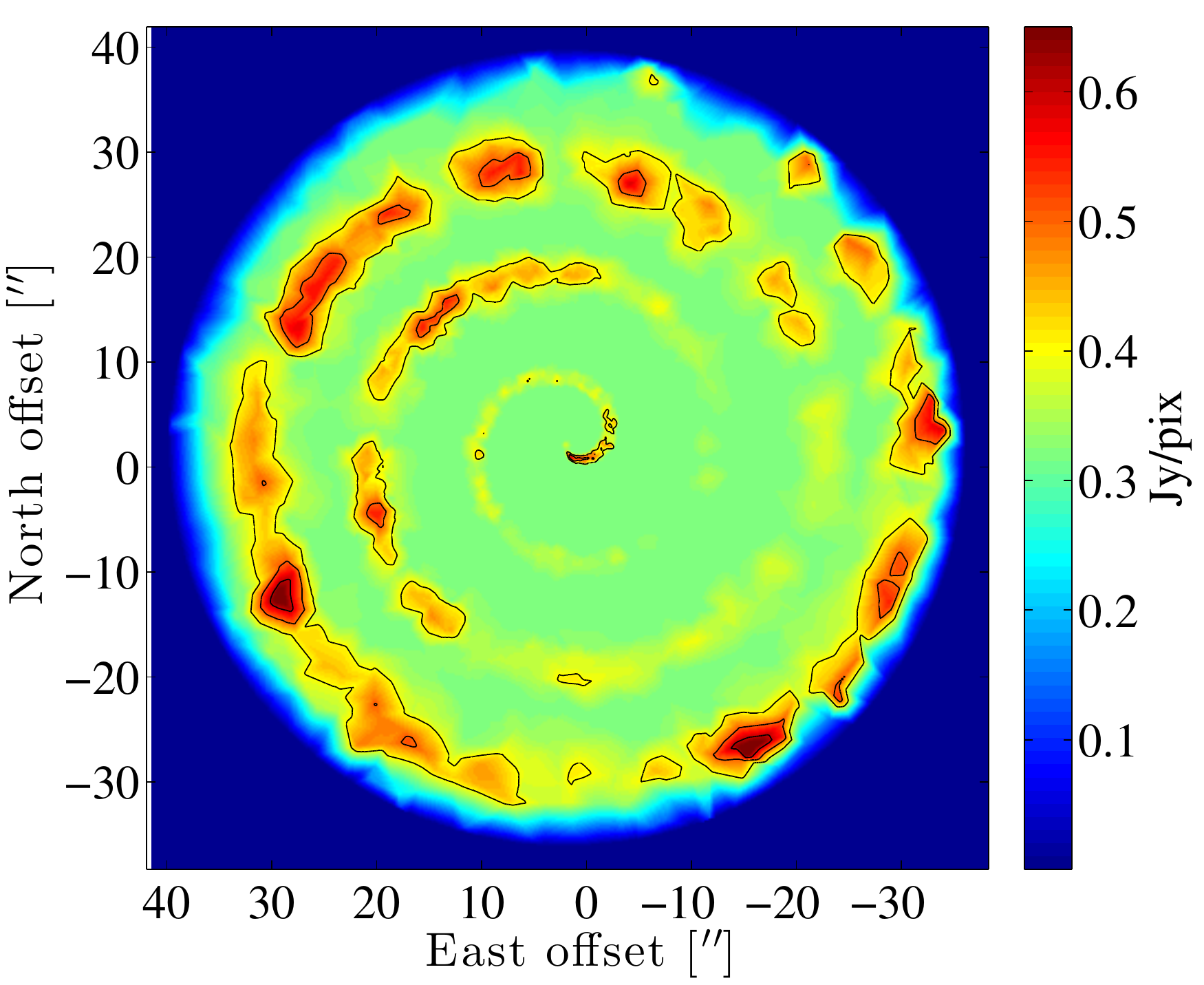}
\includegraphics[height=4.1cm]{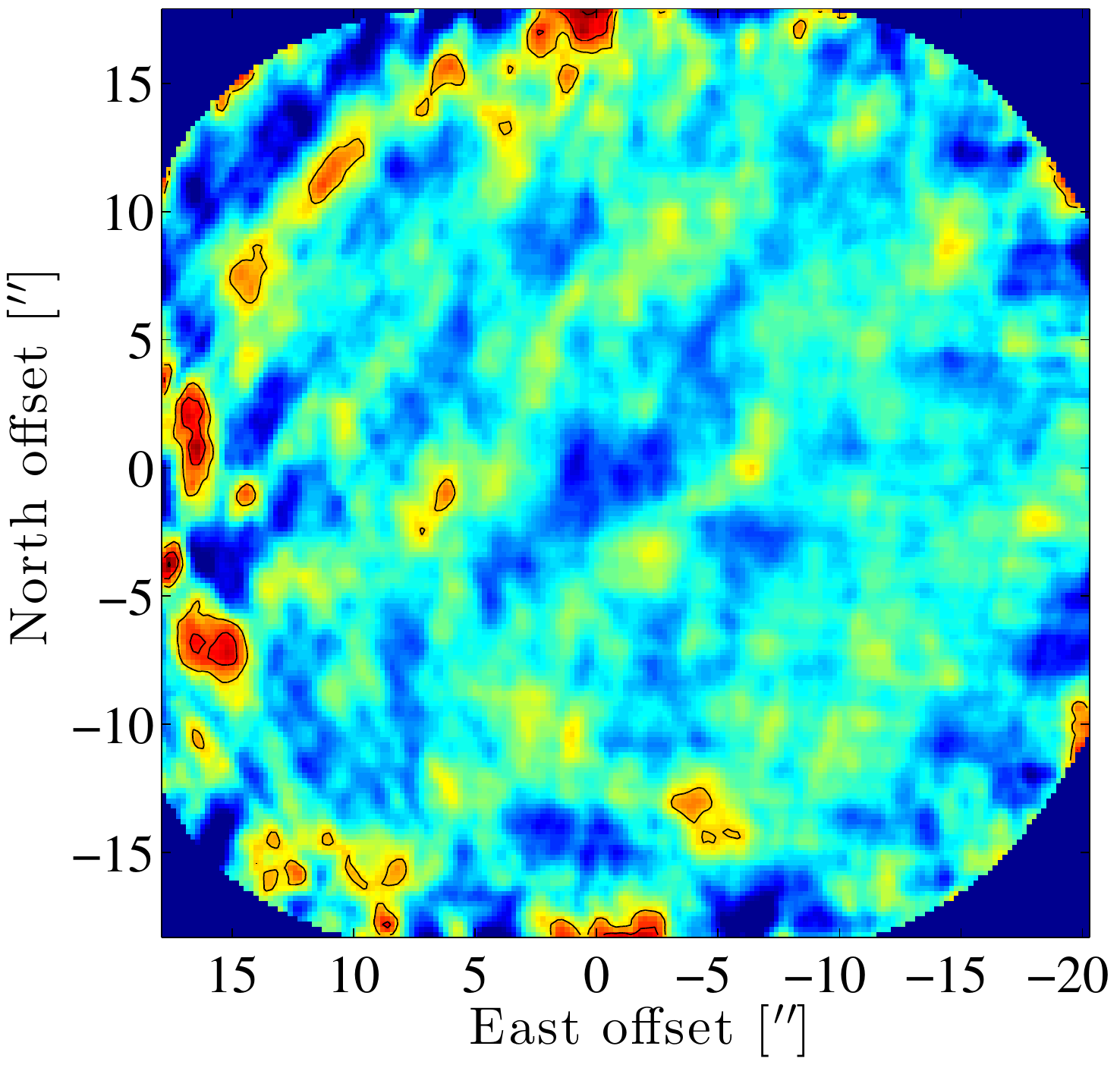}
\includegraphics[height=4.1cm]{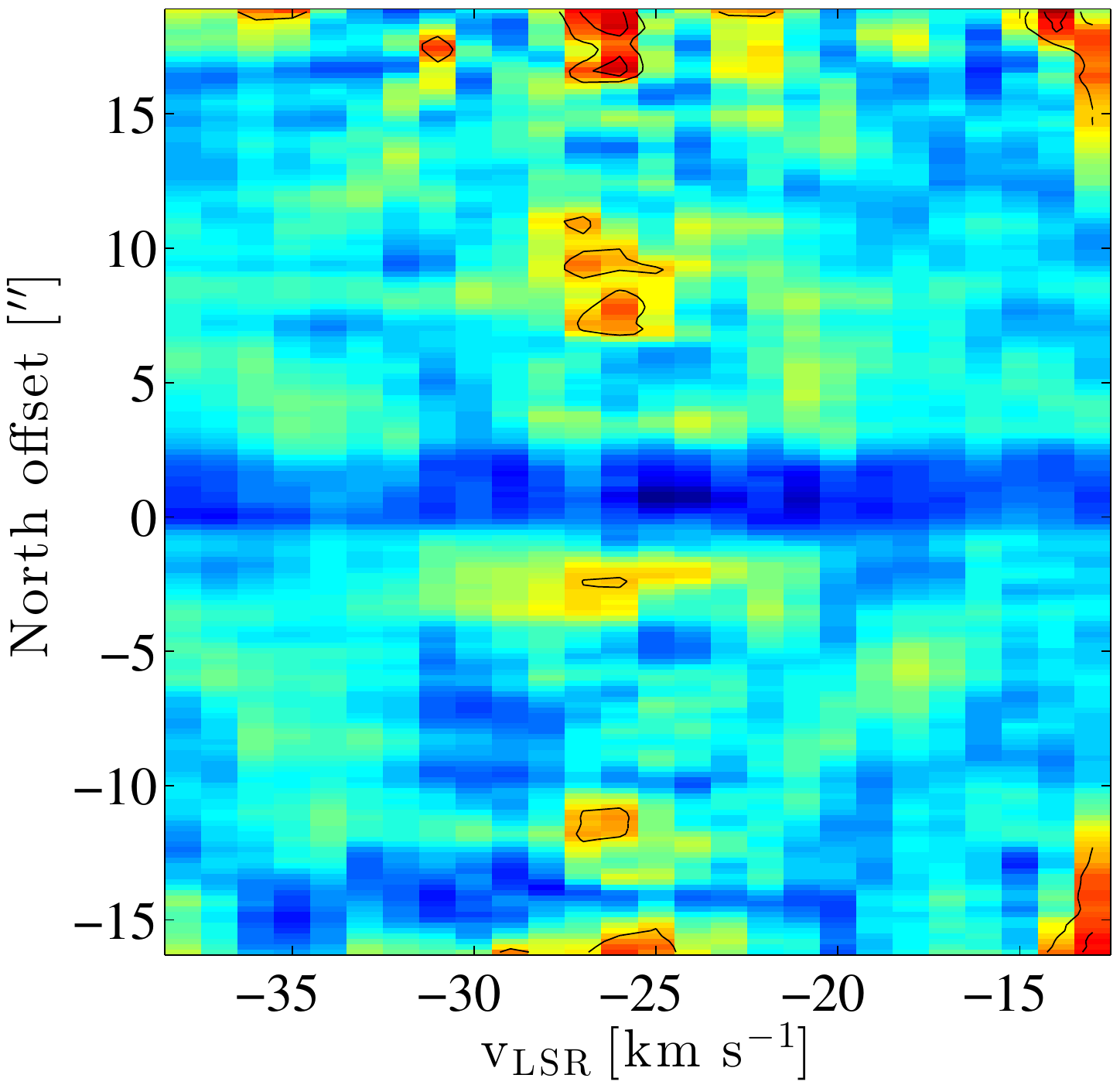}
\includegraphics[height=4.1cm]{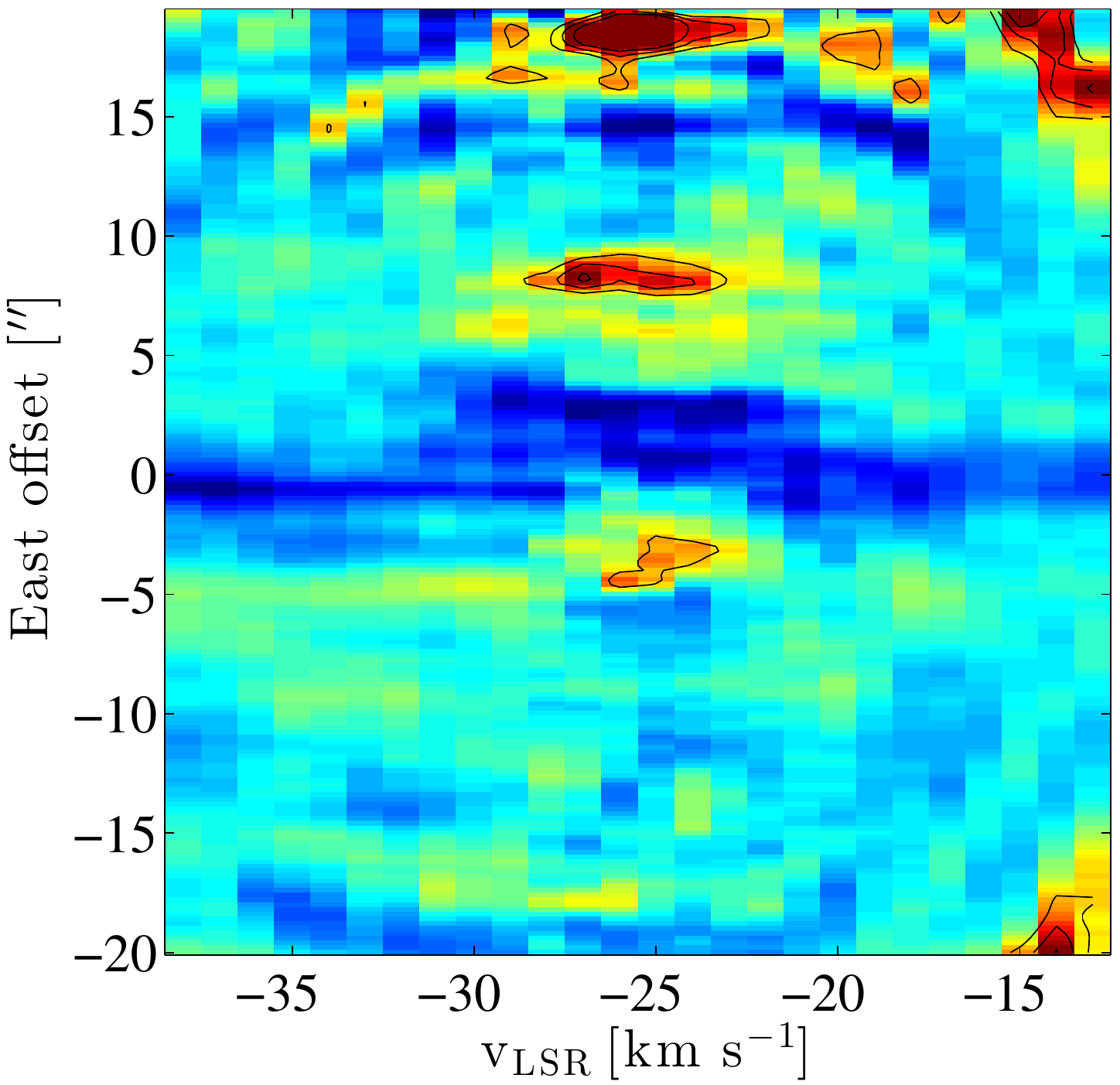}
\includegraphics[height=4.2cm]{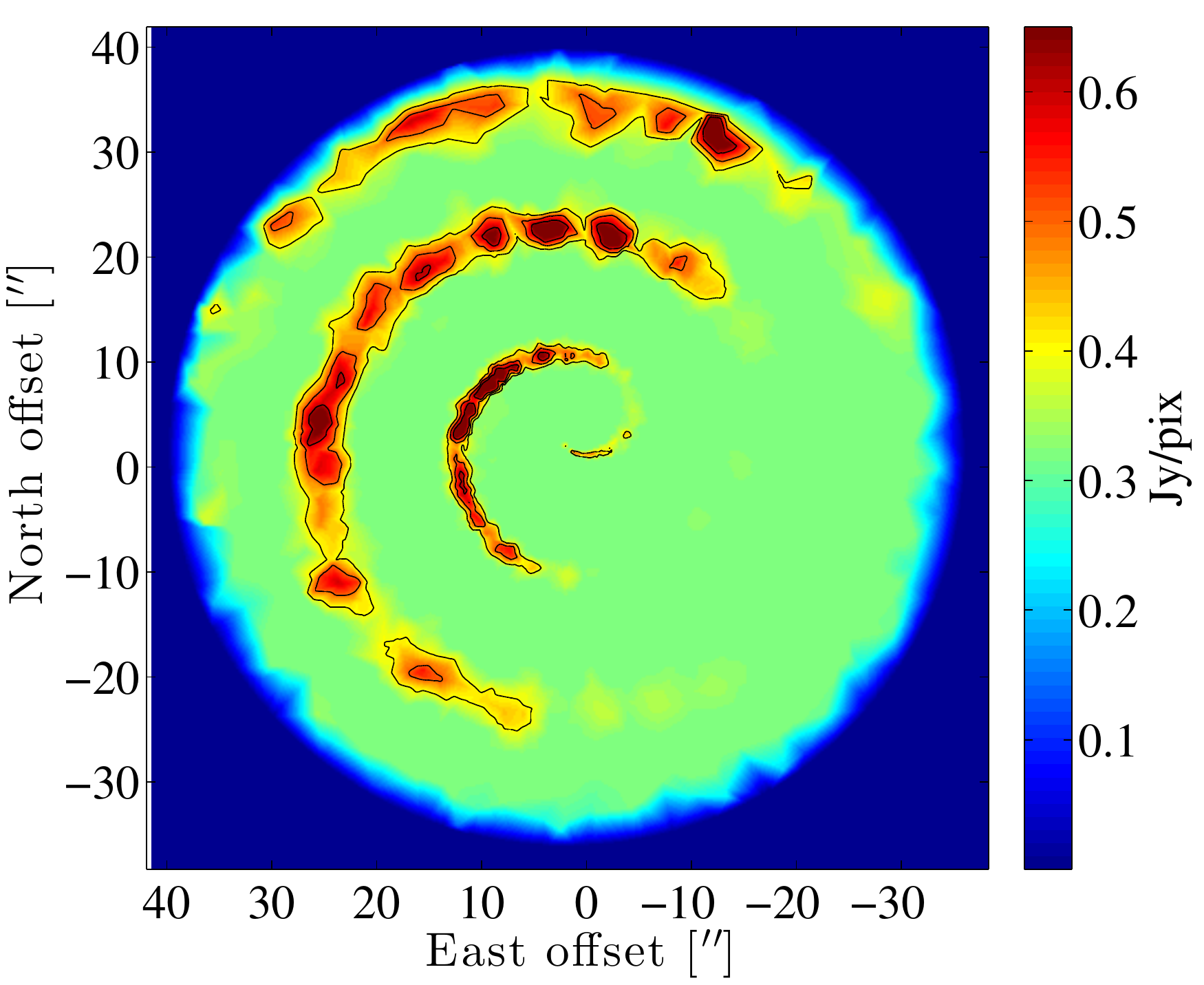}
\includegraphics[height=4.1cm]{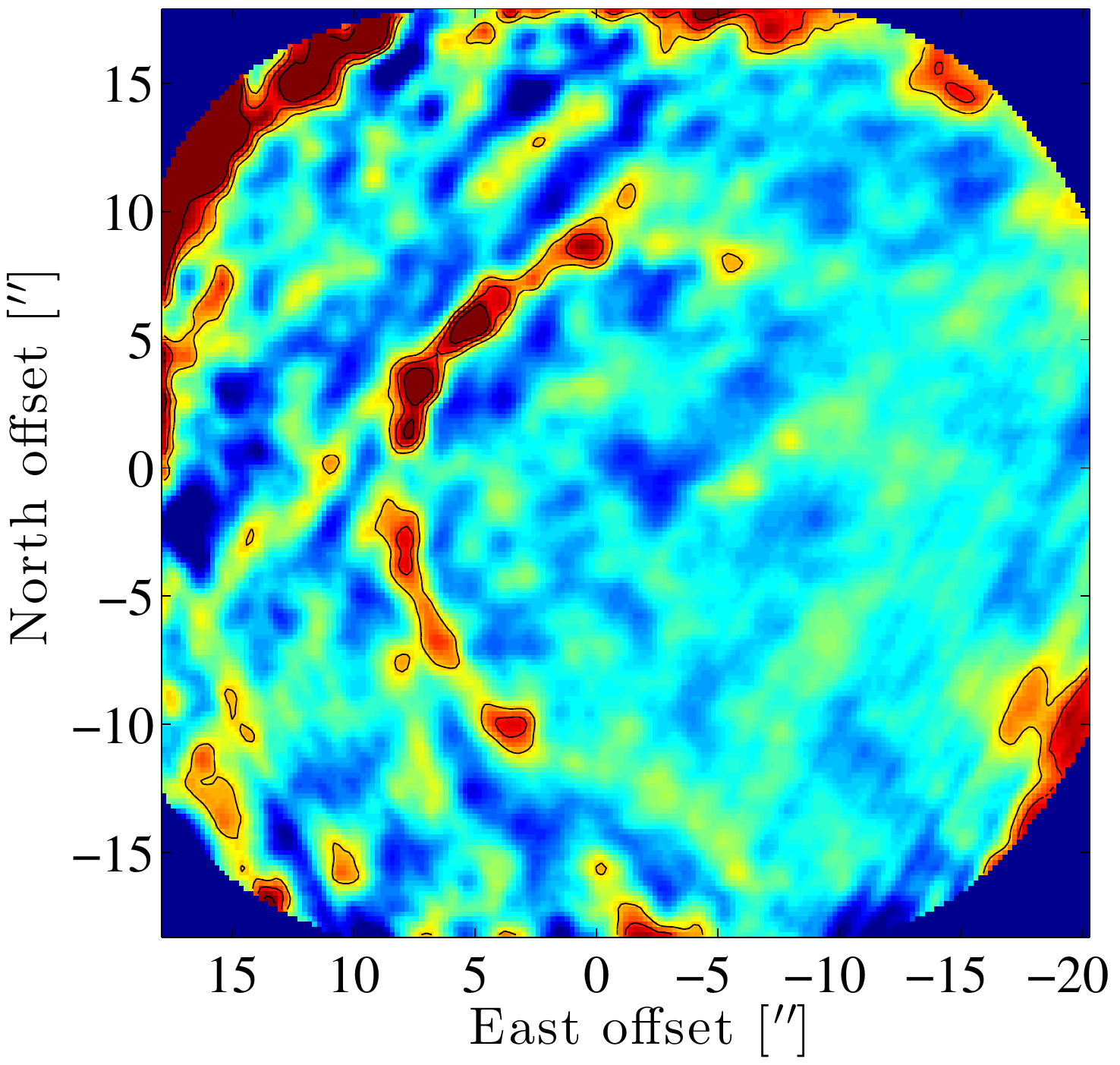}
\includegraphics[height=4.1cm]{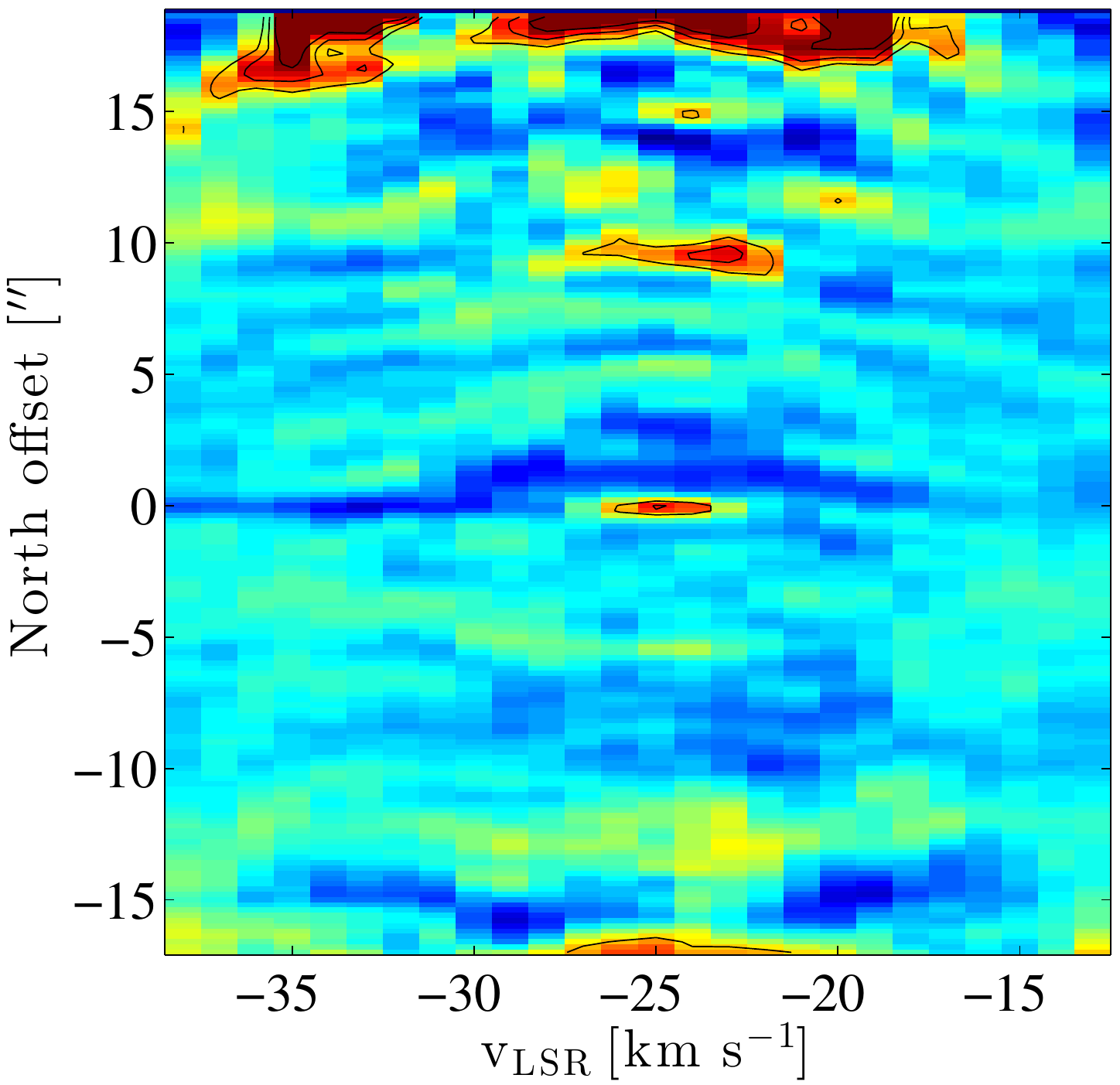}
\includegraphics[height=4.1cm]{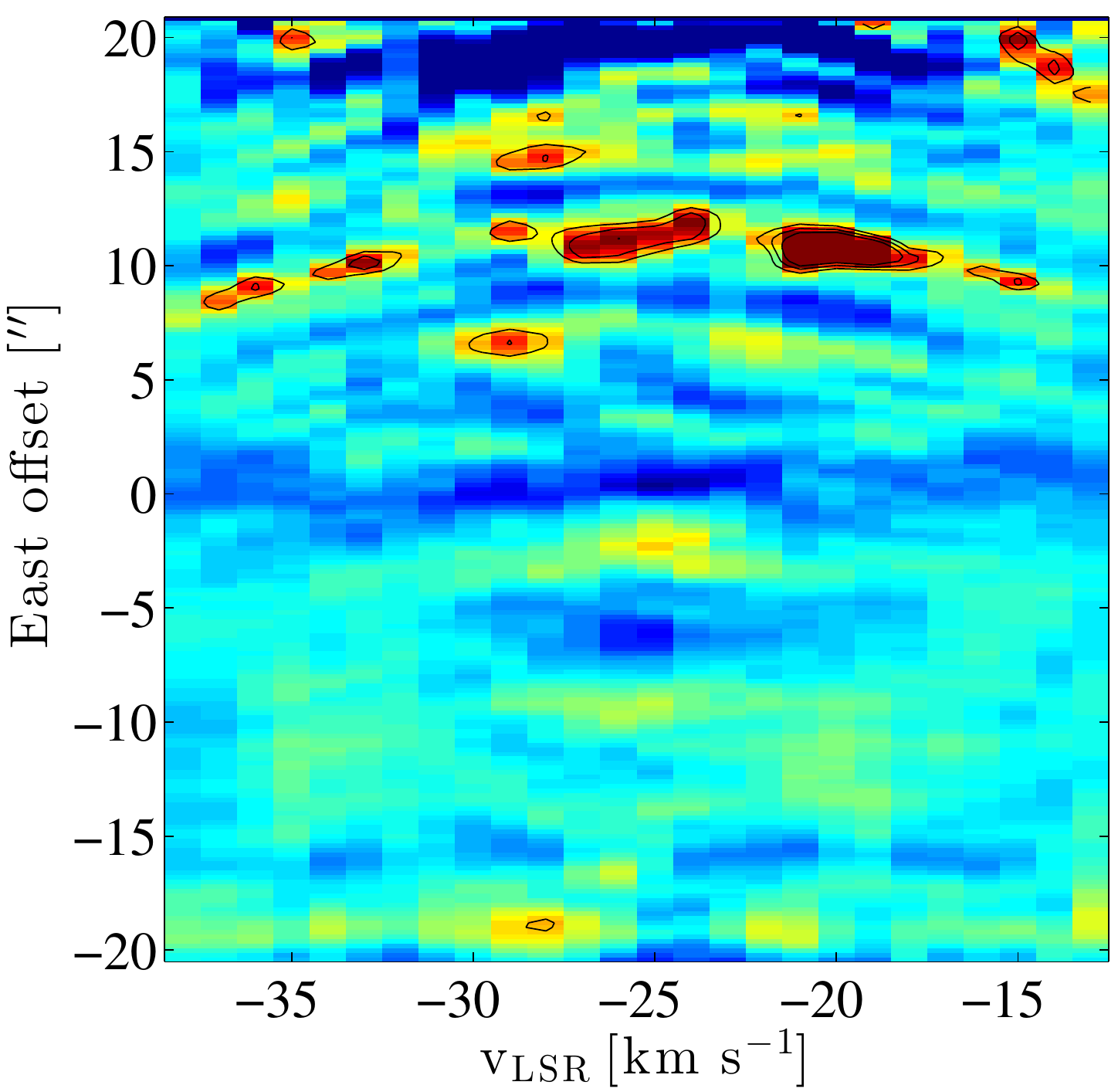}
\caption{Results from the radiative transfer modelling of the two hydrodynamical models with different orbital eccentricities. The upper row shows the results from the $e$=0.2 model. The lower row shows the results from the $e$=0.6 model. The far left image shows the output of the full model over the central channels (analogous to Fig.~\ref{lineandimage}) without taking the response of the interferometer into account. The middle left image shows the output from the ALMA simulator. The two right images show PV-diagrams generated perpendicular to the model binary axis (PA=0$^\circ$) and along it (PA=90$^\circ$), respectively. The contours are drawn at 35, 50, 75\%-fractions of the peak emission.}
\label{e0pt6}
\end{figure*}

Figure~\ref{e0pt6} shows the results from the LIME radiative transfer model. The far left image is the output image modelled at 0\farcs2 resolution before being convolved with the response of the interferometer, and generated over the central channels (analogous to Fig.~\ref{lineandimage}). As seen in previous publications \citep[][]{mastmorr98,mohapods12,kimetal15}, the binary companion will shape a spiral pattern in the outflowing CSE of the AGB star. For a system with the same properties as W~Aql (orbital period=1394/5 years, $v_{\rm{exp}}$=16.5\,km\,s$^{-1}$), the arc separation will be 10--12\arcsec. The emission contrast between the arcs and the interarc regions is a factor of $\leq$2 for the $e$=0.2 model and $\leq$3 for the $e$=0.6 model. The model binary axis is along the east-west direction.

The middle left image in Fig.~\ref{e0pt6} shows the emission as it would be seen by ALMA using the same main array, ACA, and TP configurations as used for the W~Aql observations in Cycle 1 (we note that the  image size is different to that in the far left image). The full channel maps are shown in Figs.~\ref{m02channelmaps}.2 and \ref{mchannelmaps}.3. This shows particularly well the difficulty in recovering weakly contrasting features even with ALMA, and also how insufficient cleaning can introduce false features, which  emphasizes that weak emission structures should not be overinterpreted. This image was generated using the CASA task simanalyze without the careful, iterative, multiscale cleaning used when imaging the real data, and therefore the fidelity is worse than in the W~Aql images in Figs~\ref{lineandimage}, \ref{pv90}, and \ref{channelmaps}.1.

The two right images show PV diagrams generated from the middle left image taking the interferometer response into account. The middle right diagram shows the emission distribution along an axis perpendicular to the model binary axis (PA=0$^{\circ}$);  the far right diagram shows the emission along the binary axis (PA=90$^{\circ}$). The diagrams show the brightest (``real'') arcs in the image together with some weak (false) features. When the observational set-up from Cycle~1 is used, the PV diagrams show that only one or two  arc-like structures ($\sim$10\arcsec)  with larger separations can be detected, and that a larger area would have to be mapped in order to draw firm conclusions about the eccentricity of the orbit of the two resolved companions in the W~Aql system.

The upper row of Fig.~\ref{e0pt6} shows the results from the $e$=0.2 model, and the lower row shows the results from the $e$=0.6 model. A difference compared to a circular orbit \citep[see e.g.][]{maeretal12} introduced when varying the eccentricity is that the distribution of circumstellar material across the rotation axis will be asymmetric and there will be  more material on the periastron side. The contrast across the rotation axis will grow with eccentricity, as will the contrast between the arcs and the interarc regions (see above and Fig.~\ref{e0pt6}). Also, as seen in Fig.~\ref{e0pt6}, the arc separation on the apastron side is larger when the eccentricity is larger.

%__________________________________________________________________

\section{Discussion}
\label{dis}

\subsection{Comparison to previous resolved observations of the CSE around W~Aql}
The dust shell detected by \citet{tateetal06} would not be resolved by the ALMA observations, but it is possible that it is an inner extension of the arcs seen close to the star (see  Fig.~\ref{lineandimage}).

\begin{figure}
\center
\includegraphics[width=7.5cm]{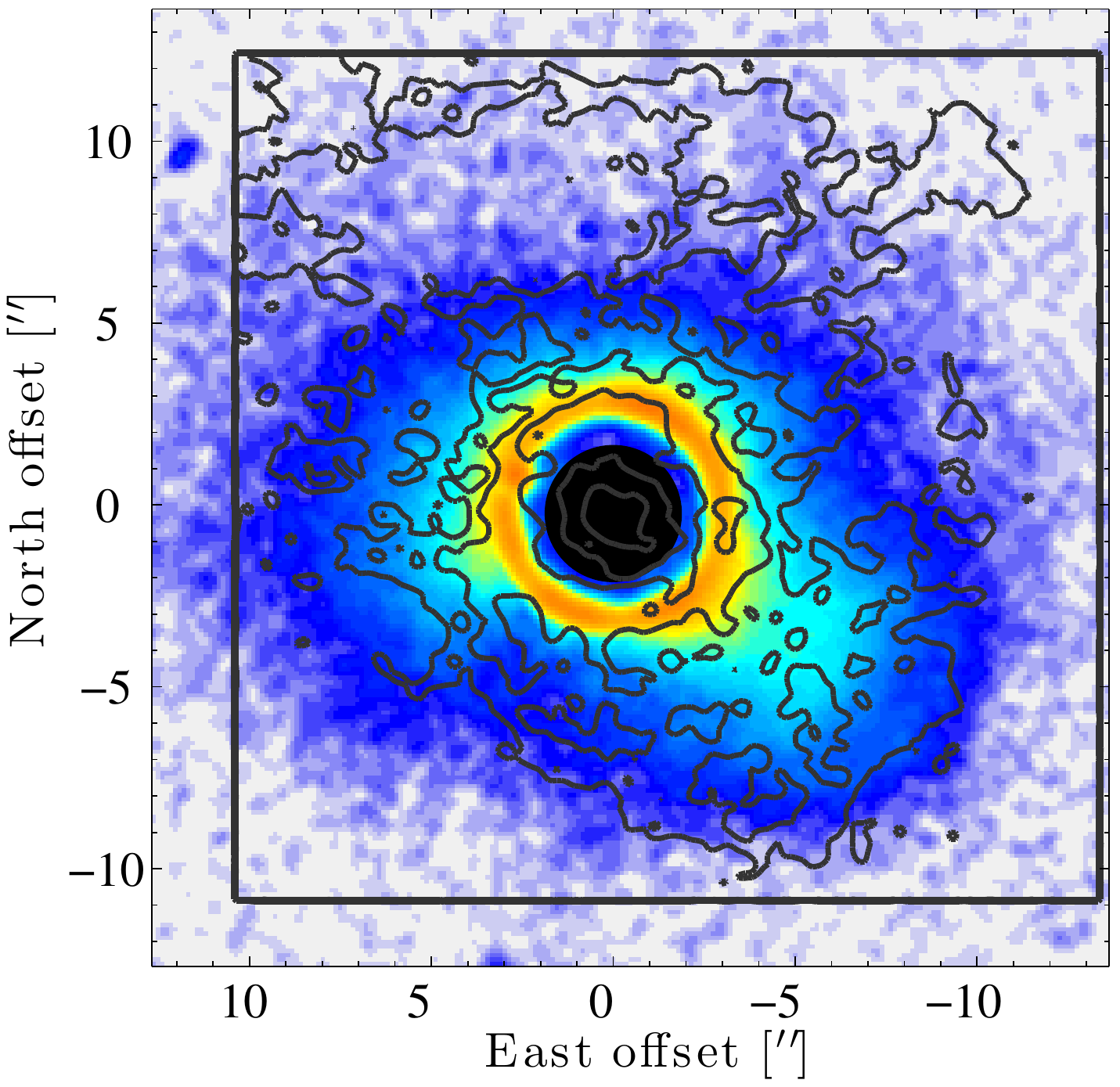}
\caption{ CO($J$=3$\rightarrow$2) emission from ALMA (grey contours) overlayed on the R-band dust-scattered emission image from PolCor \citep{ramsetal11}.}
\label{polcor}
\end{figure}

The dust-scattered emission over about 1\arcmin~around W~Aql imaged by \citet[][using the PolCor instrument]{ramsetal11} showed that the circumstellar dust is distributed asymmetrically around the star with more dust on the south-west side. Figure~\ref{polcor} shows the CO emission contours from Fig.~\ref{lineandimage} overlayed on the PolCor image.  The figure shows that the dust and CO gas distributions overlap on these scales, and that the south-west emission (density) enhancement appears in both images. The small-scale structures, or arcs, seen in the central-channel CO line image (Fig.~\ref{lineandimage}) are not seen in the scattered-light R-band image since the arcs, which appear at different velocity channels (Appendix A), are smeared in the dust emission image.

The large-scale morphology of the circumstellar dust around W~Aql has been studied in detail by \citet{mayeetal13} using images from Herschel/PACS. The dust distribution appears elliptical with brighter emission reaching further  from the star along the east--west axis (to 40--45\arcsec$\sim$16000\,AU at D=395\,pc). More diffuse emission extends further on the west side, but is not apparent on the east side where the emission appears truncated by a large arc-like feature, possibly a bow-shock \citep[although not aligned with the space motion as calculated by][]{mayeetal13}. The inner region of the PACS images (overlapping with the region mapped by ALMA) shows a slight elongation to the south-west, but there is also bright emission on the east side which does not have an apparent counterpart in the gas emission images. By assuming an orbital period $\leq$1000\,yrs and a wind velocity of 20\,km\,s$^{-1}$, \citet{mayeetal13} estimate an expected arc separation of 12\farcs5 and try to fit the PACS data with different Archimedean spirals with some success. Figure~\ref{pv90} (right) shows the PV diagram of the ALMA data along PA=0$^\circ$, i.e. the north-south direction, and although the emission contrast is very weak and the image fidelity  decreases beyond $\sim$12\arcsec, recurring arcs of a similar periodicity to that suggested by \citet{mayeetal13} are seen.

\subsection{Comparison to the binary interaction models}
The hydrodynamical models using the known parameters of the W~Aql system as input, and assuming a constant mass-loss rate and expansion velocity, produce a large spiral pattern with $\sim$10\arcsec~spiral-arm separation. The main difference between the two models with different eccentricities is the contrast between the arcs of the spiral on the apastron versus periastron side. From the $e$=0.2-model there is almost no contrast between the two sides, while the $e$=0.6 model essentially lacks arcs on the apastron side (Fig.~\ref{e0pt6}). The inner 10\arcsec~of the W~Aql CSE show a clear contrast between the west and east sides, with smoother declining emission on the east side and arc-like structures at separations of 2-3'' perpendicular to the apparent binary axis (Figs.~\ref{lineandimage}, middle and right, and \ref{pv90}, left and right). This close arc pattern is not seen in the models because no process that could produce density variations on a timescale of $\sim$200 yrs was included. 

There is a hint of arc-like structures at larger separations in the observations (best seen in the PA=90$^{\circ}$ PV diagram, Fig.~\ref{pv90}, middle), but the contrast between the smooth component and the arcs is very weak (about a factor of 2, in agreement with the $e$=0.2 model) and they appear in the outer regions where the image is less reliable. There is no apparent asymmetry between the brightness of arcs with larger separations when comparing the north and south sides, but it is difficult to evaluate using the current data. As already mentioned above, a deeper and larger image would be necessary in order to draw firm conclusions about the orbit eccentricity of the known companion in the W~Aql system.

\subsection{Circumstellar gas envelope of W~Aql}
To summarize, the CO(3-2) line emission from the W~Aql system is entirely dominated by a smooth symmetric component (Fig.~\ref{bright}) from the gas envelope generated by an average mass-loss rate estimated to $\dot{M}$\,=\,3.0$\times$10$^{-6}$\,M$_{\odot}$\,yr$^{-1}$, i.e. rather high for an S-type star \citep{ramsetal09}. Superimposed on  the smooth component, there appears to be a double arc pattern in the CSE of W~Aql. It is likely that the  arcs at greater separations (10\arcsec)  are caused by the gravitational pull on the circumstellar gas by the known companion at 0\farcs46. There is no apparent asymmetry between the arc-like structures at greater separations on different sides of the system, as seen in the $e$=0.6 model, and the contrast between the arcs and interarc region agrees with the output from the $e$=0.2 model, but this is also difficult to evaluate with the current data set. The closer arcs,  marked in Fig.~\ref{lineandimage} (right), could be formed via a number of different processes: a second closer companion (or a massive planet) with
an eccentric orbit, a recent change in the wind velocity (which would have to
decrease by a factor of 3-4), mass-loss-rate variations, etc.. Density variations over the same timescale are also seen in the carbon star IRC+10216 \citep[e.g.][]{leaoetal06}. In that star, the optically thick CSE could block heat from escaping the dust formation region and thereby stop grain formation temporarily, causing episodic mass loss on the same timescales \citep{eriketal14}; however, in the S-type CSE of W~Aql, this is a less likely explanation. The pulsation/convection timescale is much shorter than the arc-formation timescale, and therefore  is probably not the cause of the arcs at smaller separations. The inner arc structure is clearly asymmetric, and co-incident with the previously imaged dust emission enhancement. It is possible that the disappearing SiO masers of W~Aql \citep{ramsetal12} hint that a disruption of the inner CSE ($<$ 5 R$_{*}$) is occurring, which could lend weight to the presence of an additional, closer companion. If this is the case, then the third body approached sufficiently close to the star to disrupt the SiO maser zone during 2010-11, but this is highly speculative. Closer imaging and monitoring of the system to look for dust formation variations \citep{ohnaetal16} or a third system body would be very interesting and would help to understand the shaping of the W~Aql CSE. 

%__________________________________________________________________

\section{Conclusions}
\label{conc}
The observations of W~Aql are part of a project with the aim of achieving a better understanding of how a binary companion will shape the expanding CSE around a mass-losing AGB star by observing stars with a known binary companion. The idea is that this would provide better constraints for the models of the interaction, as opposed to the more common strategy of inferring the companion from the shape of the CSE alone. However, even when the separation is known, this has proven perhaps more complicated than initially hoped  \citep{ramsetal14,doanetal17}, and additional information about the systems,  orbits, more exact distances, properties of the companion, etc., are required to fully understand the systems and to explain the shaping processes. 

In the W~Aql system, the circumstellar gas distribution is moderately affected by the interaction with the companion, which is also to be expected given the relatively large separation between the two stars (close to 200\,AU at 395\,pc). Instead, the distribution of the circumstellar material is mostly smooth, and the envelope has been formed by an average mass-loss rate of $\dot{M}$\,=\,3.0$\times$10$^{-6}$\,M$_{\odot}$\,yr$^{-1}$. The arcs with larger separations ($\sim$10\arcsec) are probably due to the known companion and, within the limitations of the data, consistent with a circular or possibly low-eccentricity orbit. An even weaker density contrast, as would be expected from a companion at  an even larger separation, would be very difficult to detect, even with ALMA. In addition, the inner region of the CSE around the W~Aql system is asymmetric and shows an arc pattern with a separation of 2-3\arcsec, predominately on the south-west side of the sources and overlapping with the previously mapped dust emission. The physical processes behind the south-west asymmetry and the arcs at smaller separations are not known, and cannot be easily linked to any known properties of the system.

\begin{acknowledgements}
The authors would like to thank the staff of the Nordic ALMA ARC node for their support, availability, and continuous efforts in helping produce maximum quality data products. This paper makes use of the following ALMA data: ADS/JAO.ALMA\#2012.1.00524.S. ALMA is a partnership of ESO (representing its member states), NSF (USA), and NINS (Japan), together with NRC (Canada) and NSC and ASIAA (Taiwan), in cooperation with the Republic of Chile. The Joint ALMA Observatory is operated by ESO, AUI/NRAO, and NAOJ. SM is grateful to the South African National Research Foundation (NRF) for a research grant. WV acknowledges support from ERC consolidator grant 614264. TD acknowledges support from the ERC consolidator grant 646758 AEROSOL and the FWO Research Project grant G024112N. GQL acknowledges support from the European Research Council under the European Union's Seventh Framework Programme (FP/2007-2013) / ERC Grant Agreement n. 610256 (NANOCOSMOS).

\end{acknowledgements}

\bibliographystyle{aa}
\bibliography{waql}

\appendix
\section{Channel maps}
\label{apenchan}

\begin{figure*}
\flushleft
\label{channelmaps}
\includegraphics[height=3.9cm]{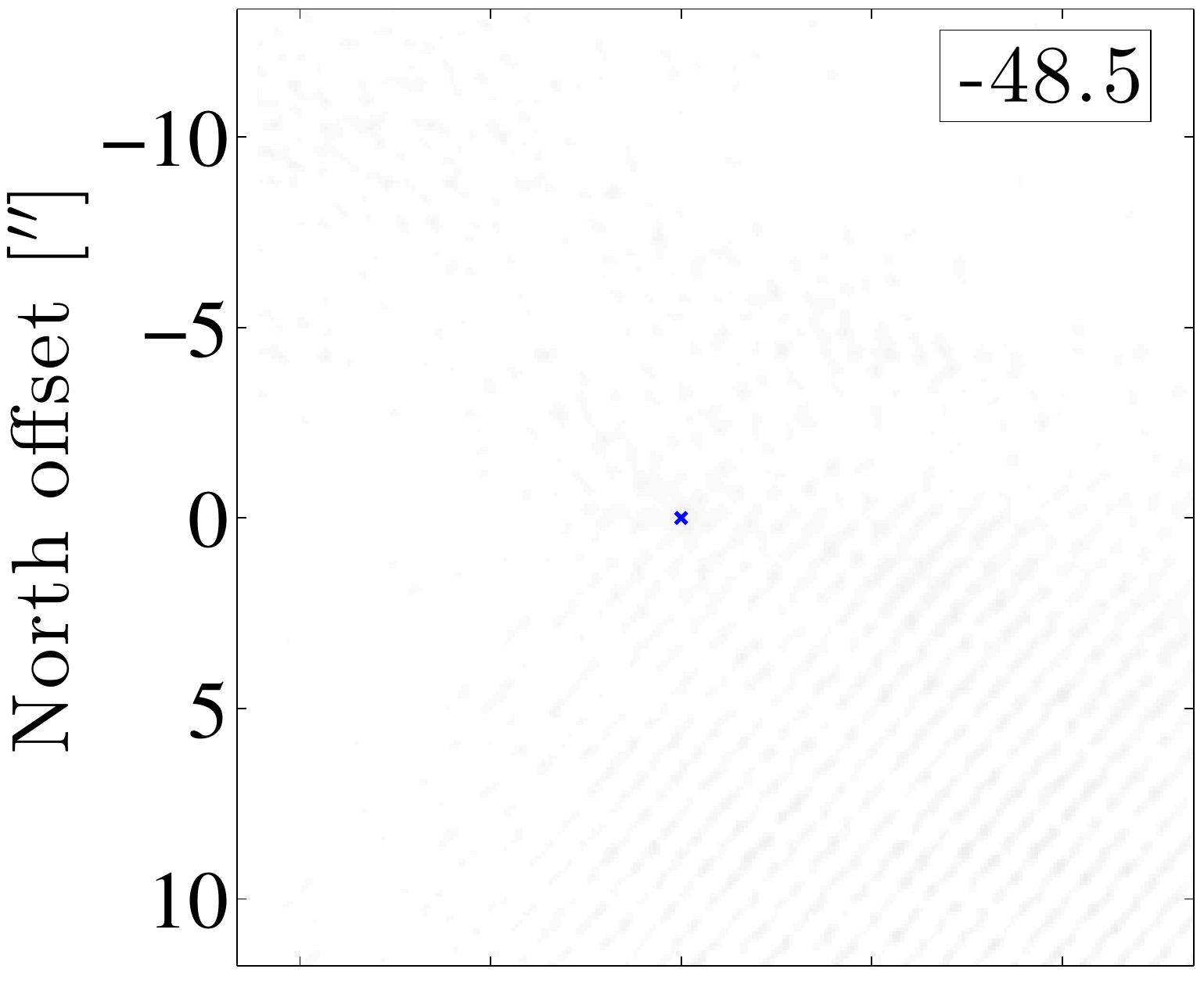}
\includegraphics[height=3.9cm]{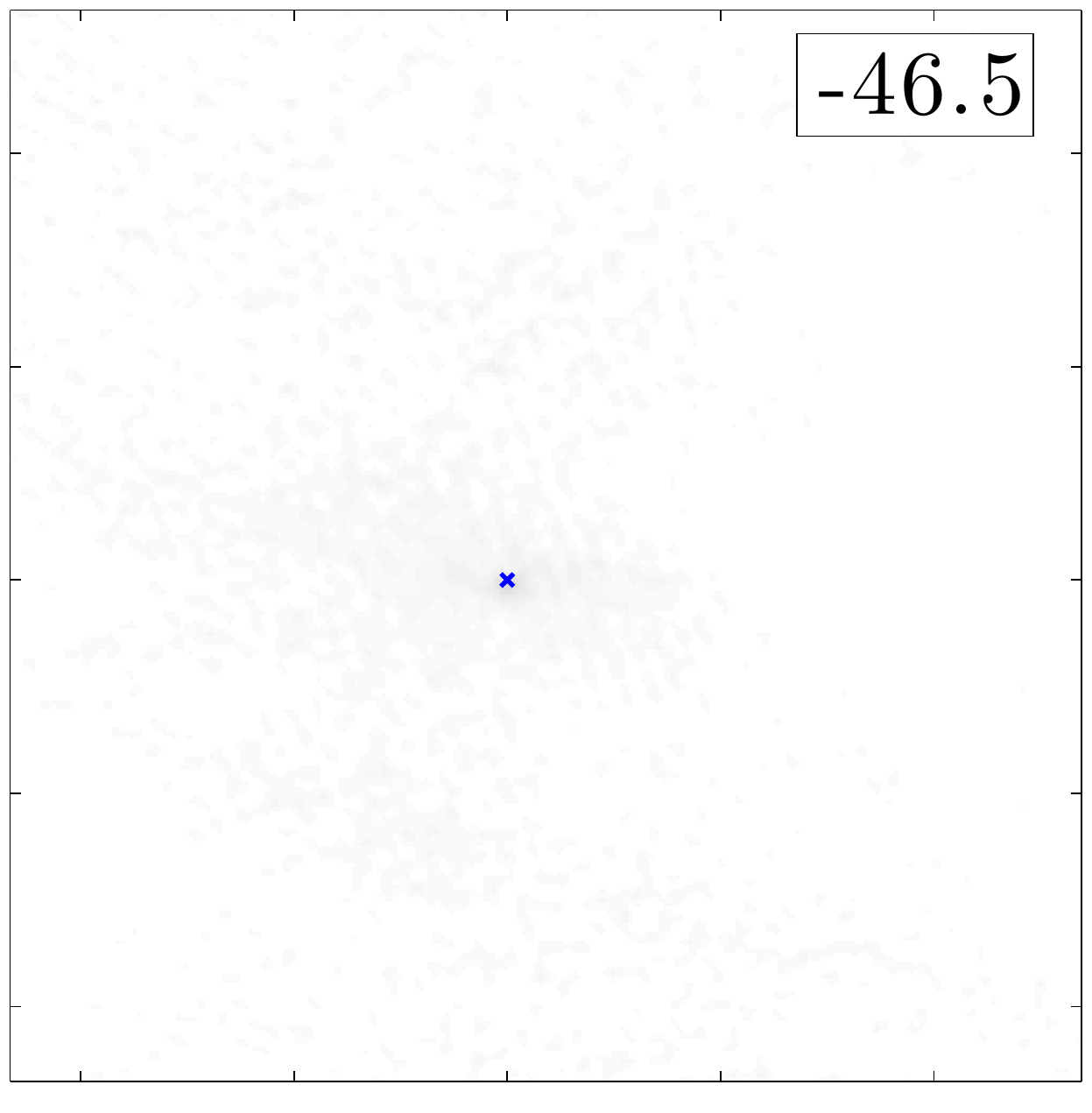}
\includegraphics[height=3.9cm]{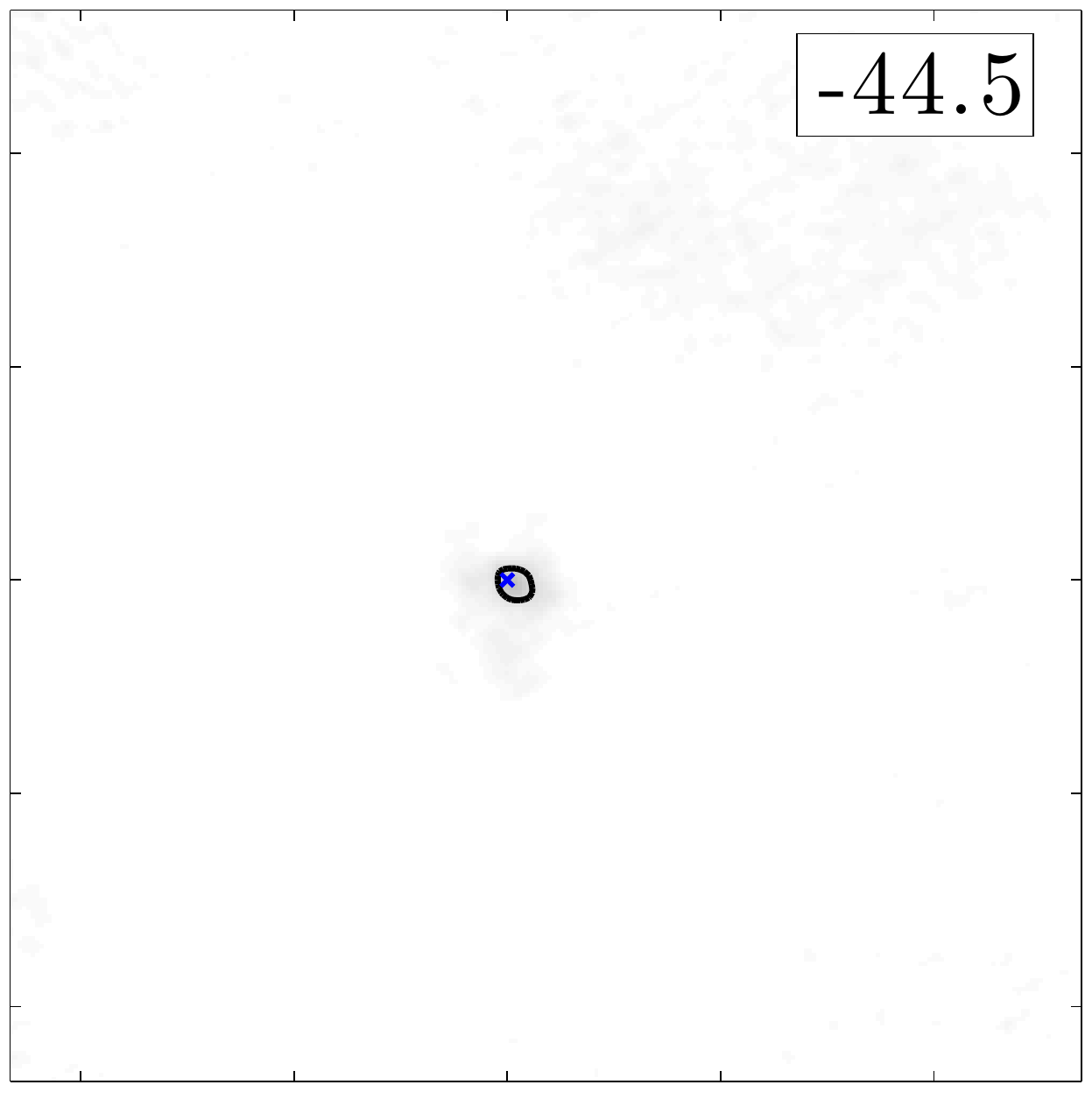}
\includegraphics[height=3.9cm]{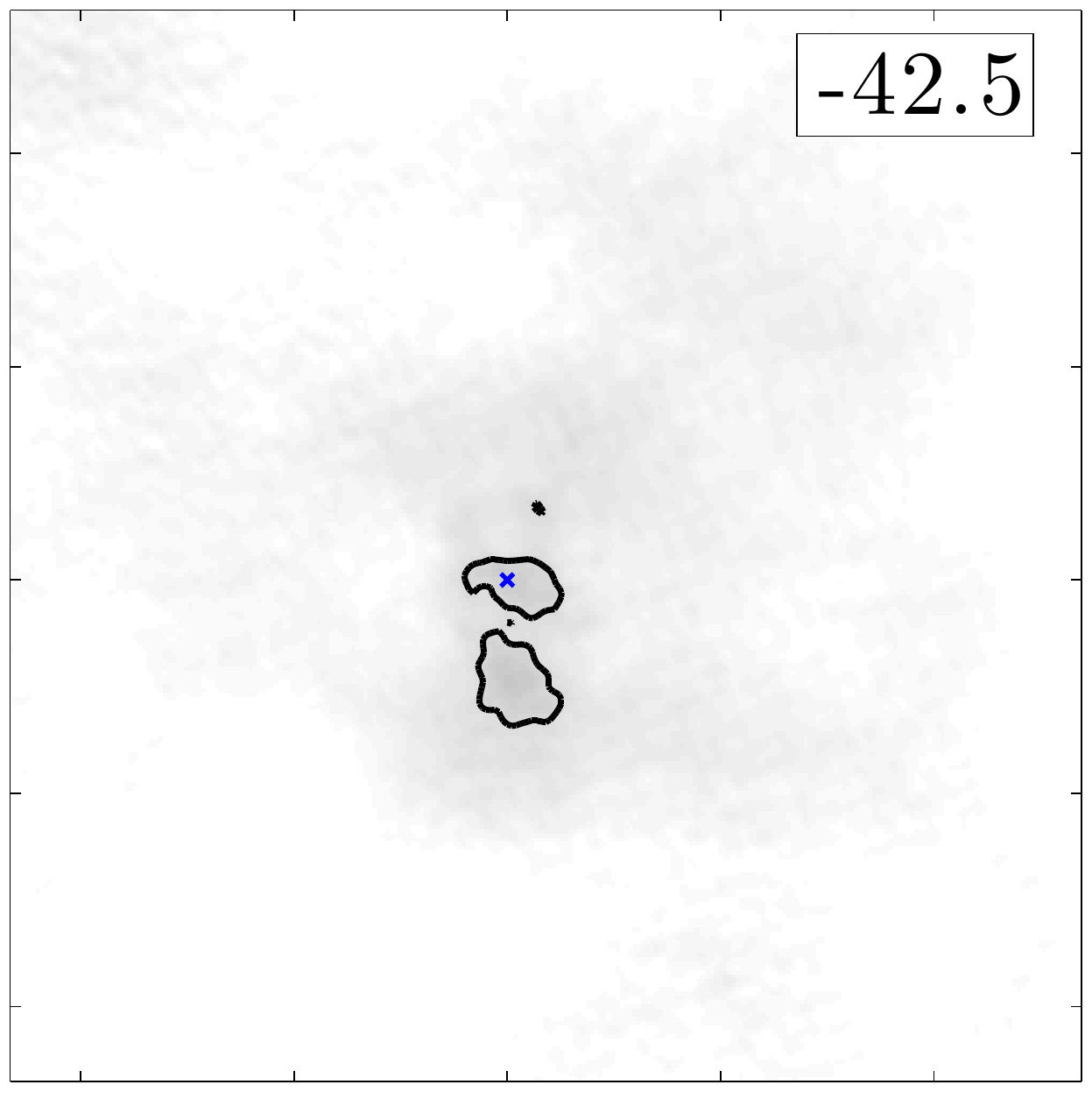}
\includegraphics[height=3.9cm]{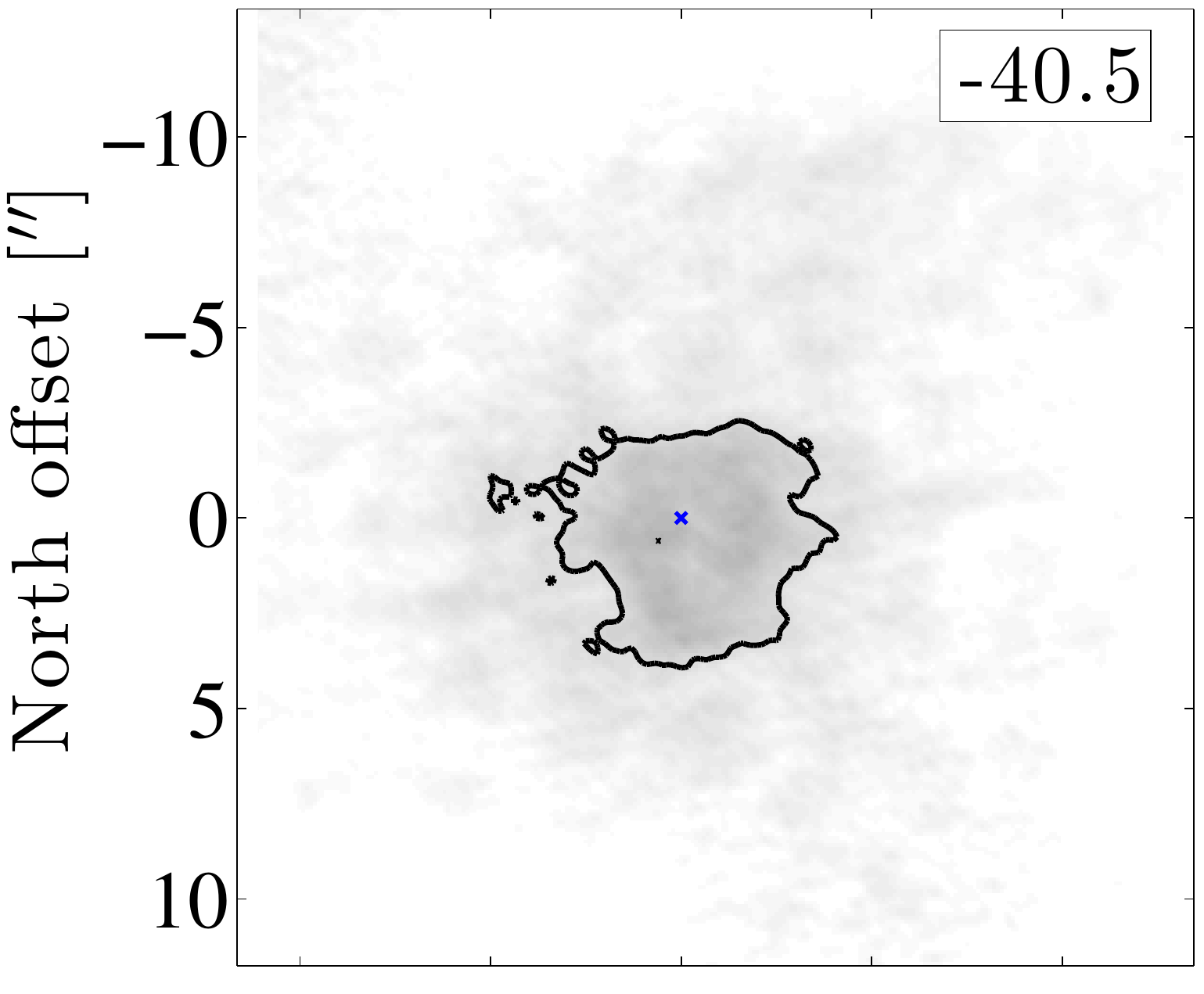}
\includegraphics[height=3.9cm]{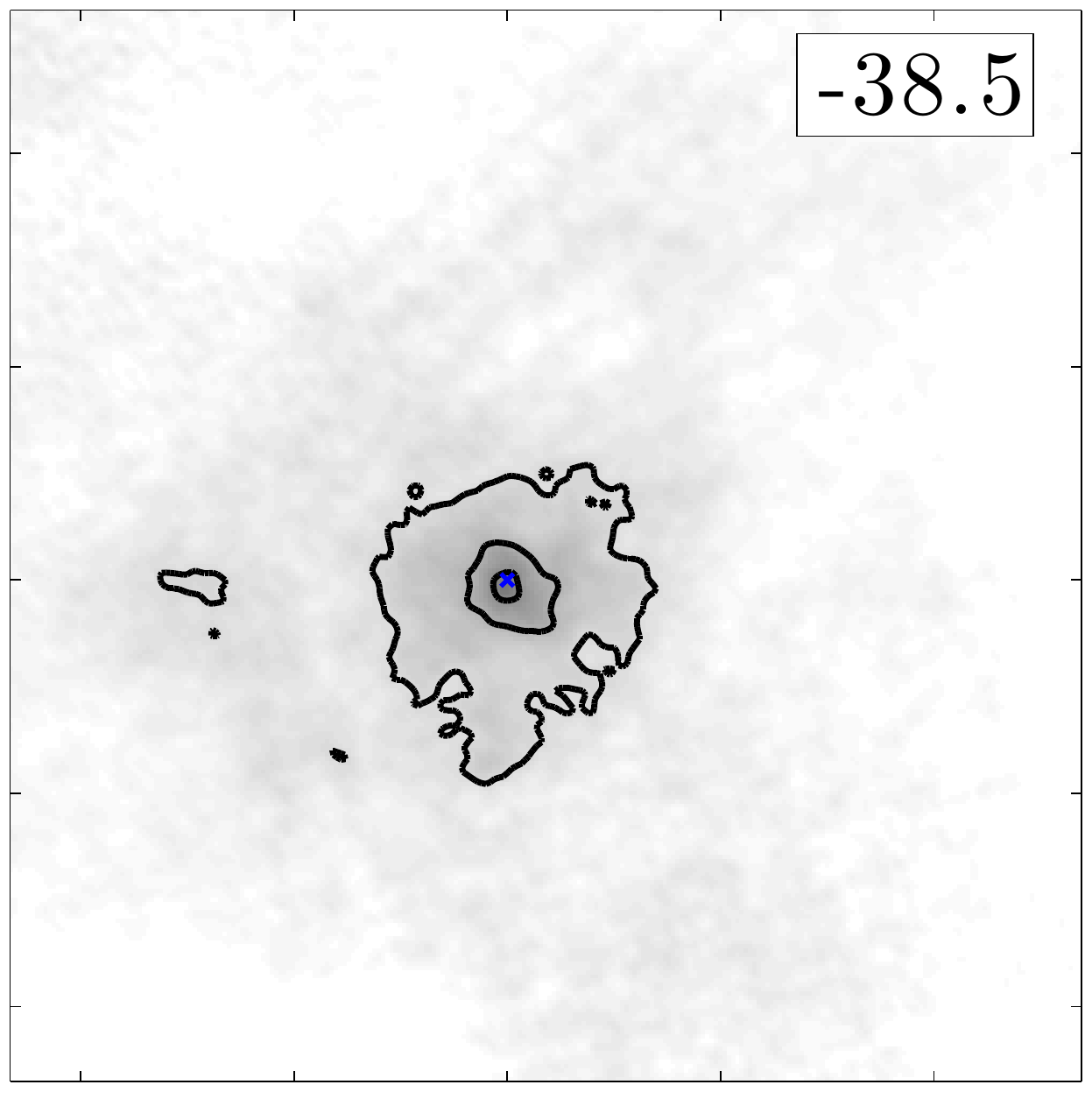}
\includegraphics[height=3.9cm]{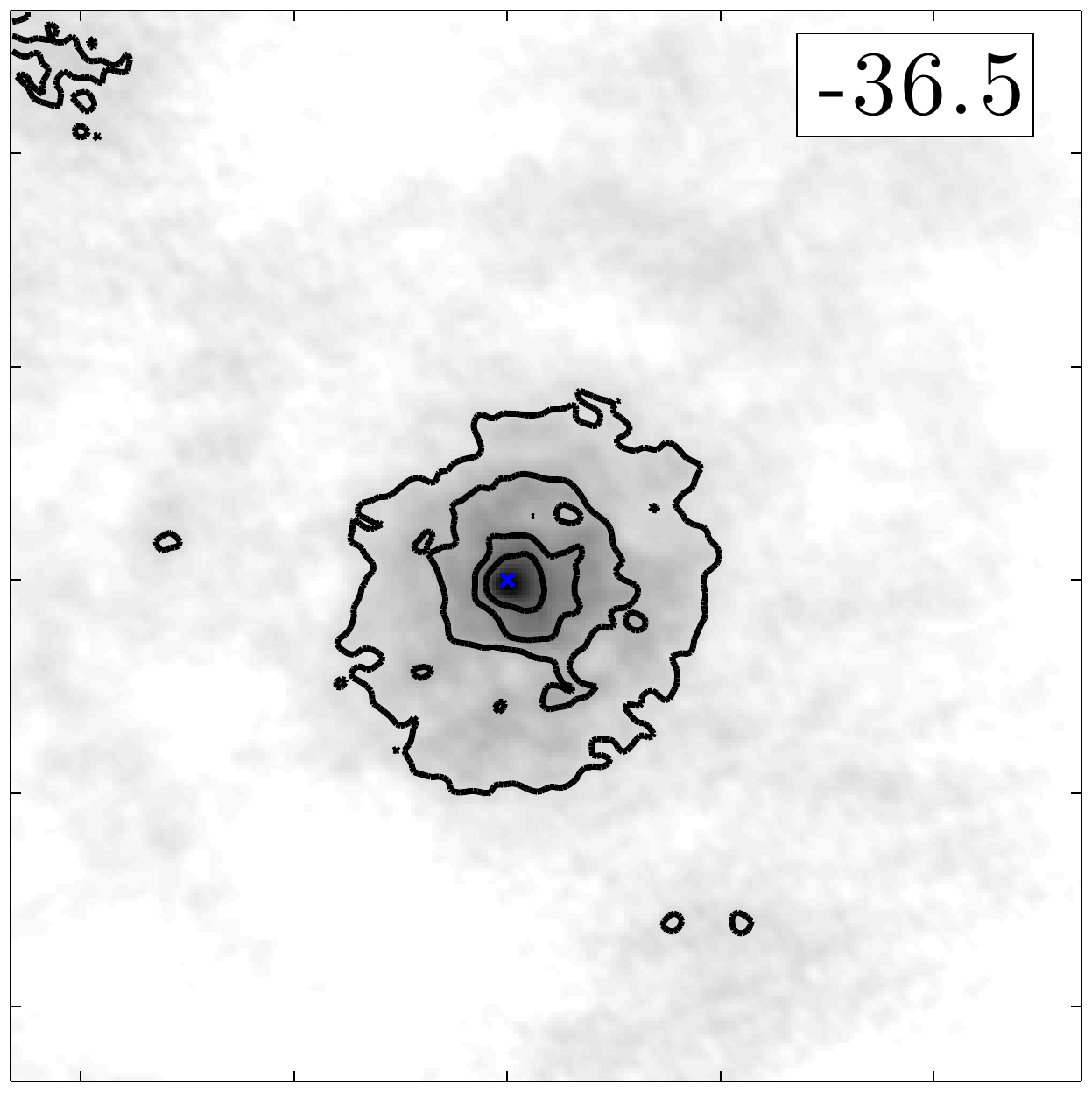}
\includegraphics[height=3.9cm]{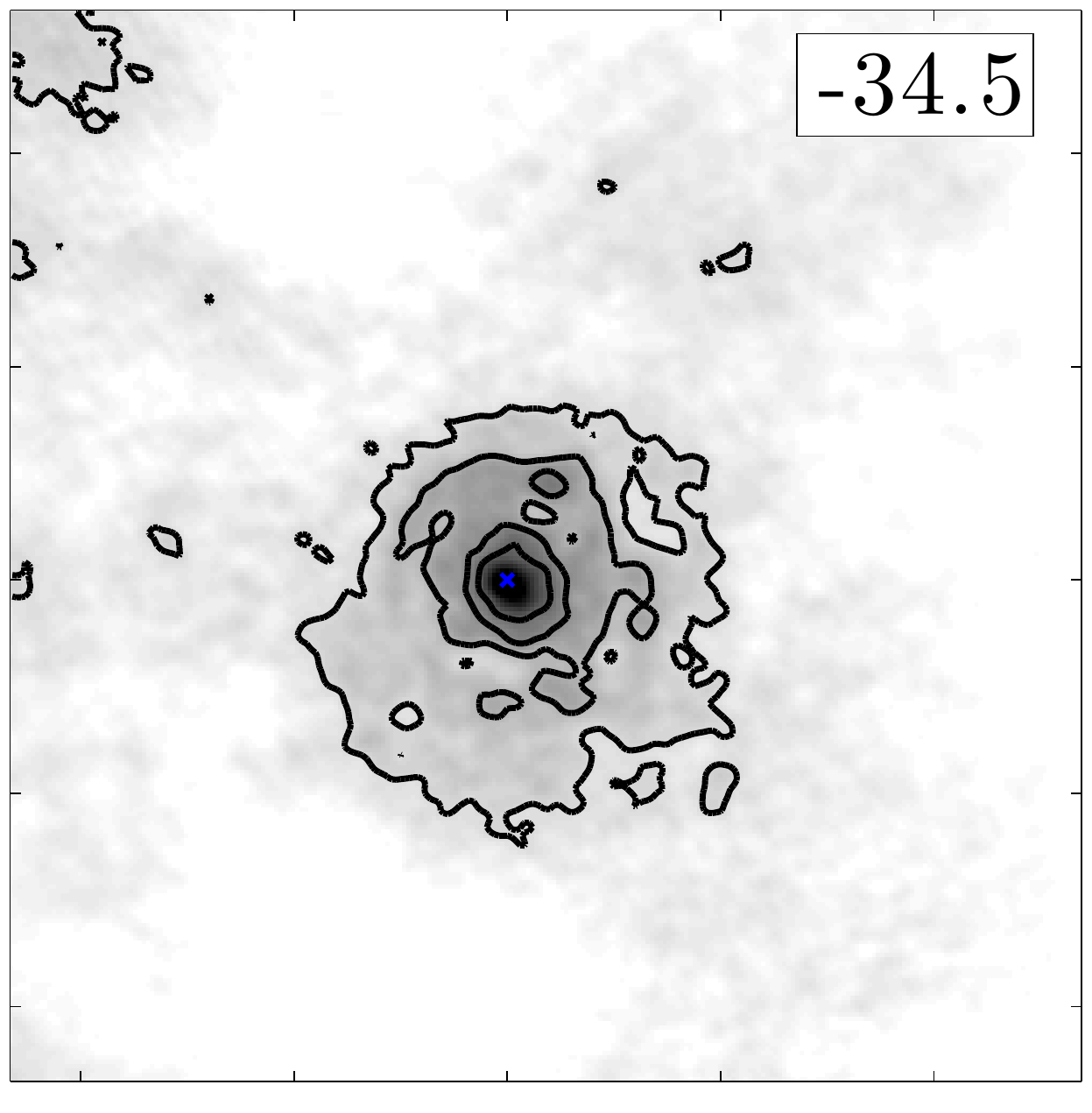}
\includegraphics[height=3.9cm]{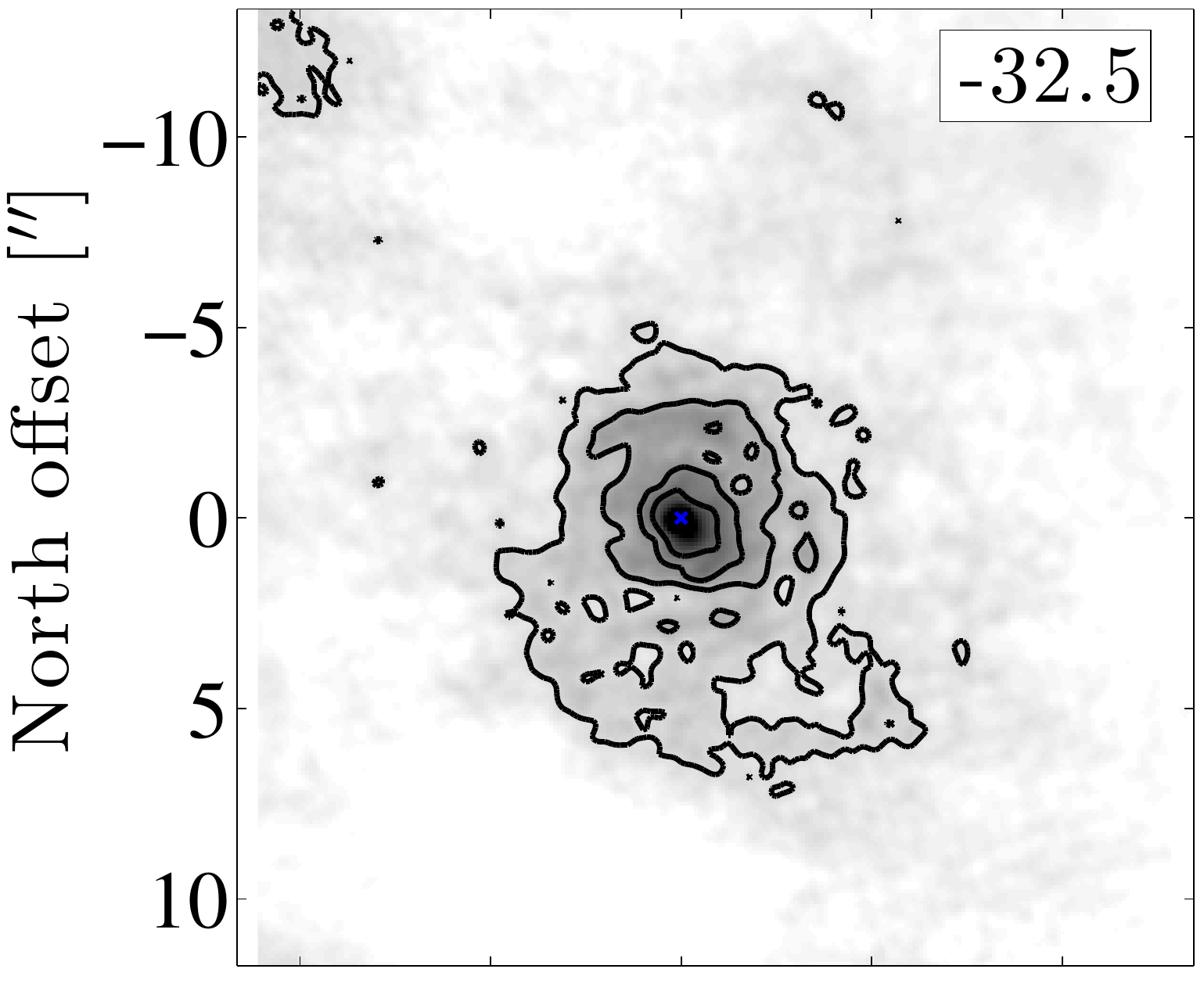}
\includegraphics[height=3.9cm]{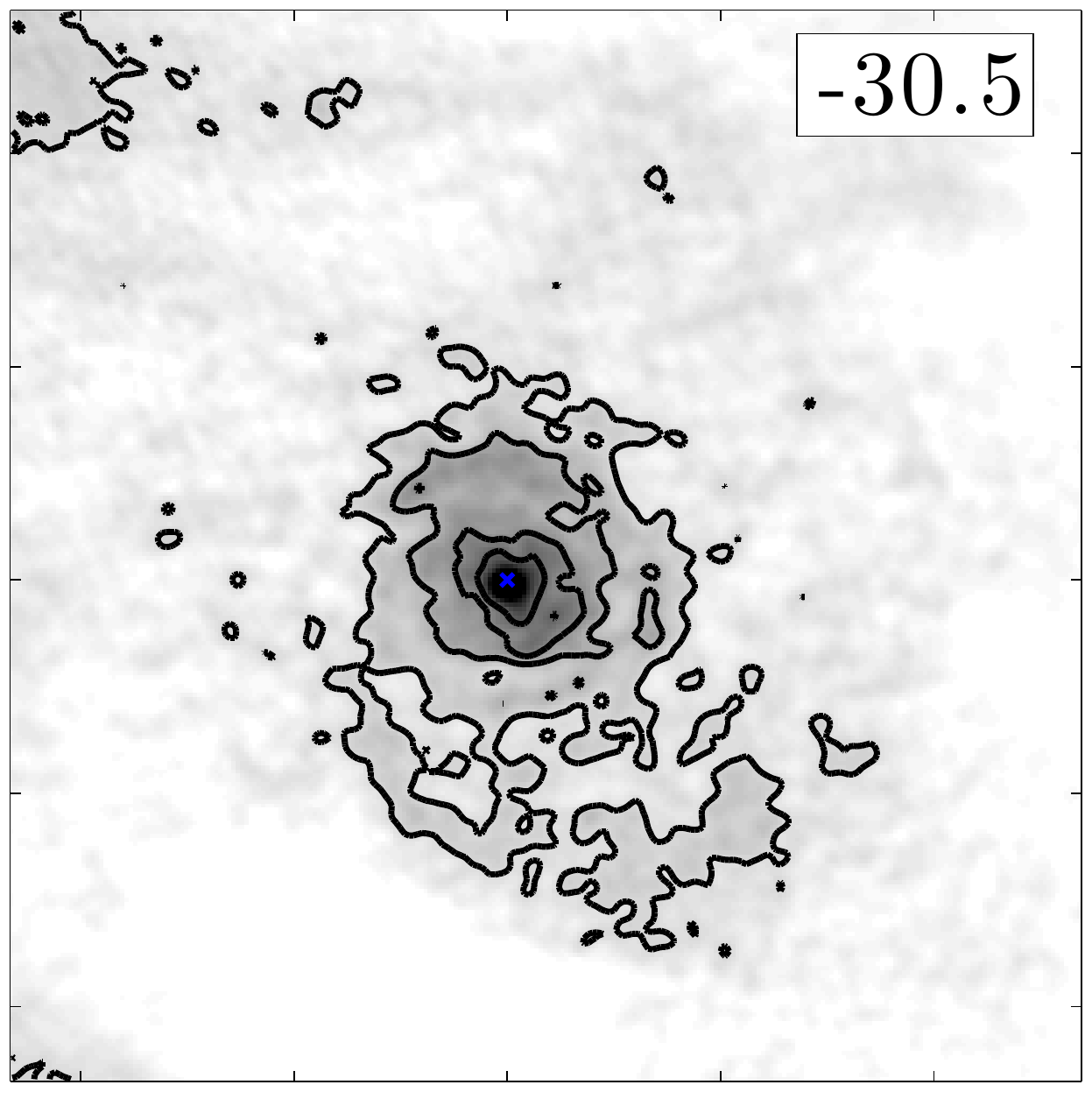}
\includegraphics[height=3.9cm]{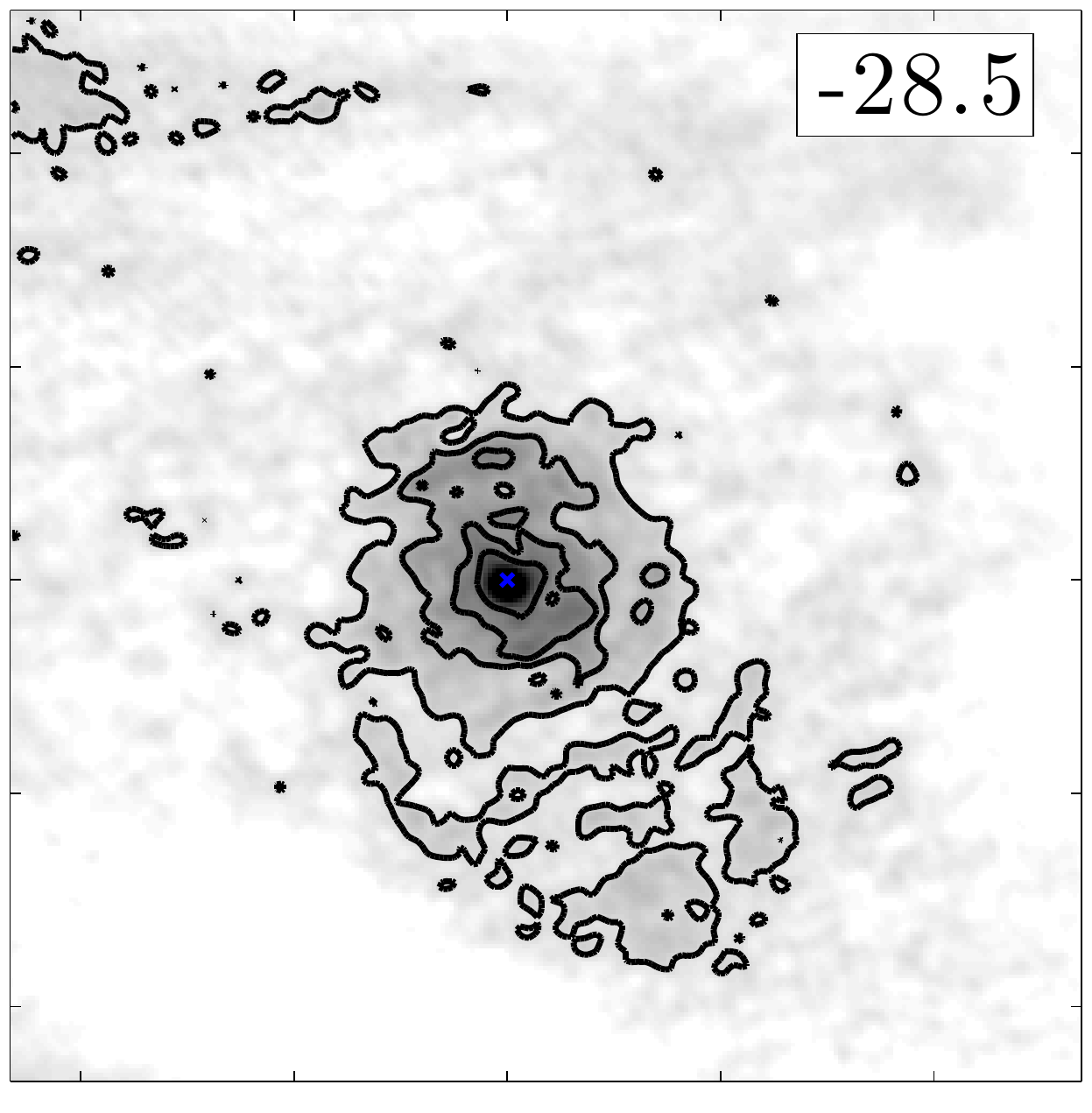}
\includegraphics[height=3.9cm]{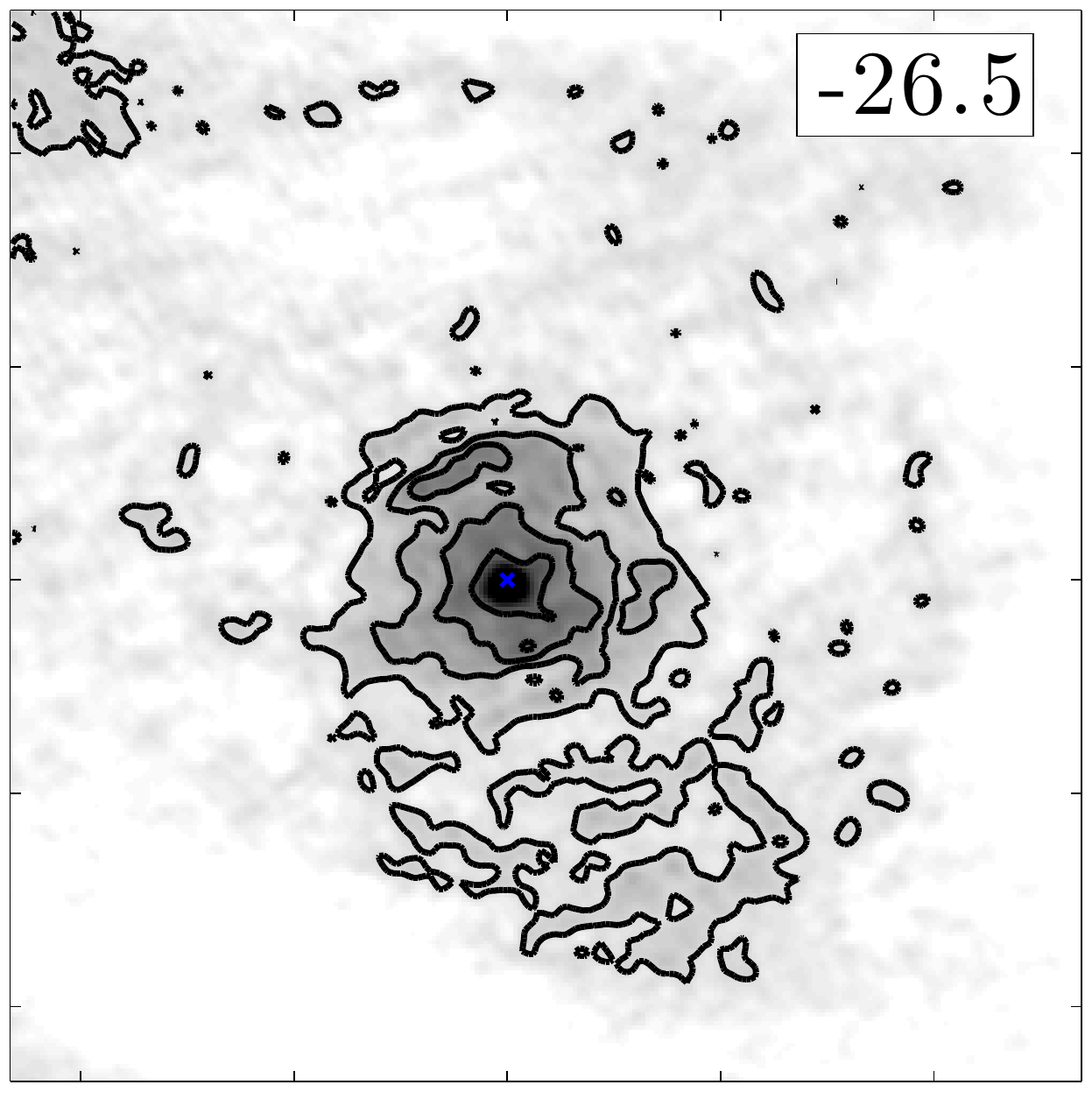}
\includegraphics[height=3.9cm]{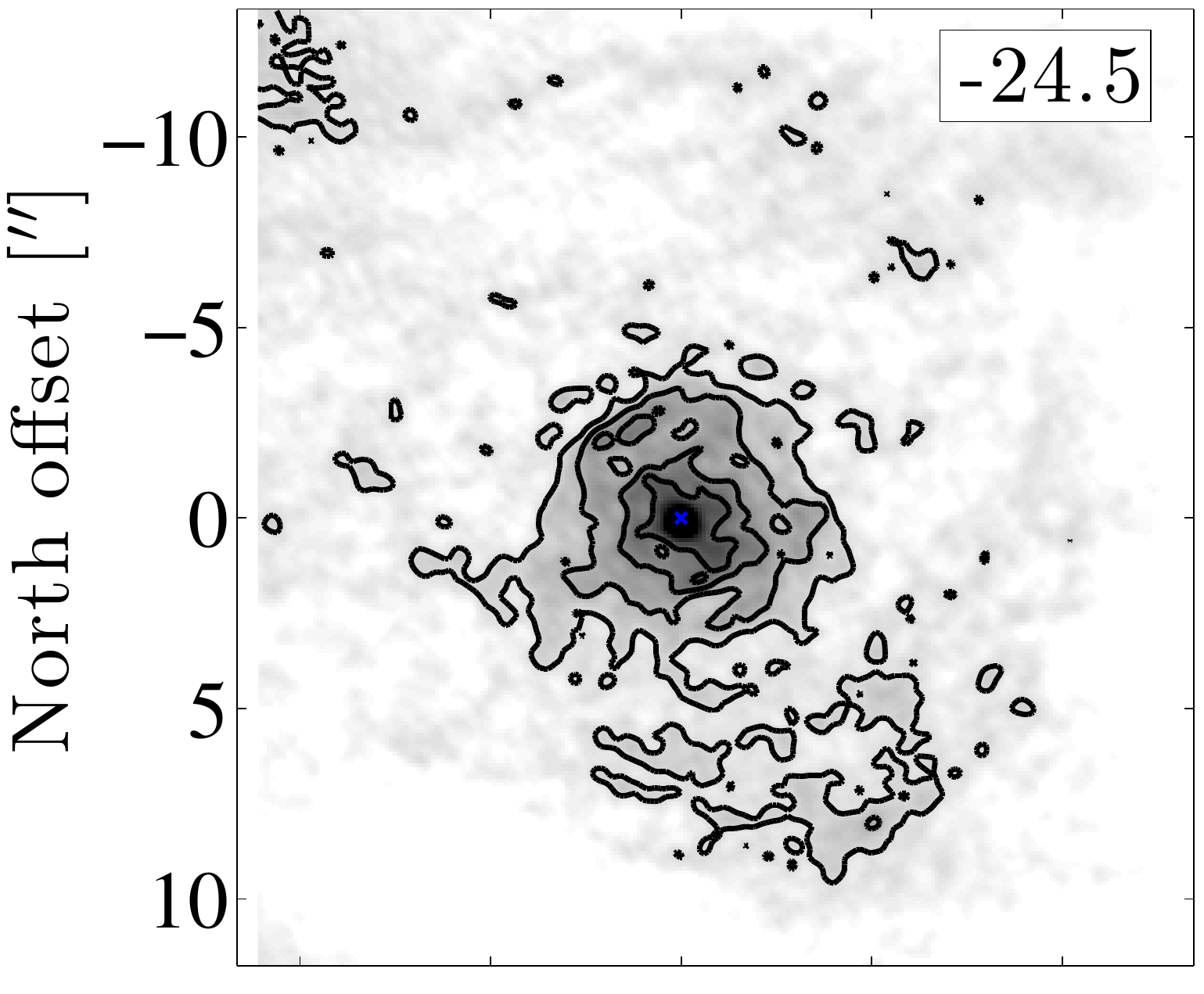}
\includegraphics[height=3.9cm]{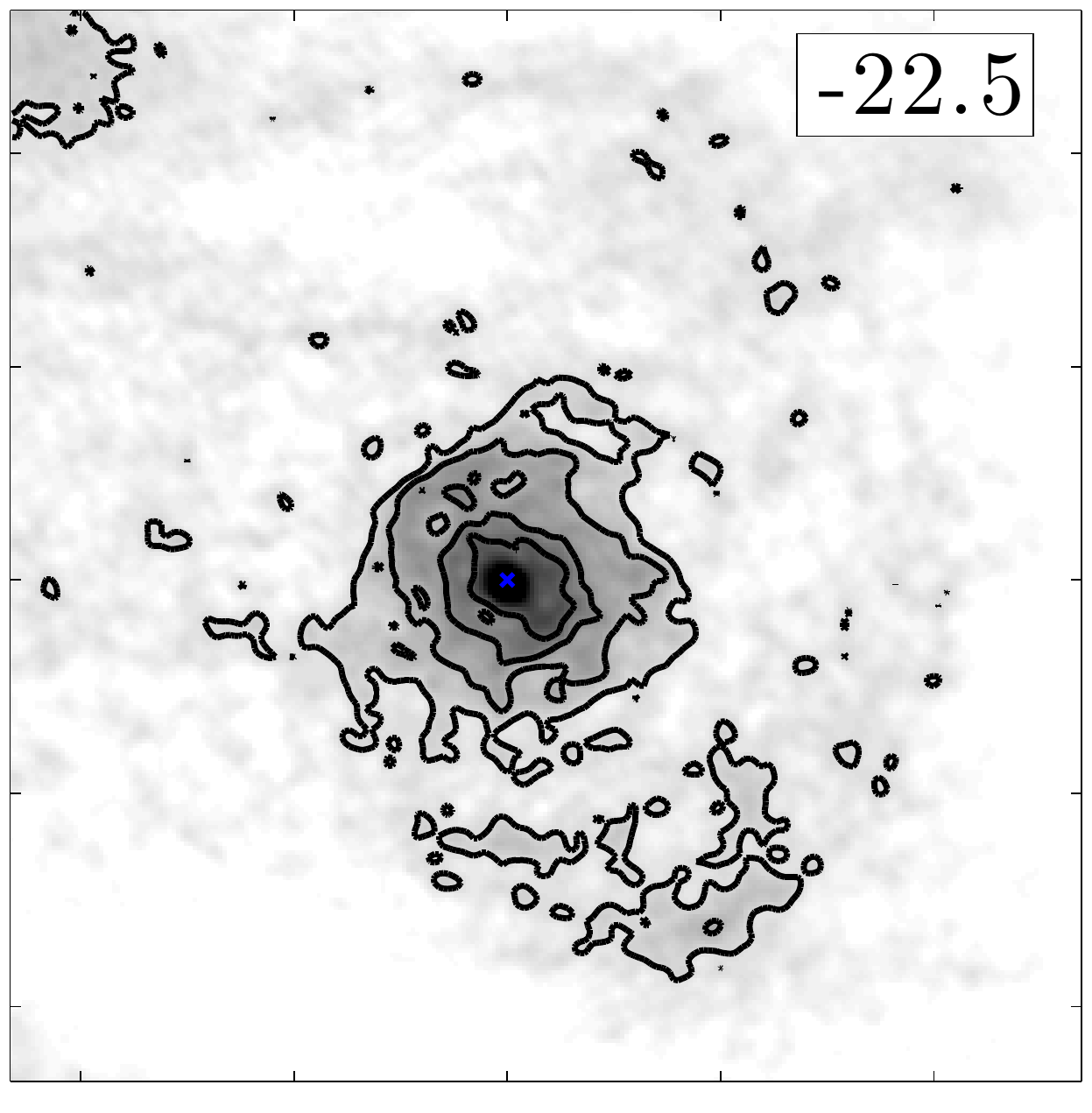}
\includegraphics[height=3.9cm]{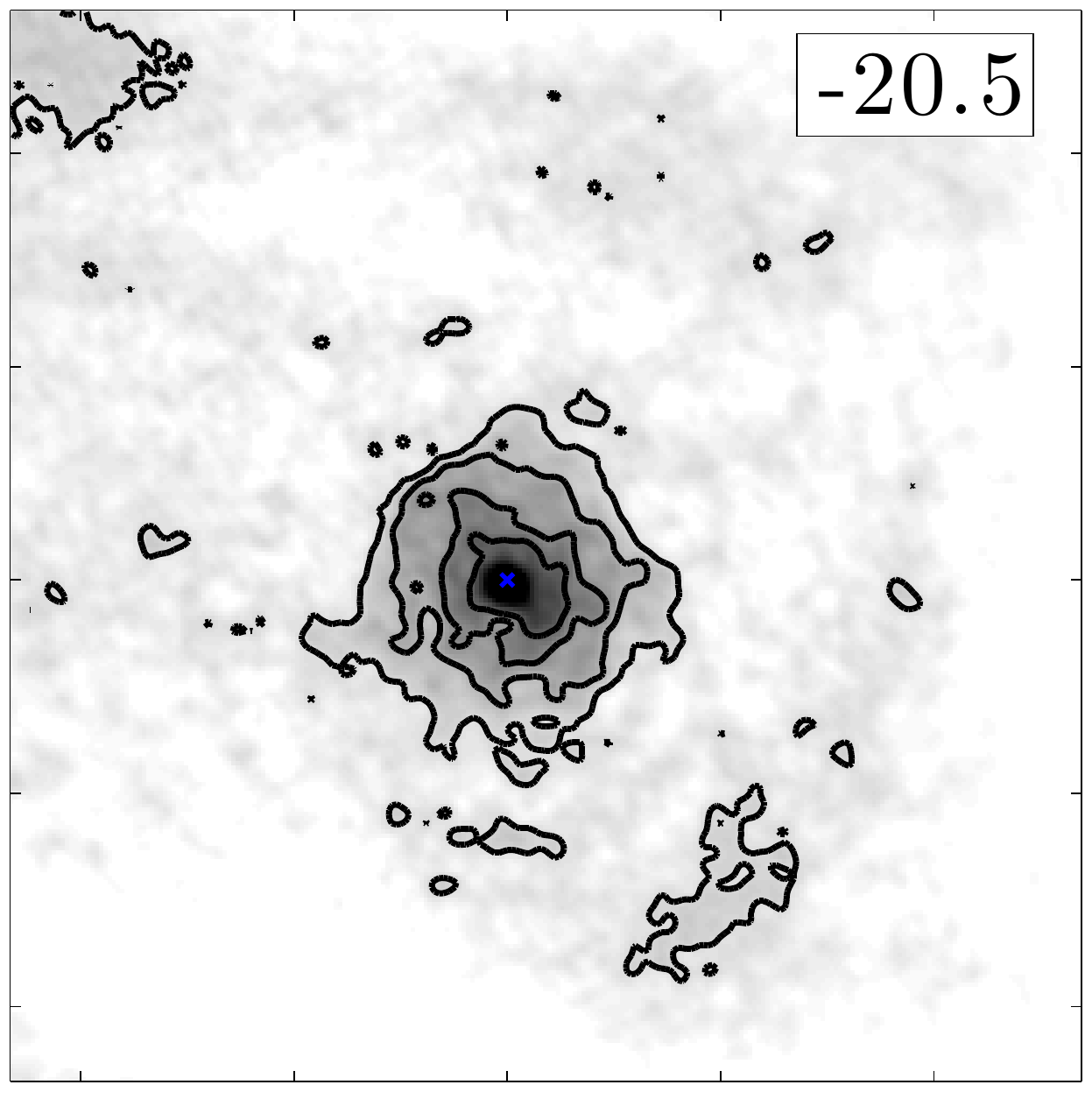}
\includegraphics[height=3.9cm]{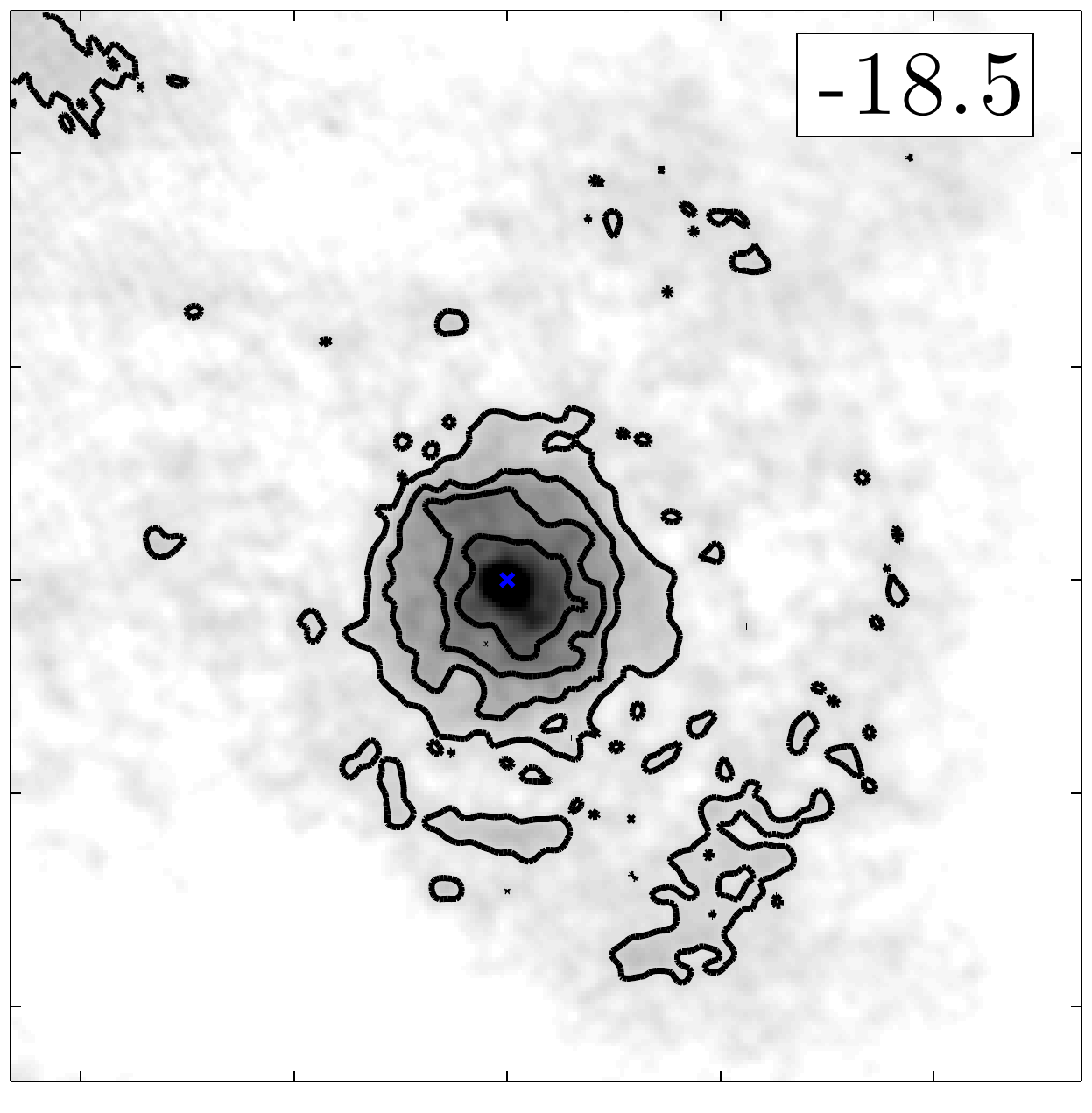}
\includegraphics[height=3.9cm]{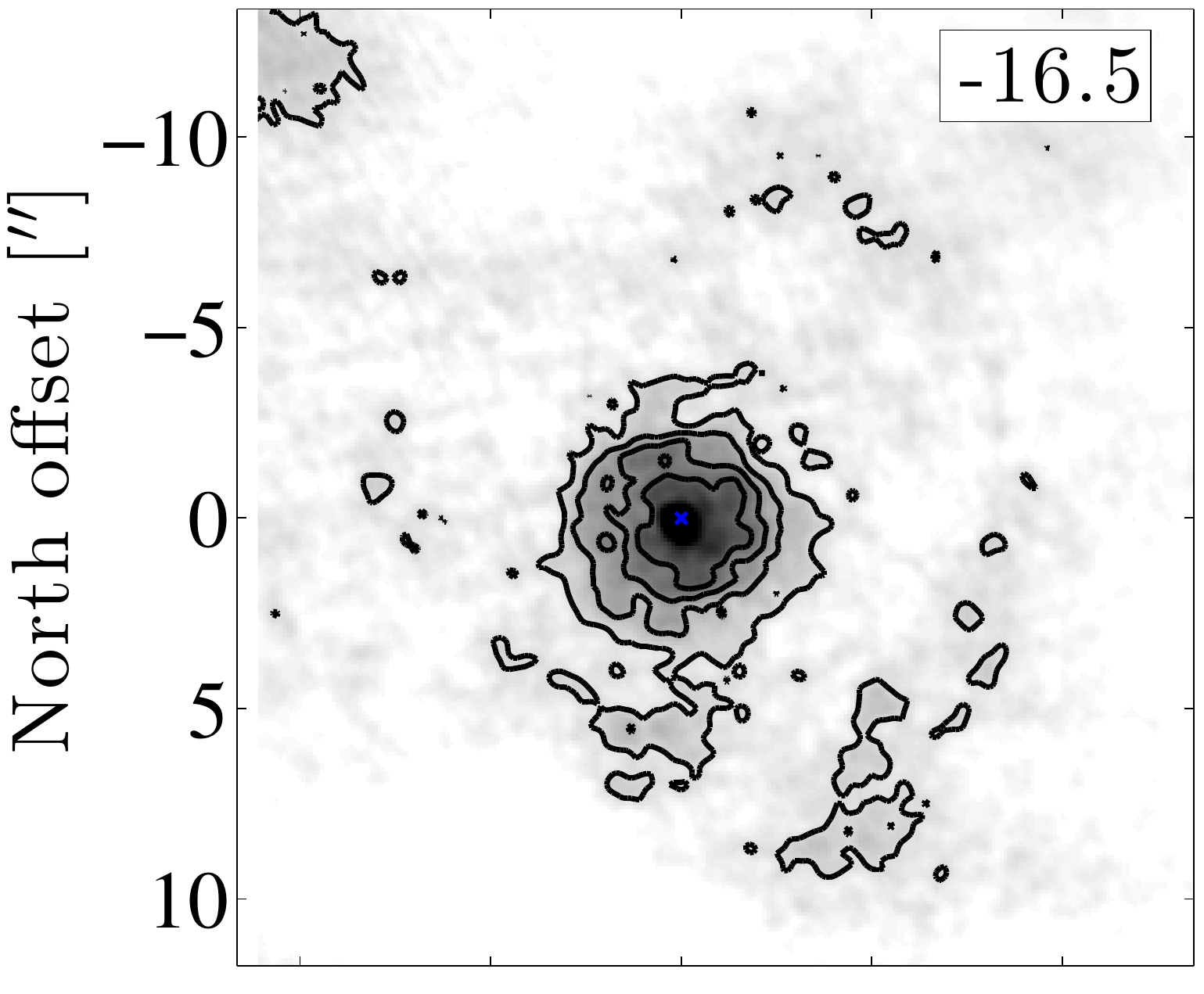}
\includegraphics[height=3.9cm]{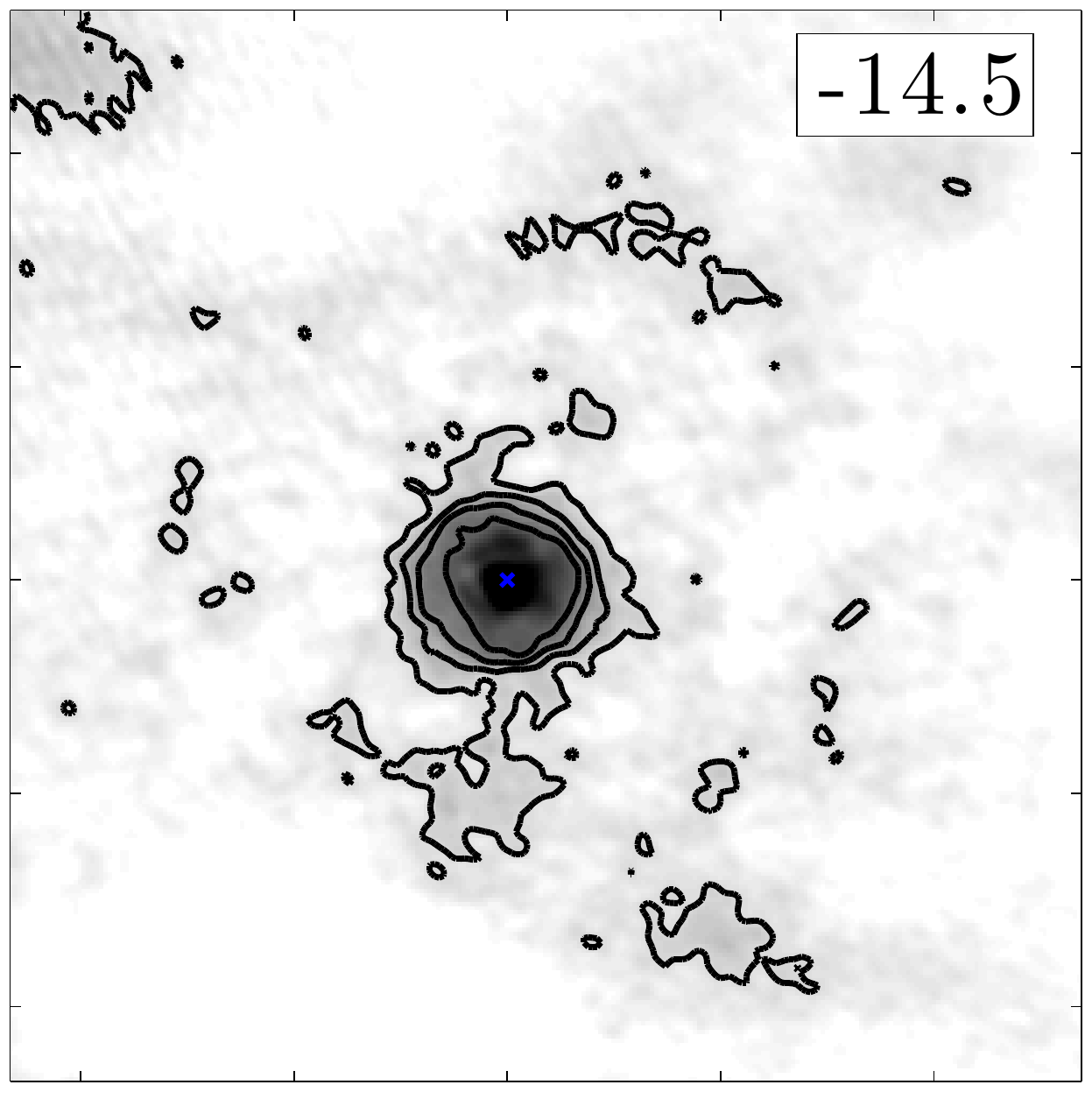}
\includegraphics[height=3.9cm]{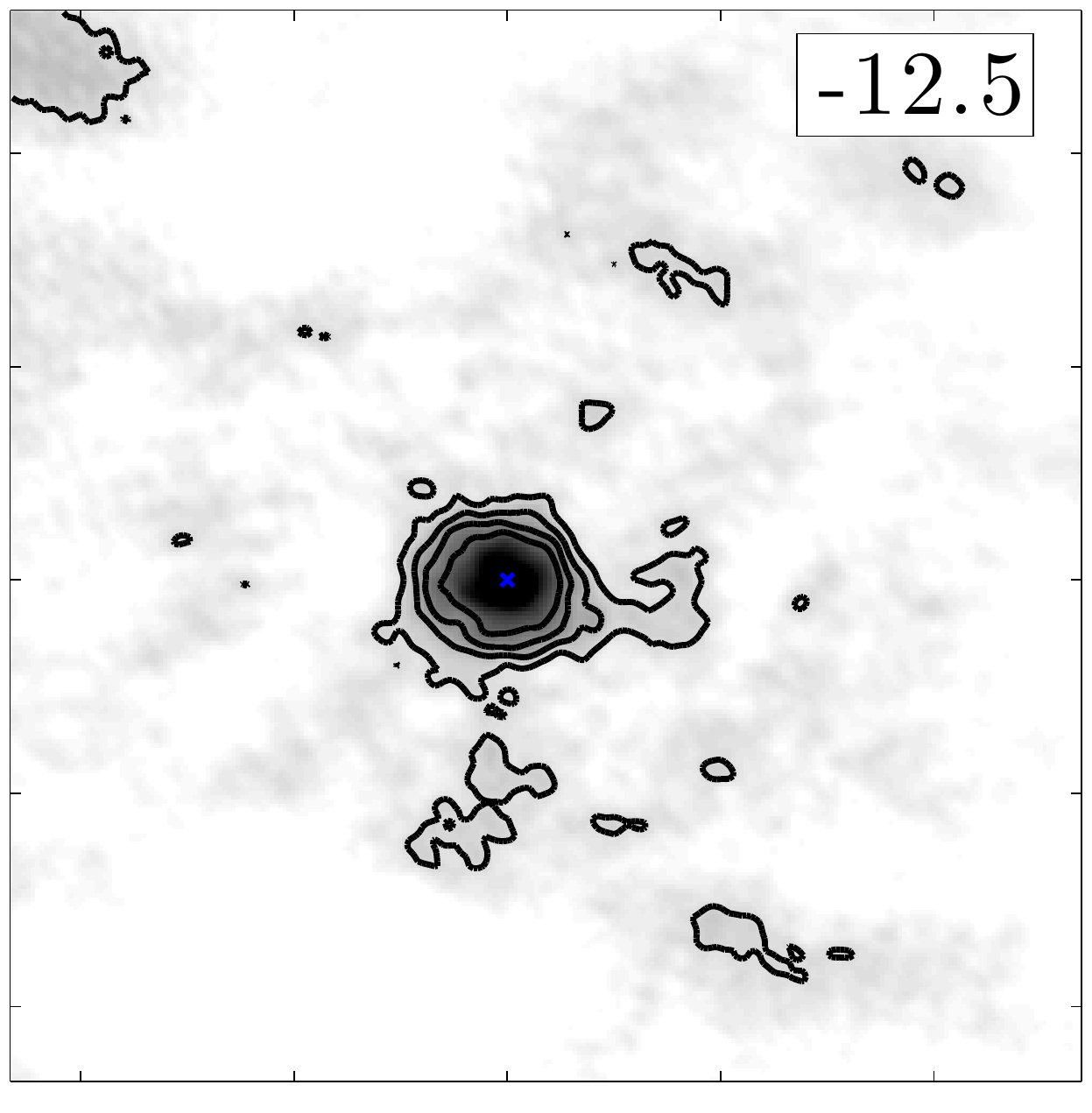}
\includegraphics[height=3.9cm]{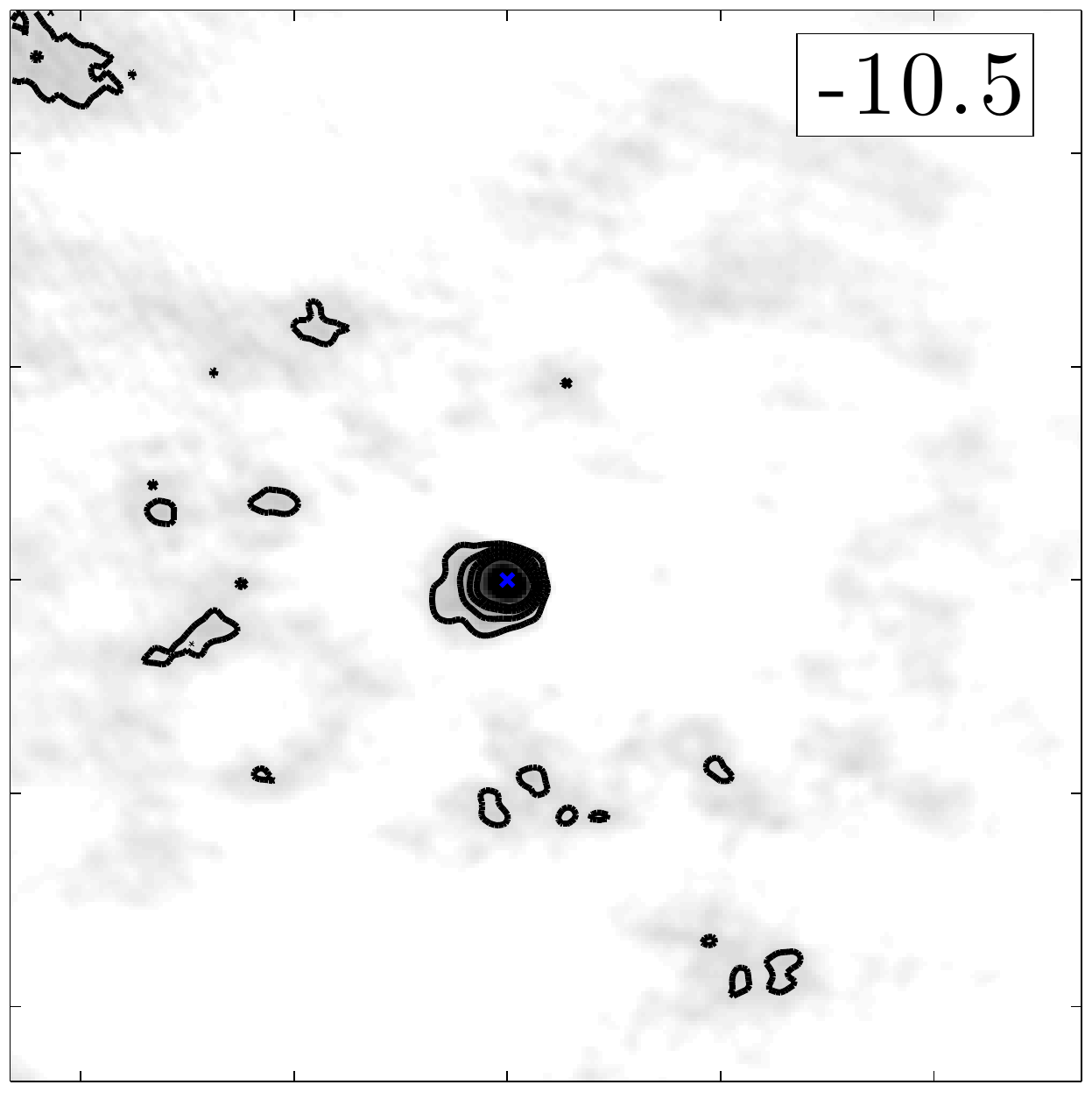}
\includegraphics[width=4.72cm]{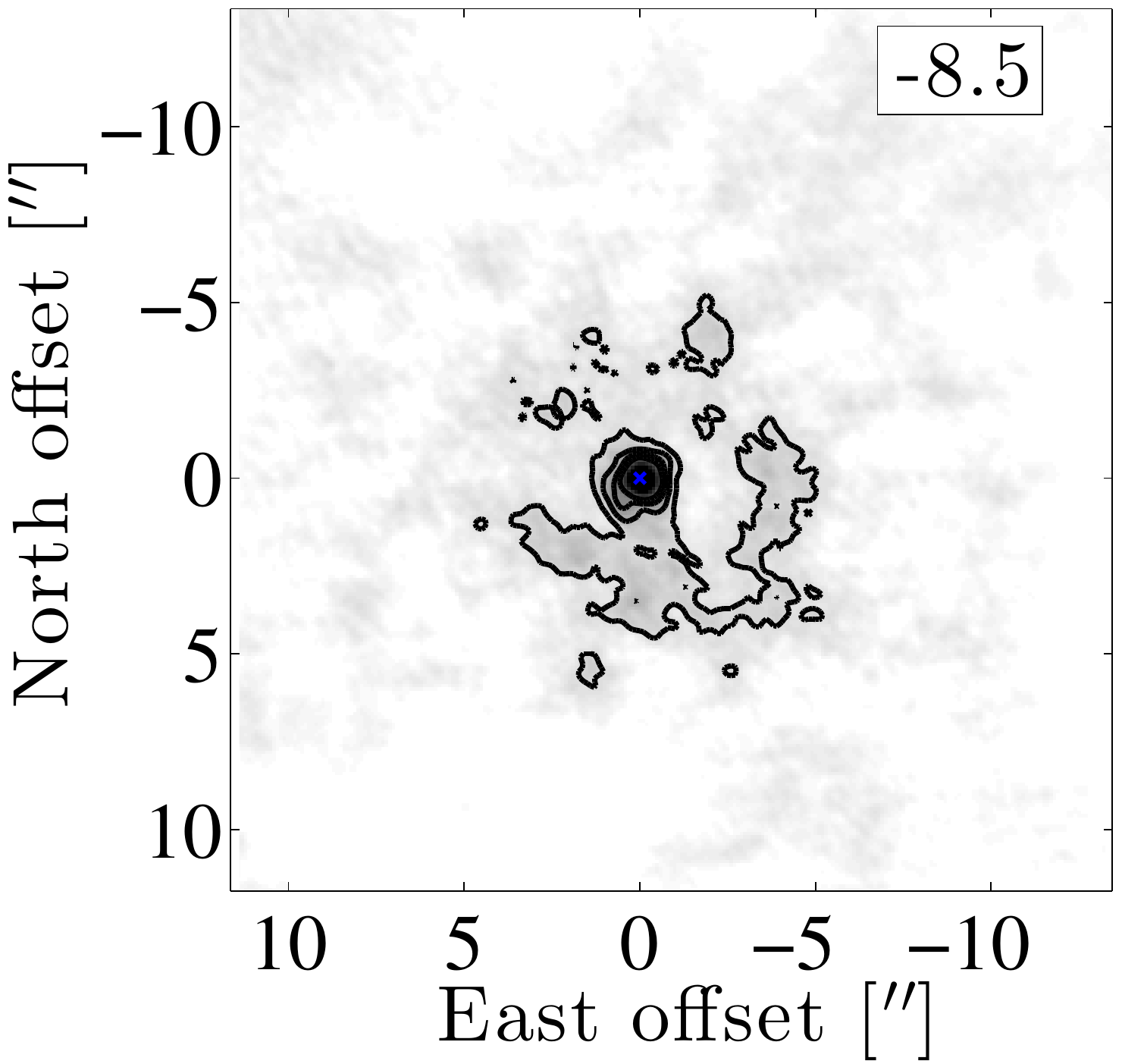}
\includegraphics[width=3.83cm]{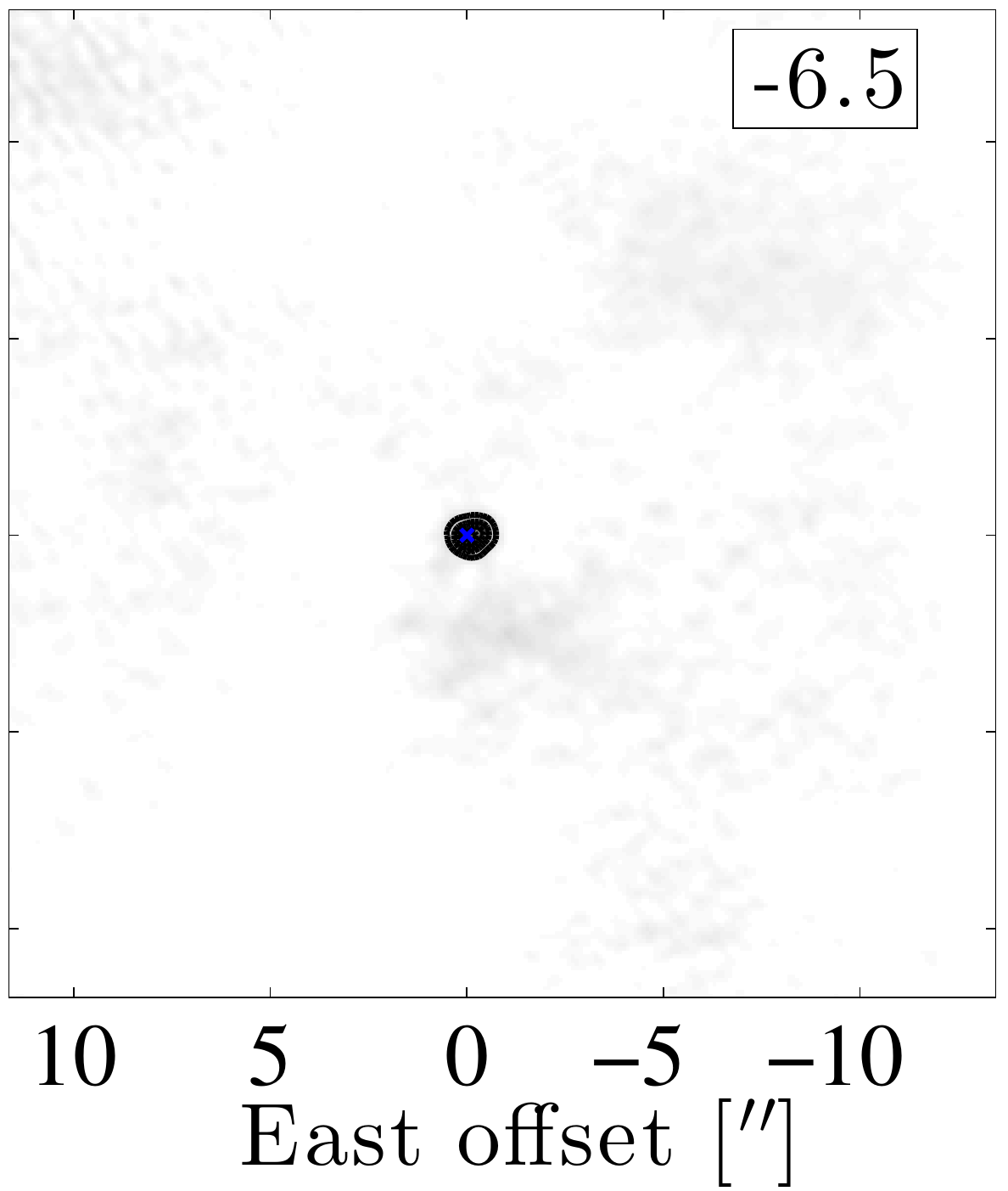}
\includegraphics[width=3.83cm]{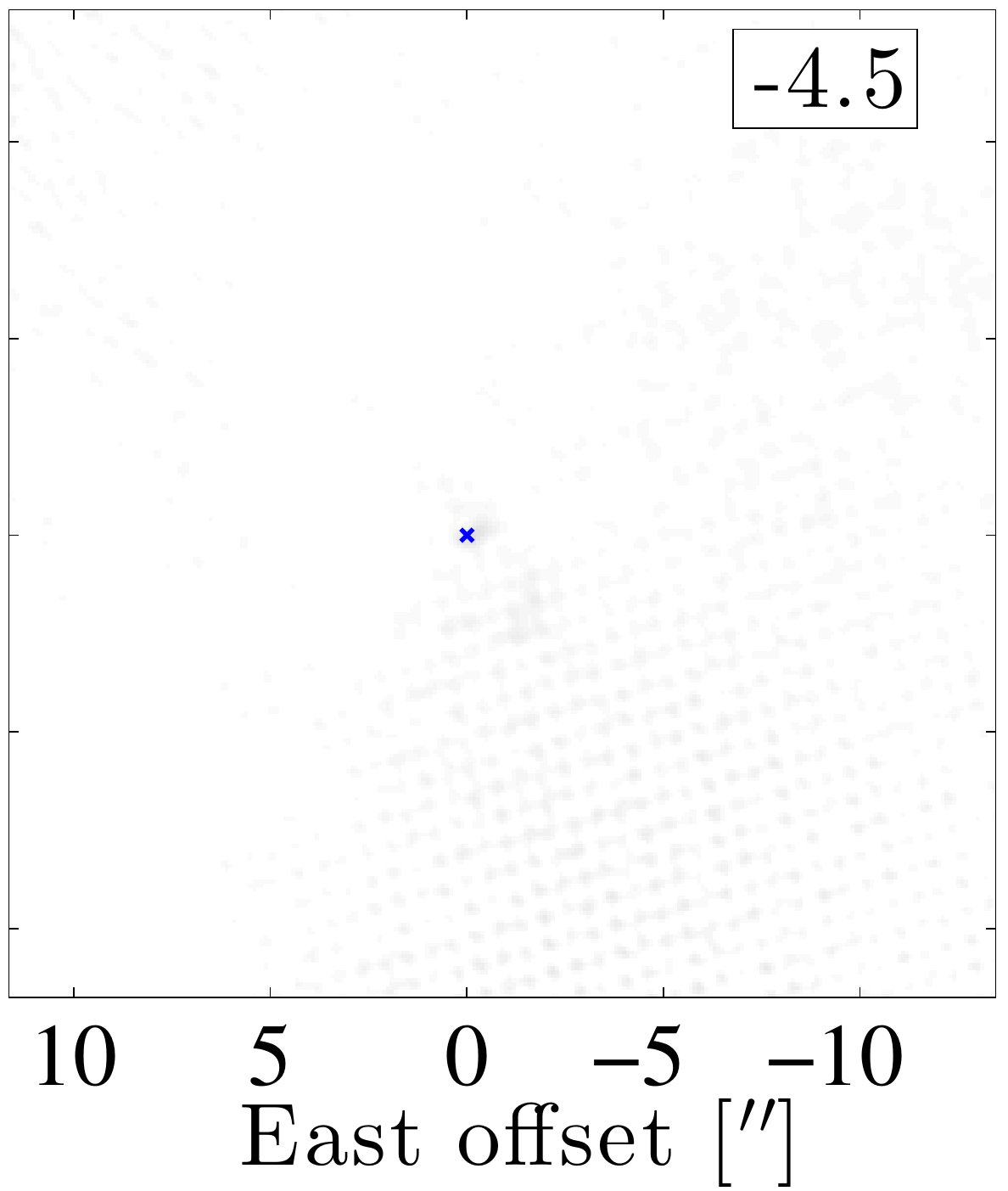}
\includegraphics[width=3.83cm]{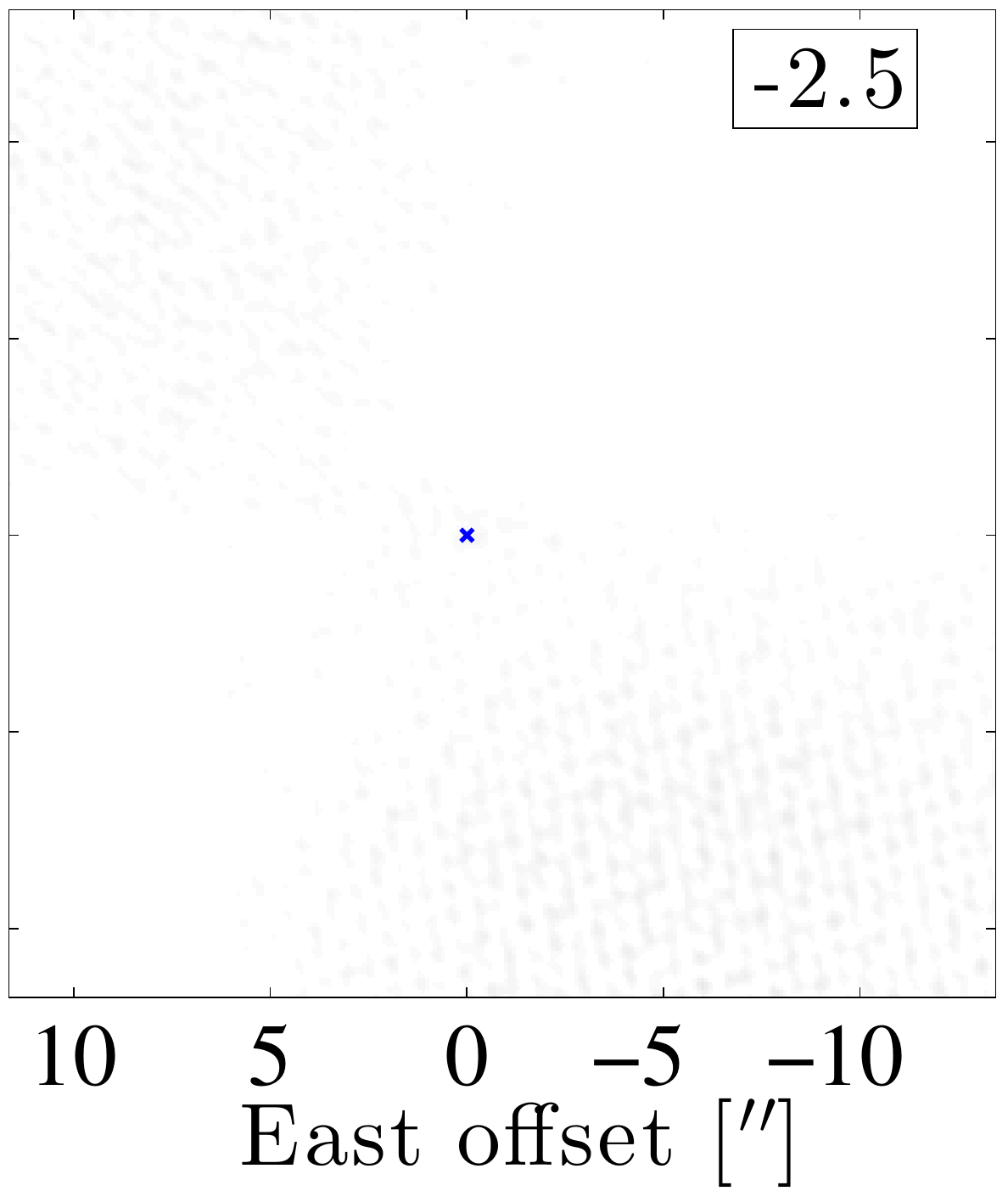}
\caption{ ALMA CO(3-2) channel maps. The LSR velocity of each channel is given in the upper right corner legend. Contours are given at 10, 20, 30, and 40$\sigma$, where $\sigma$ has been measured in the emission-free channels. The blue cross marks the peak of the continuum emission.}
\end{figure*}

\begin{figure*}
\flushleft
\label{m02channelmaps}
\includegraphics[height=3.9cm]{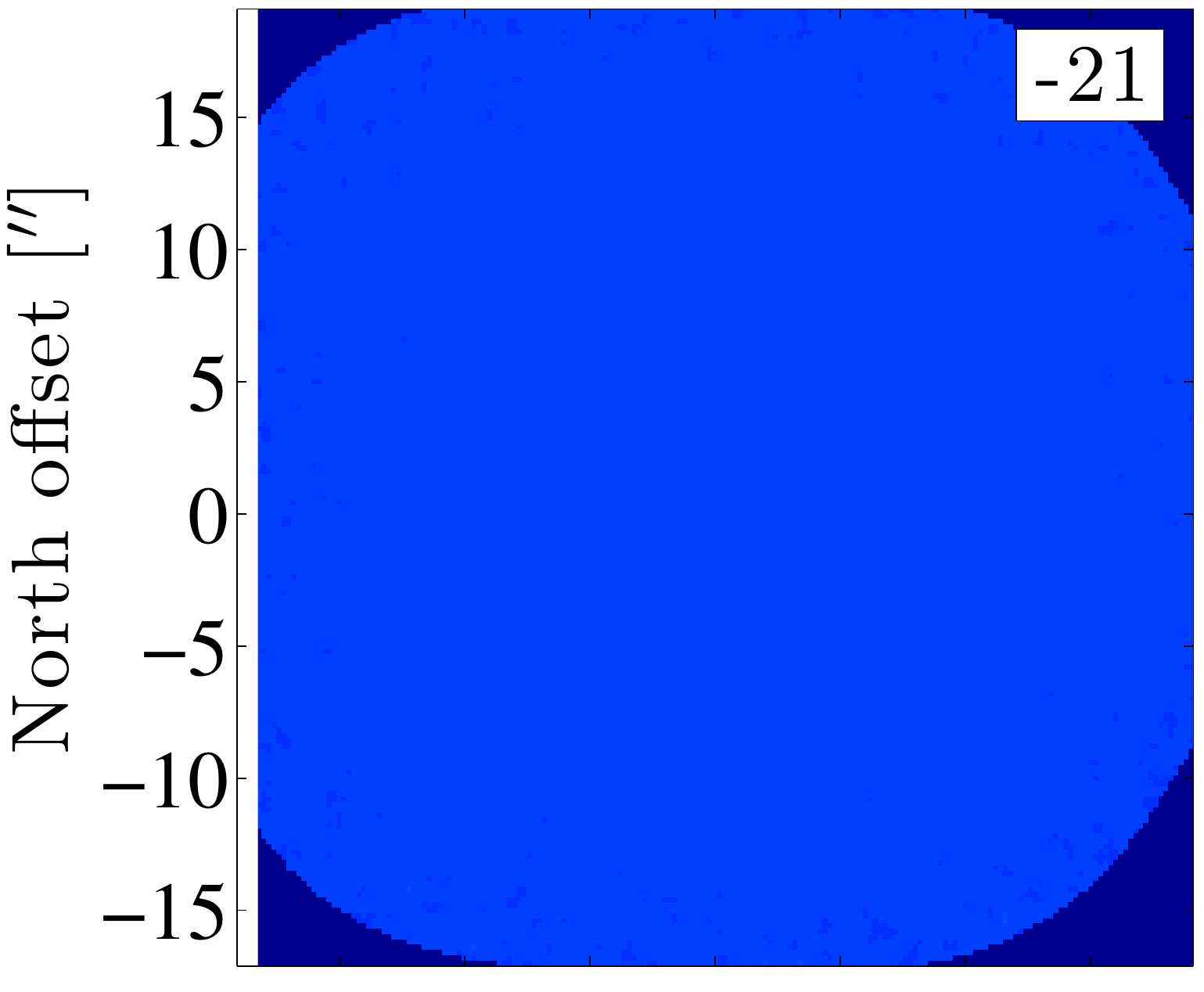}
\includegraphics[height=3.9cm]{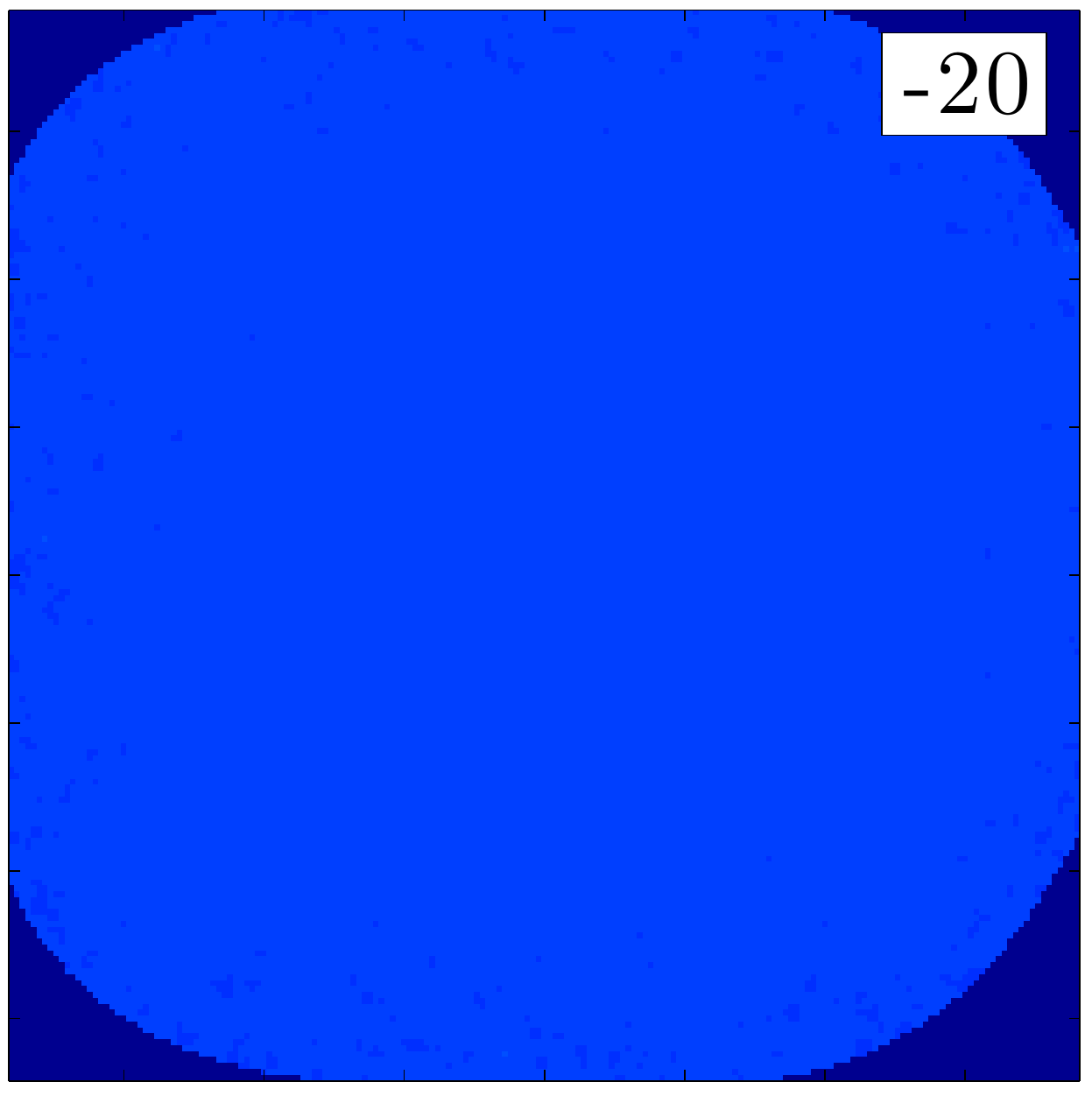}
\includegraphics[height=3.9cm]{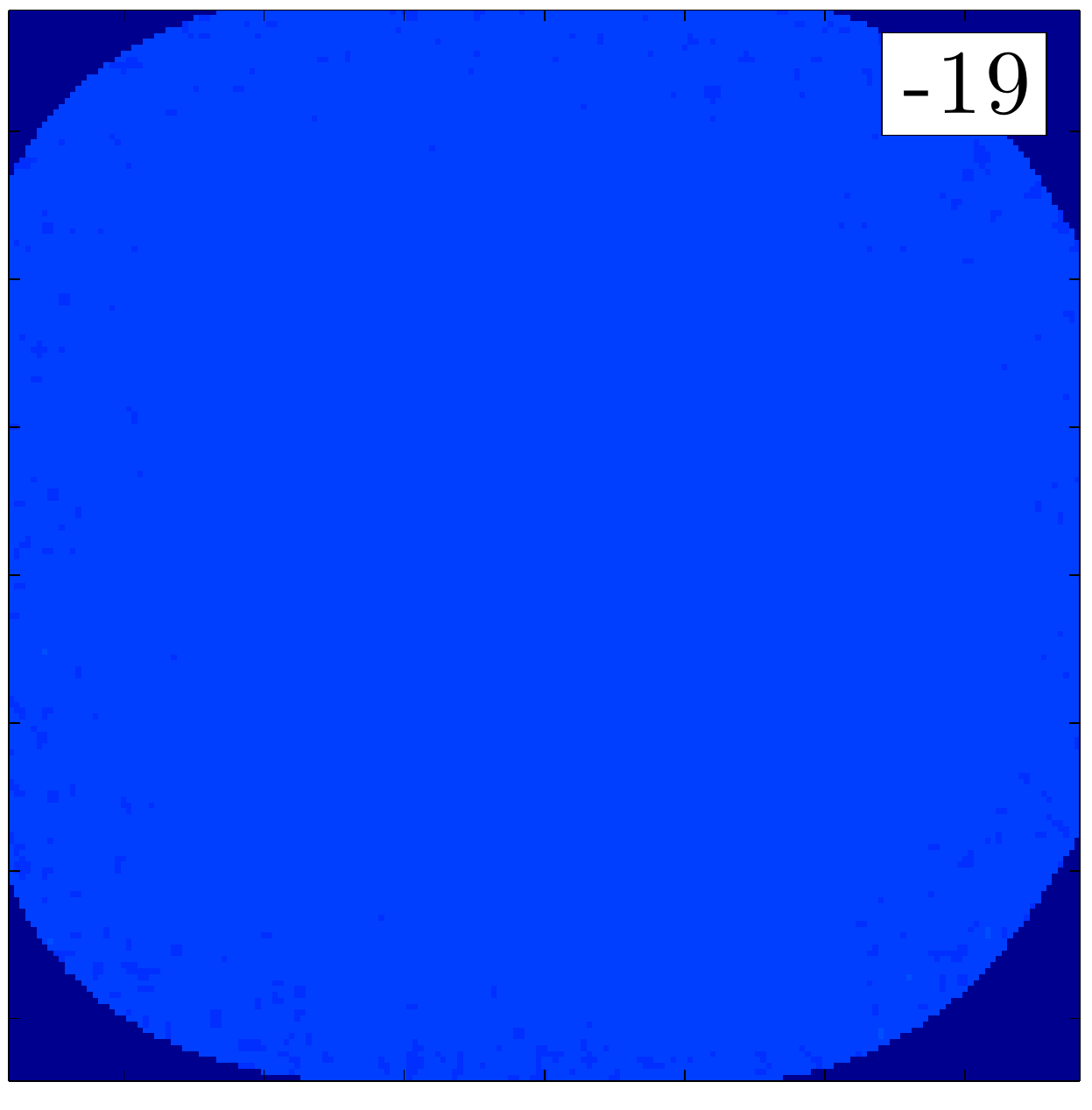}
\includegraphics[height=3.9cm]{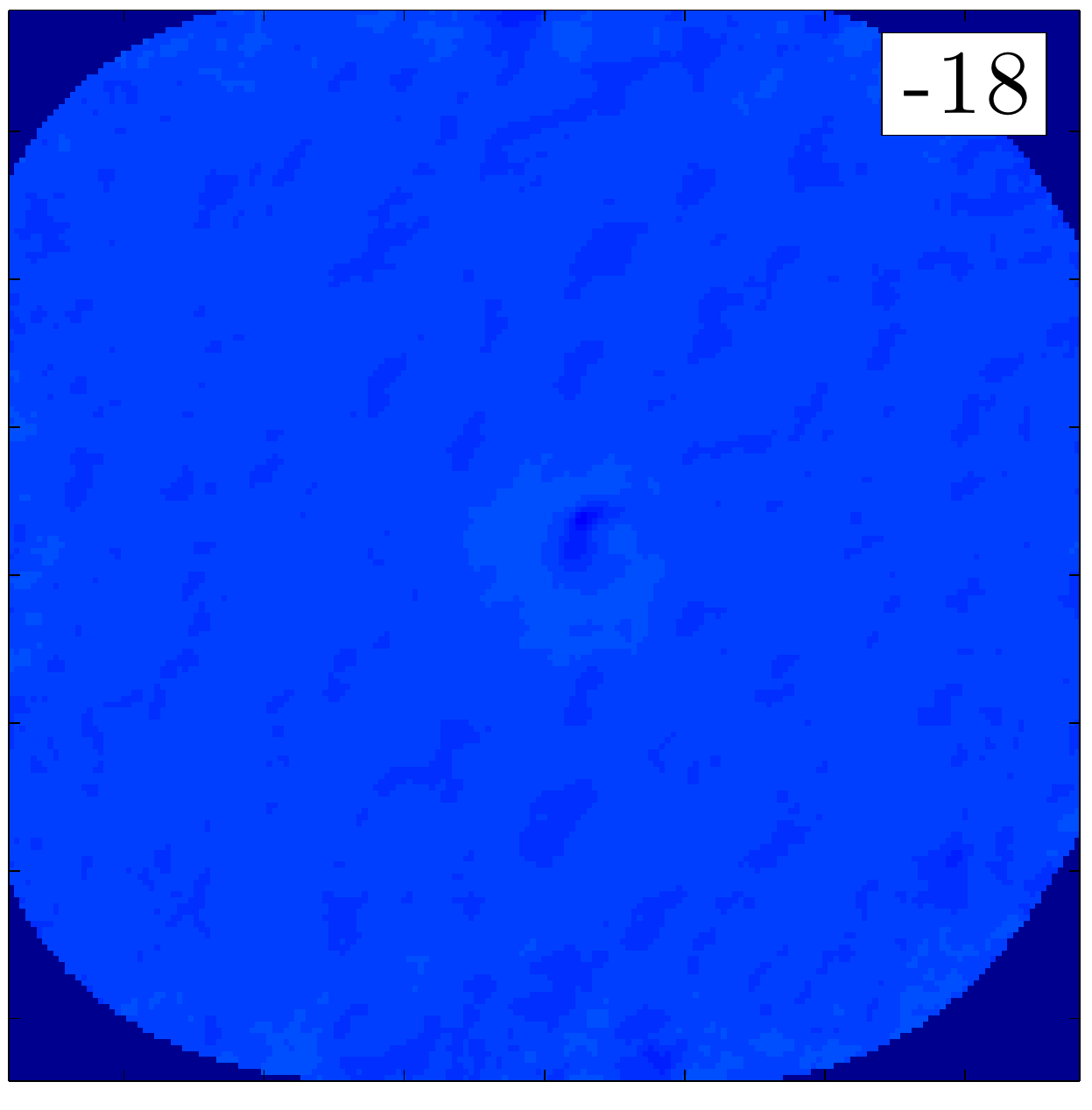}
\includegraphics[height=3.9cm]{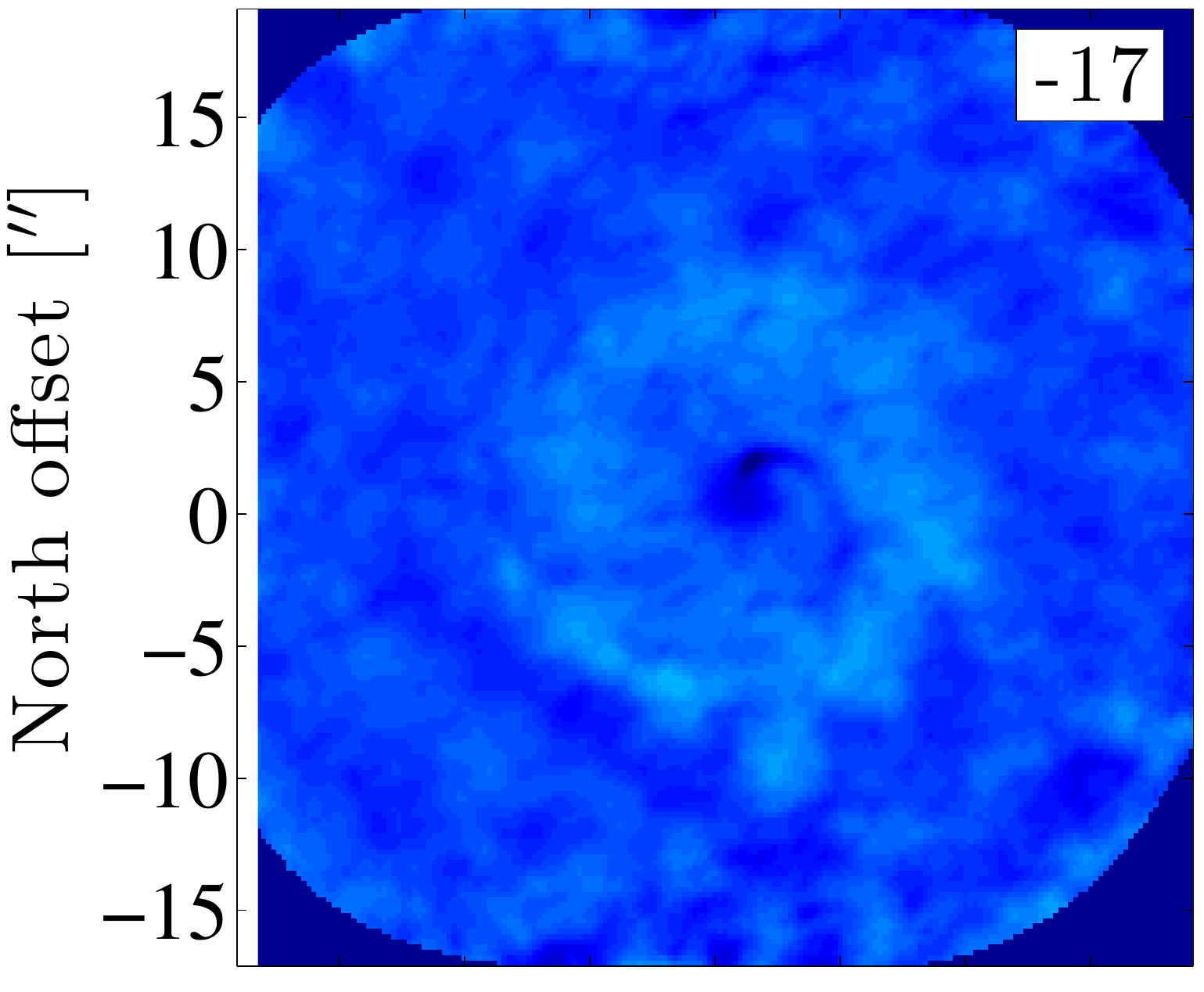}
\includegraphics[height=3.9cm]{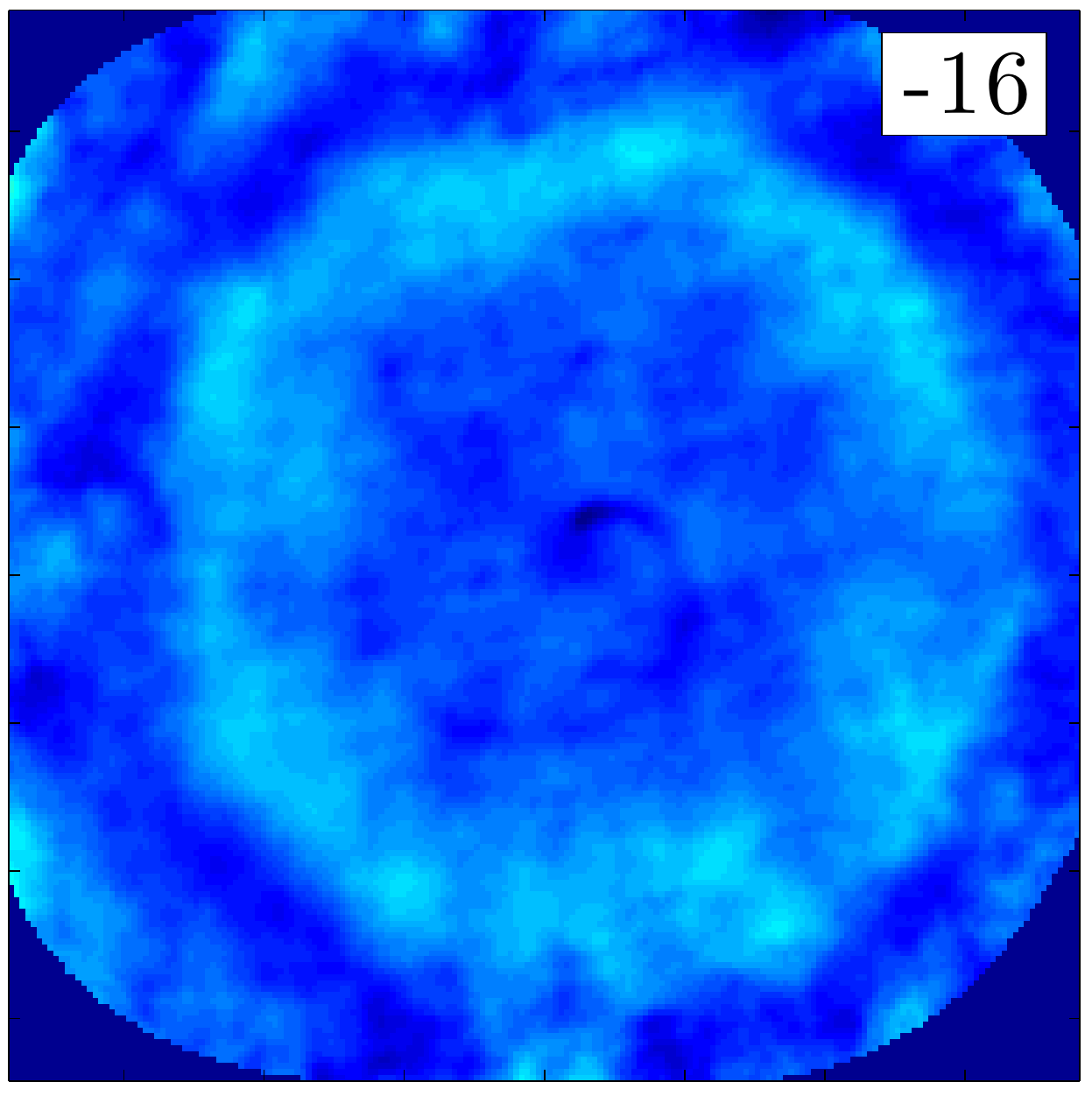}
\includegraphics[height=3.9cm]{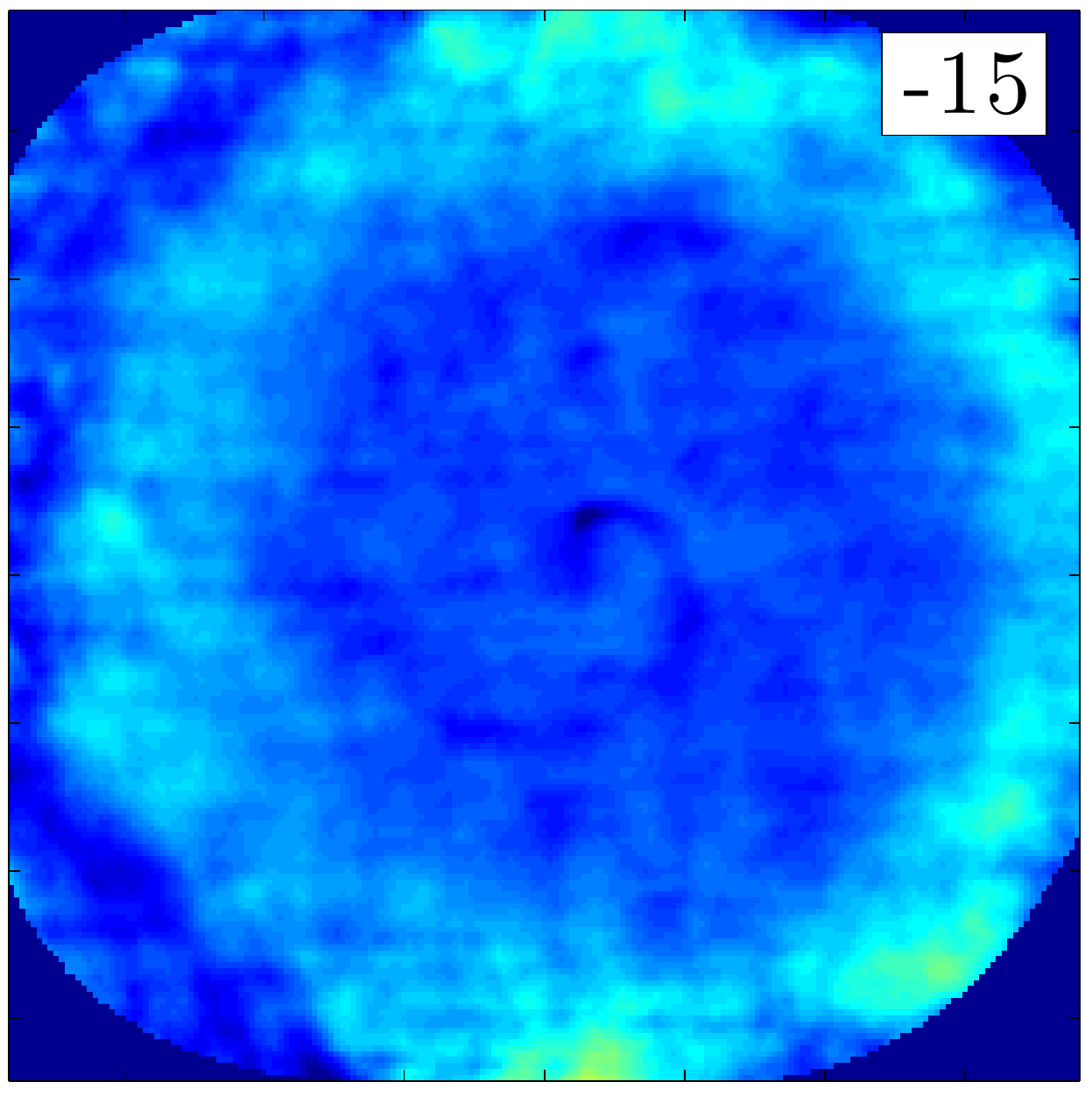}
\includegraphics[height=3.9cm]{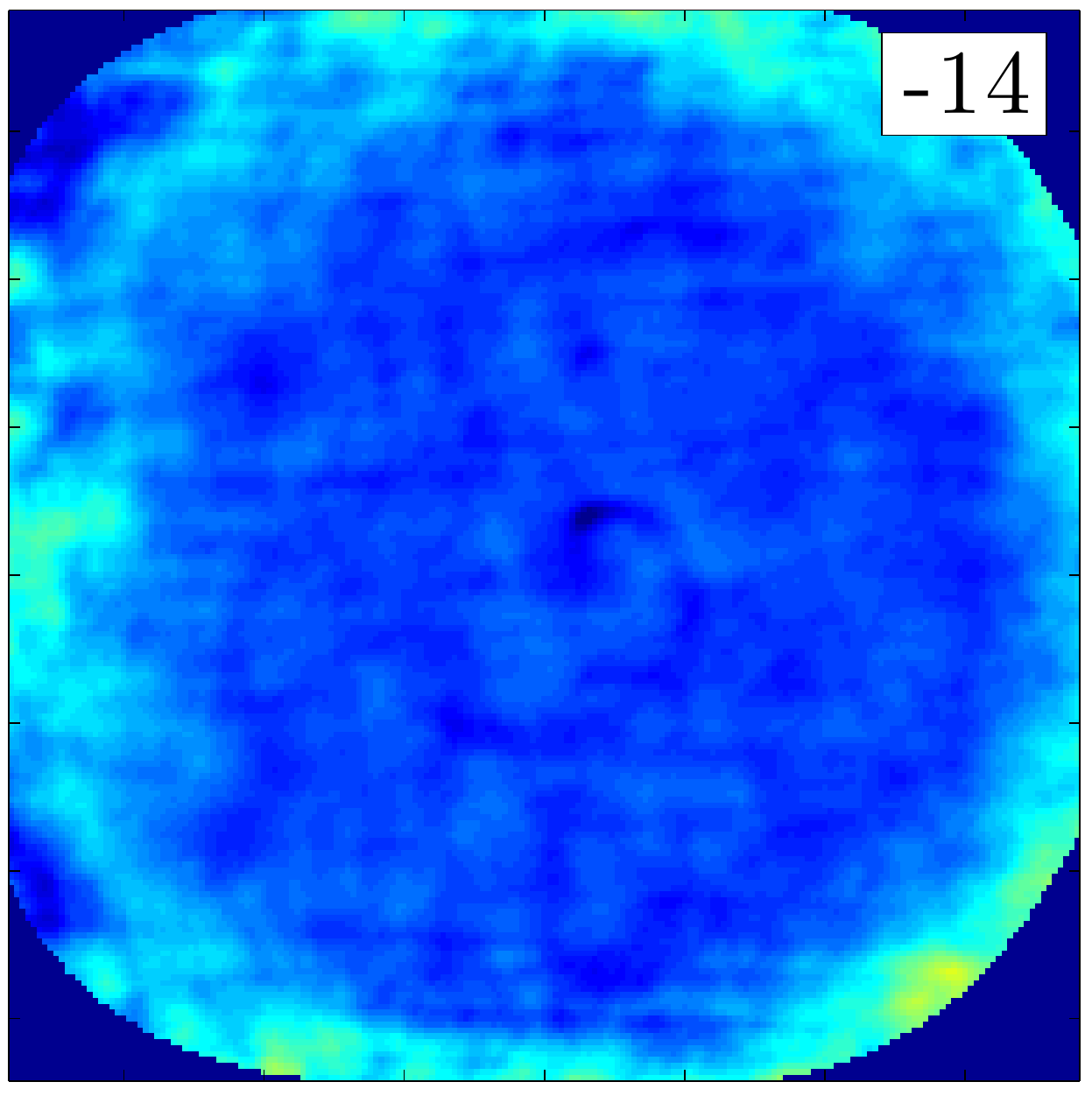}
\includegraphics[height=3.9cm]{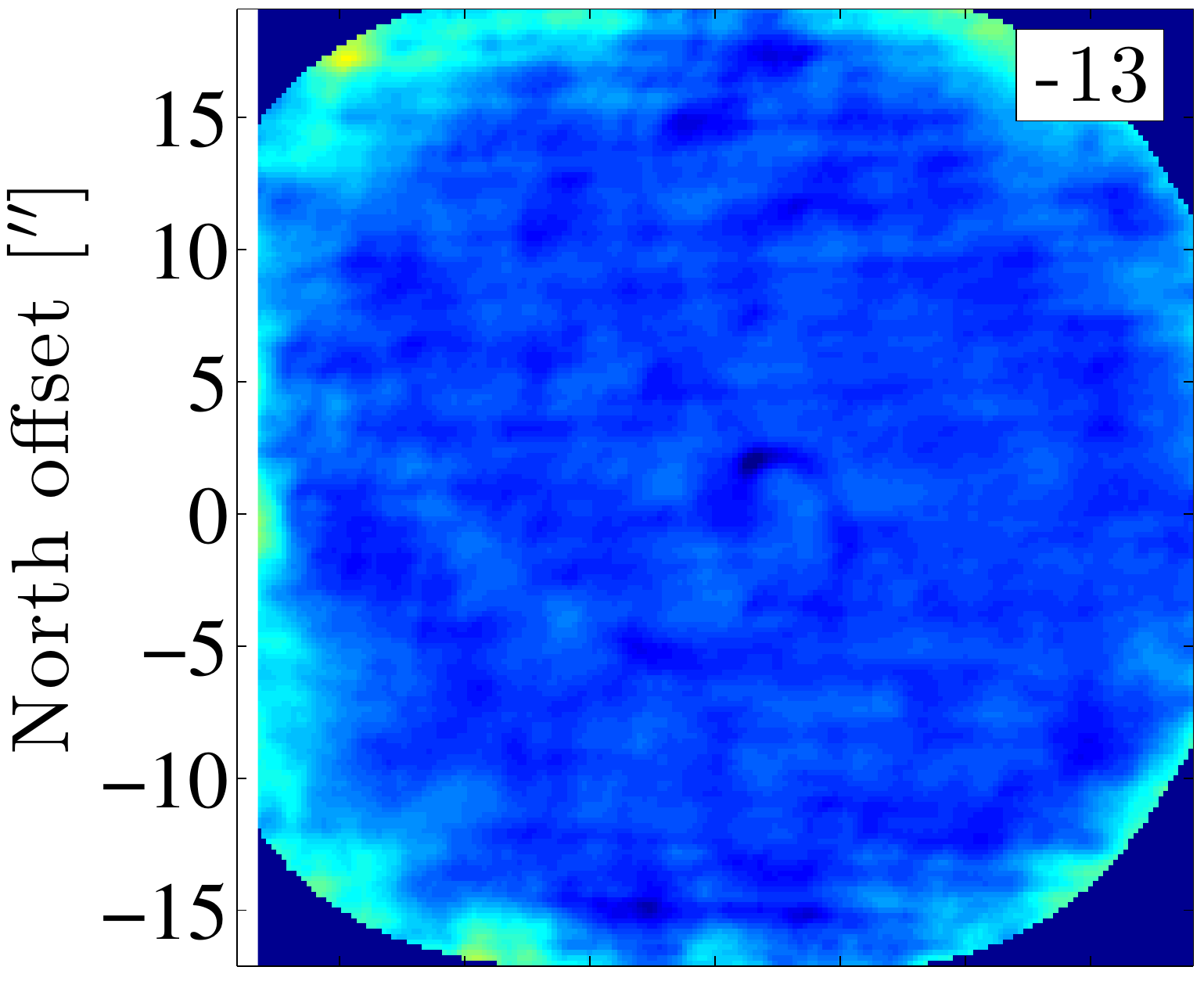}
\includegraphics[height=3.9cm]{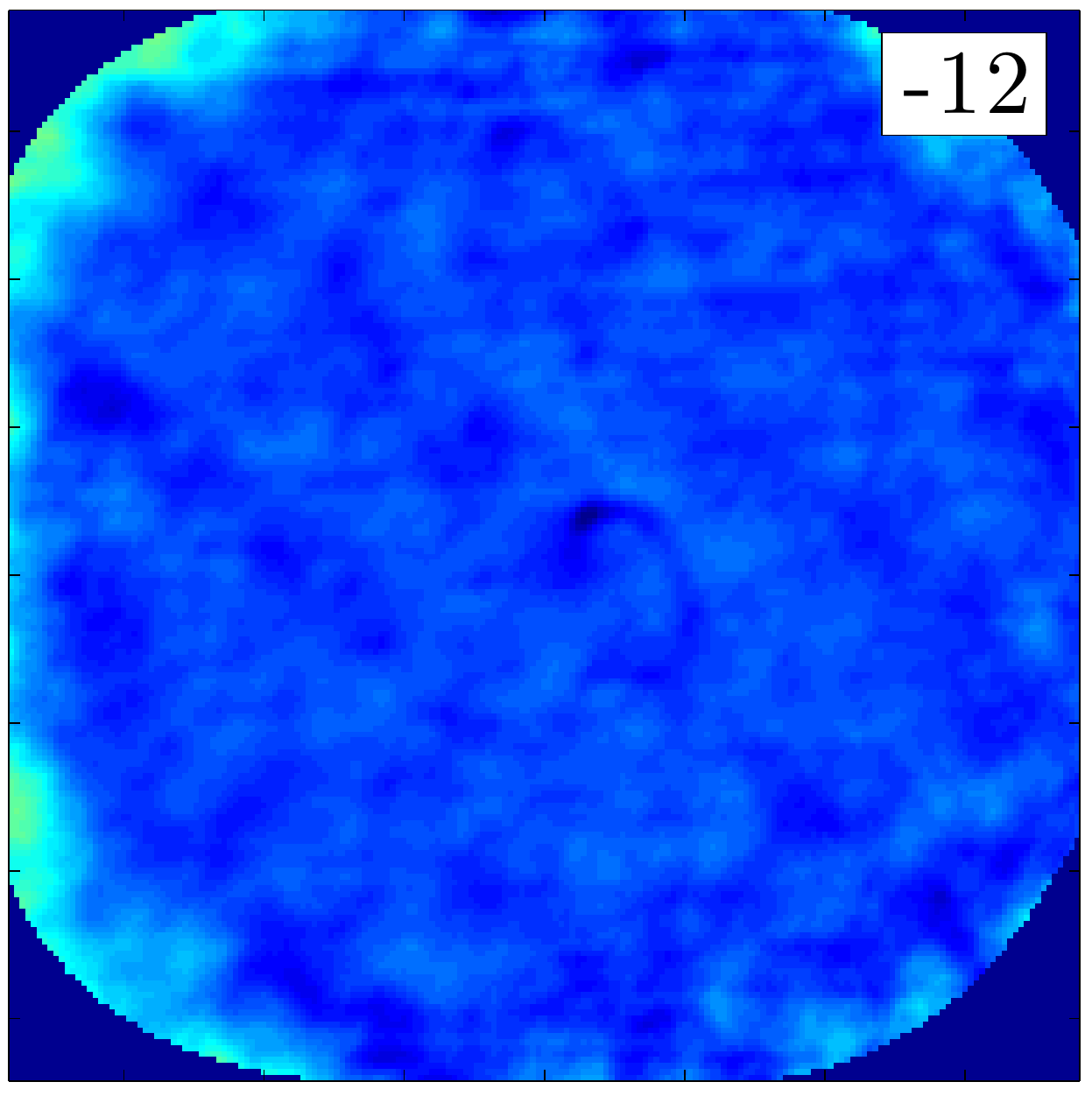}
\includegraphics[height=3.9cm]{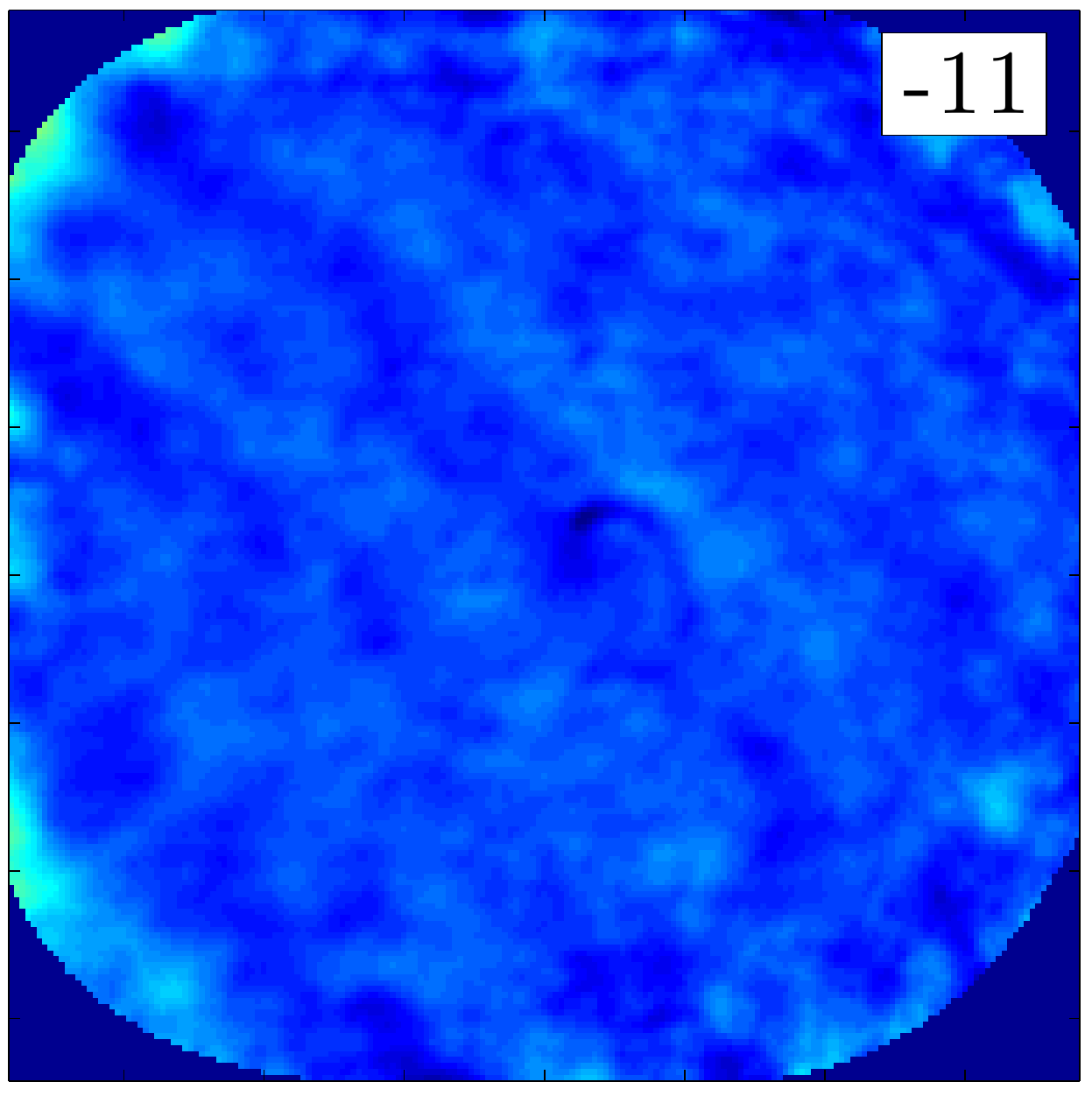}
\includegraphics[height=3.9cm]{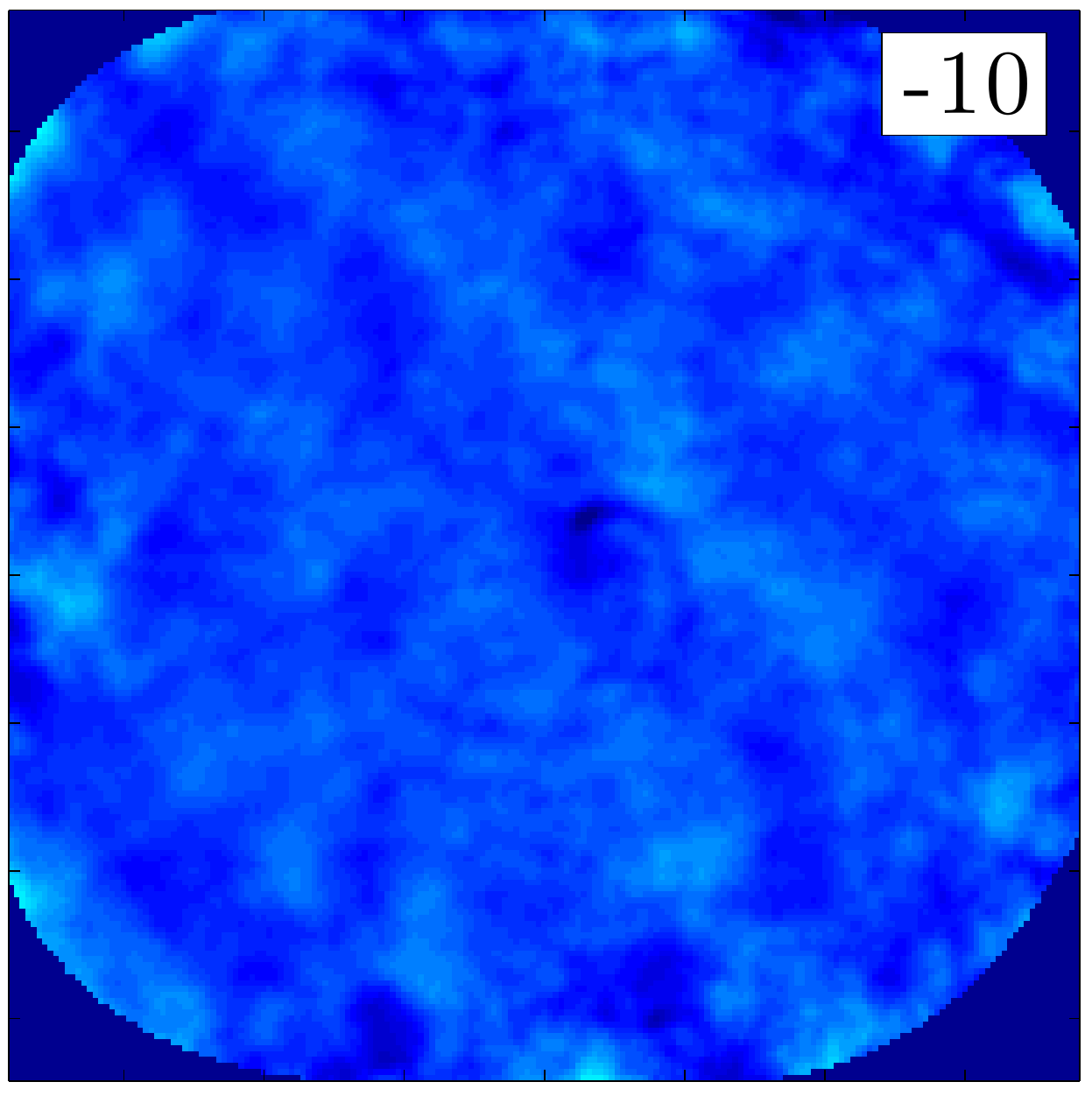}
\includegraphics[height=3.9cm]{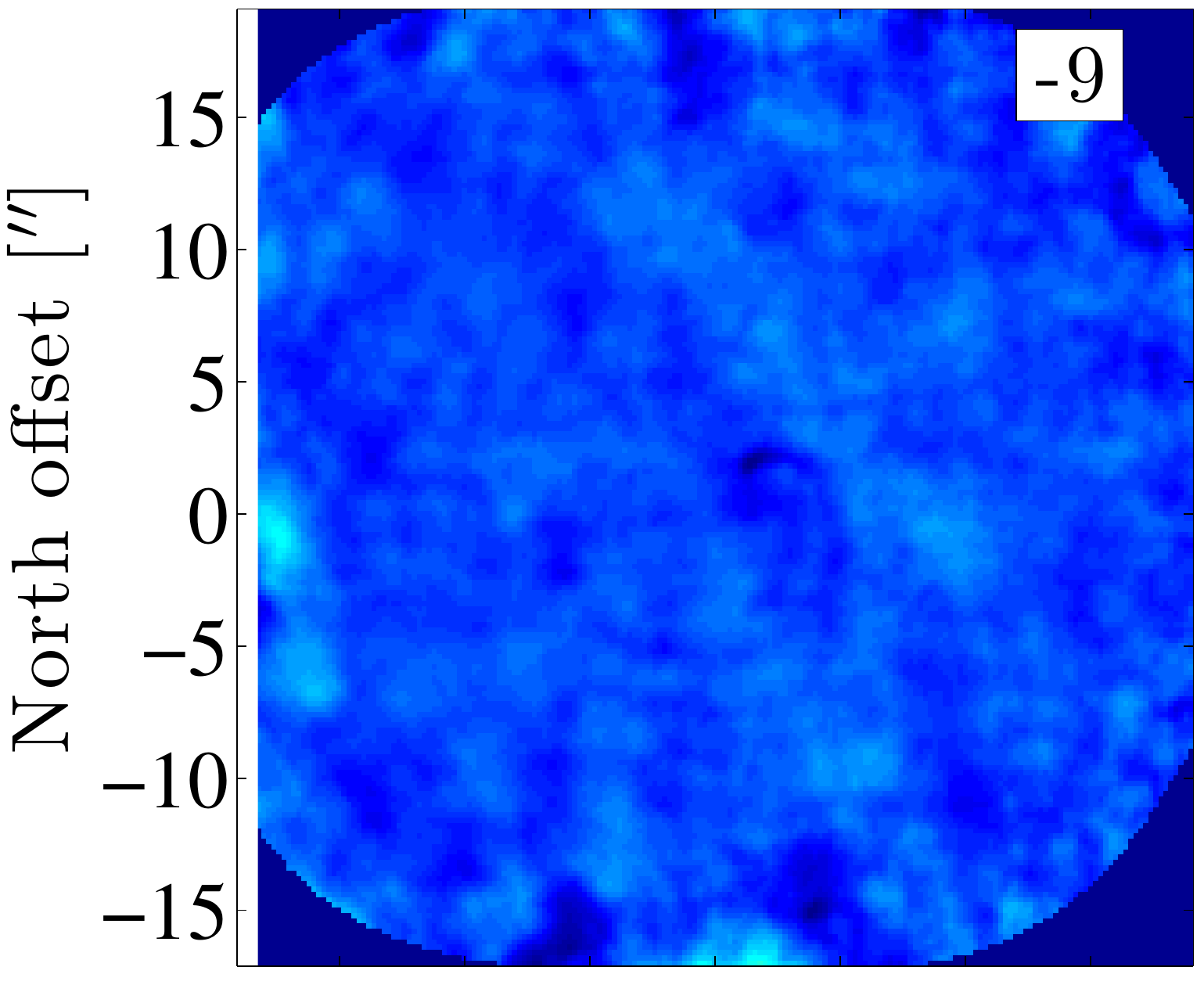}
\includegraphics[height=3.9cm]{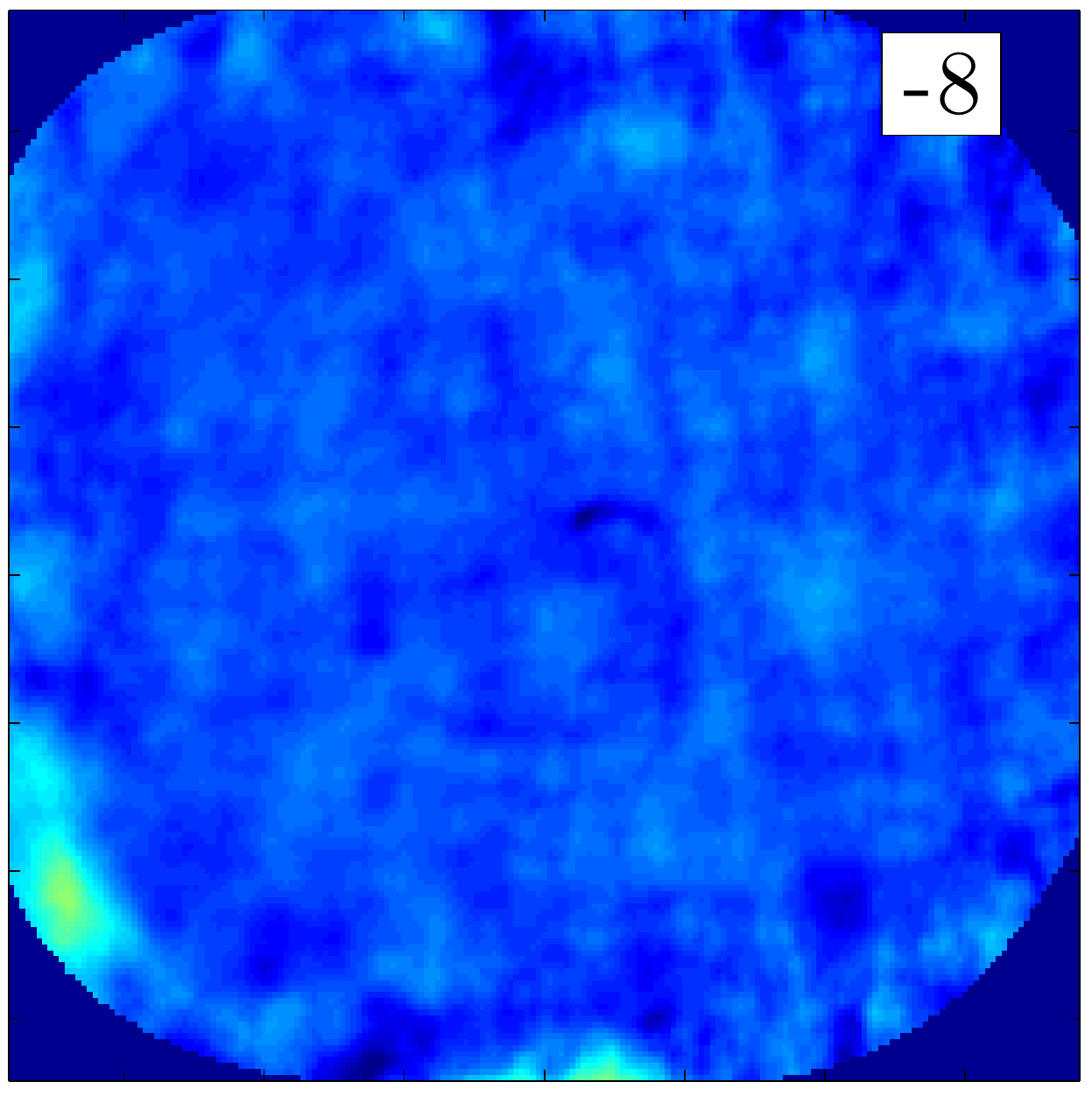}
\includegraphics[height=3.9cm]{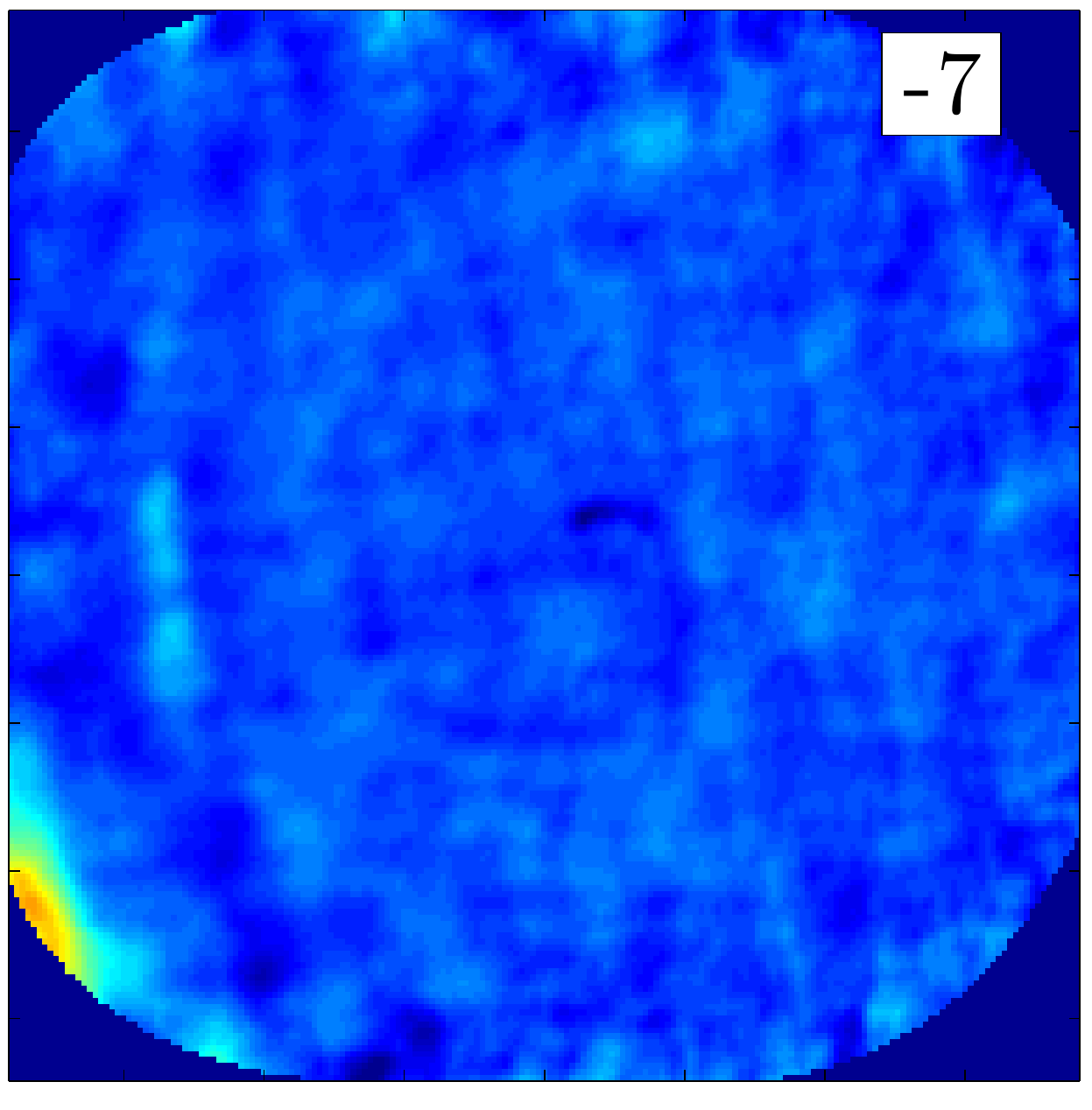}
\includegraphics[height=3.9cm]{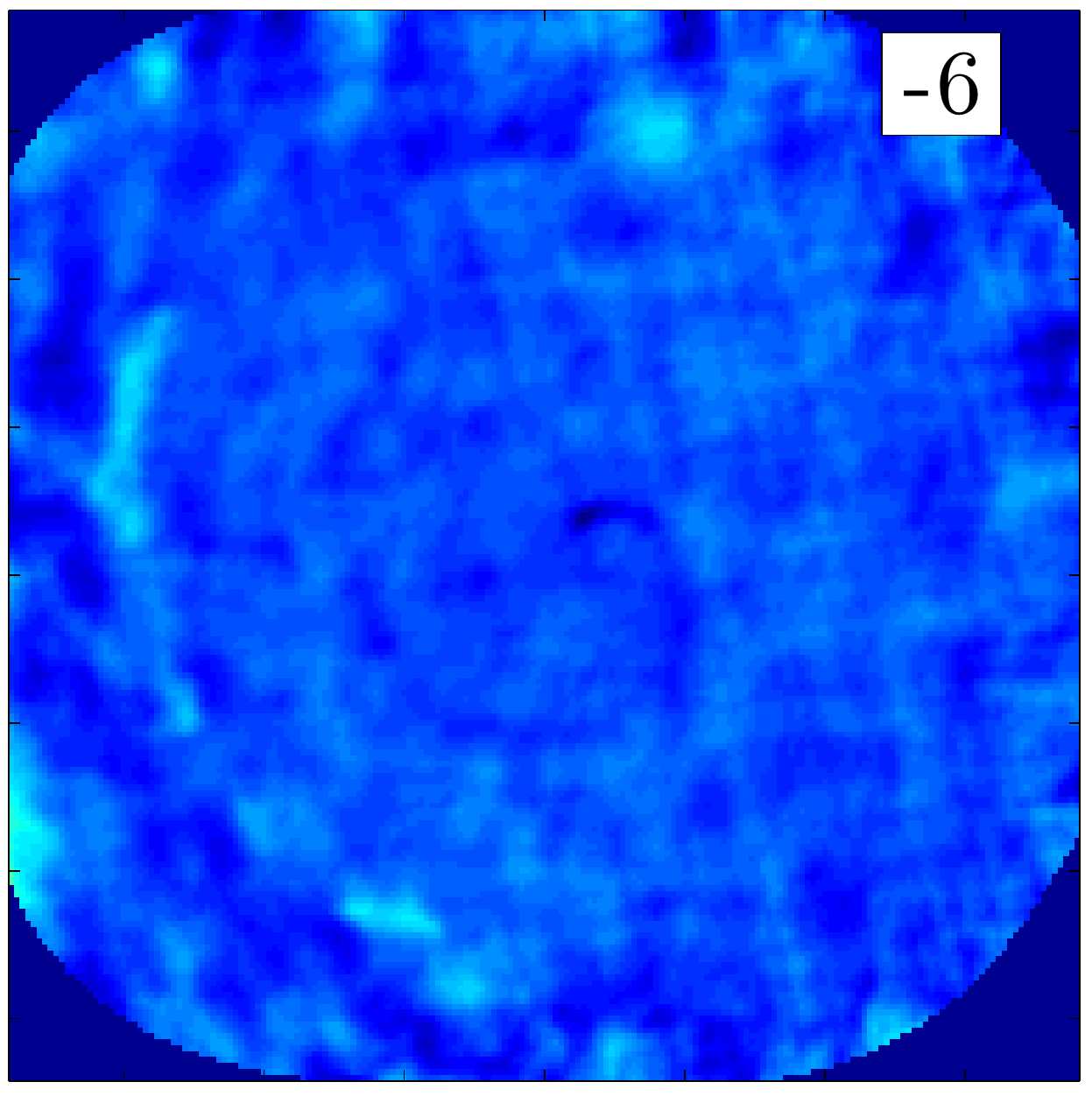}
\includegraphics[height=3.9cm]{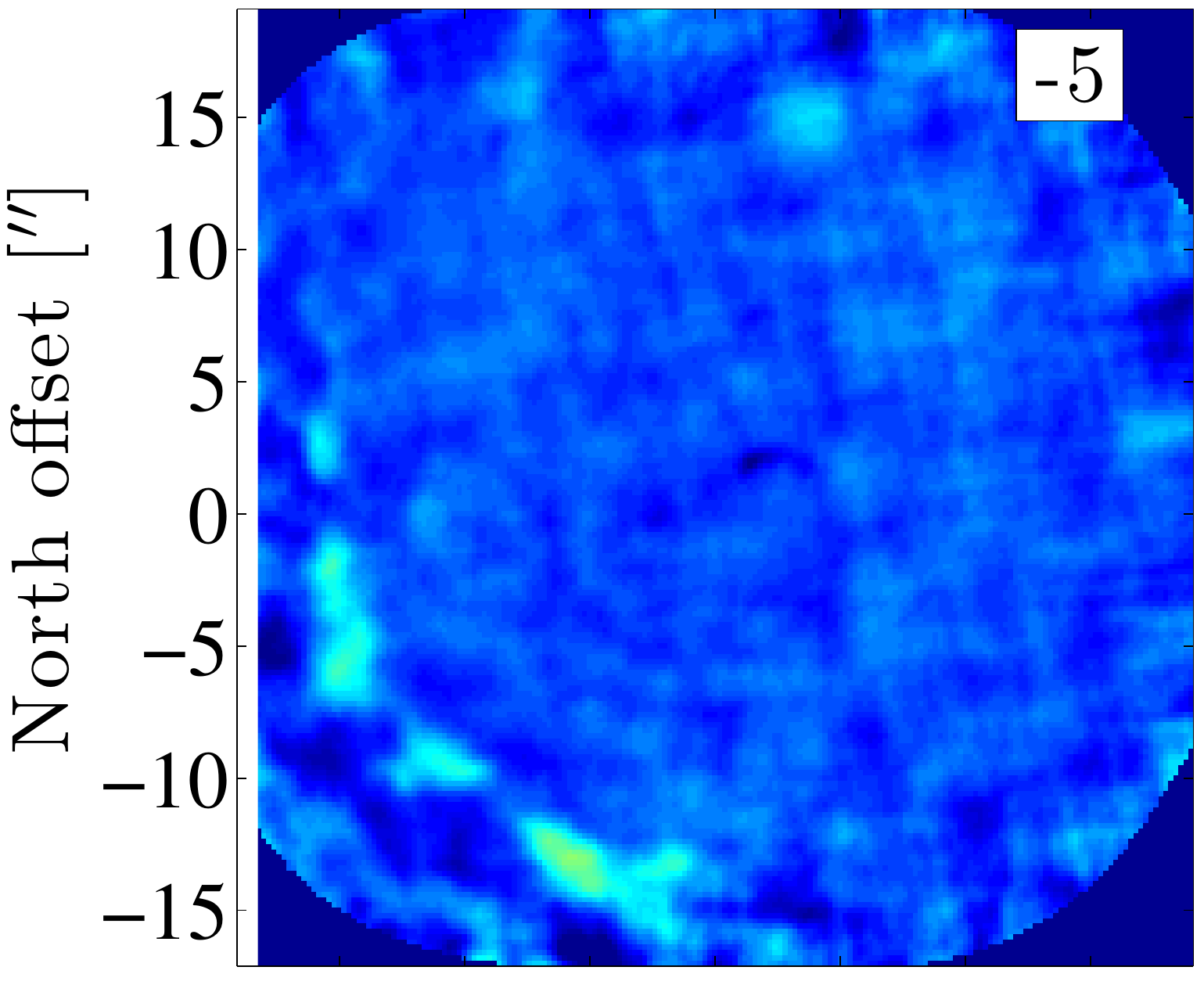}
\includegraphics[height=3.9cm]{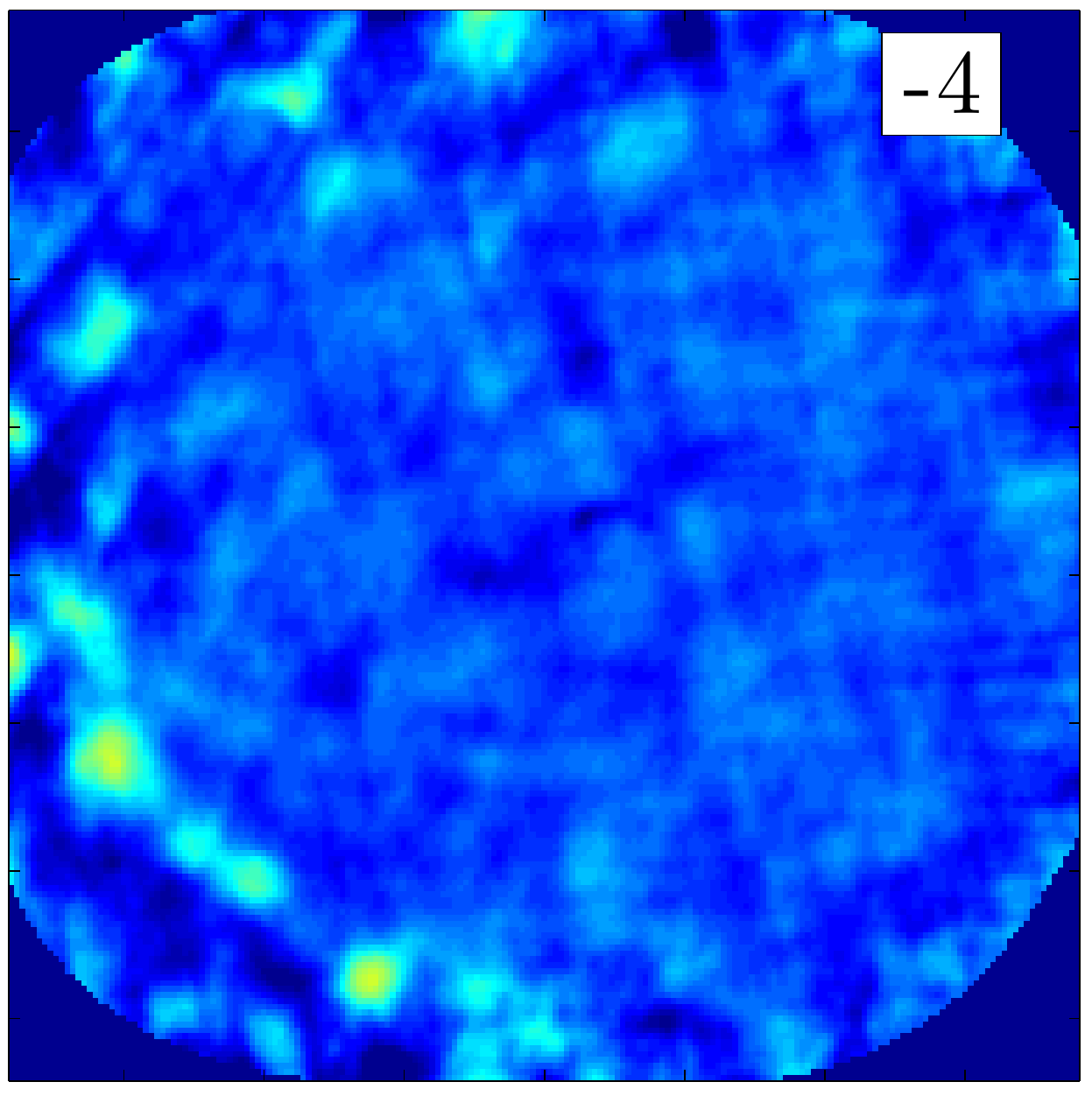}
\includegraphics[height=3.9cm]{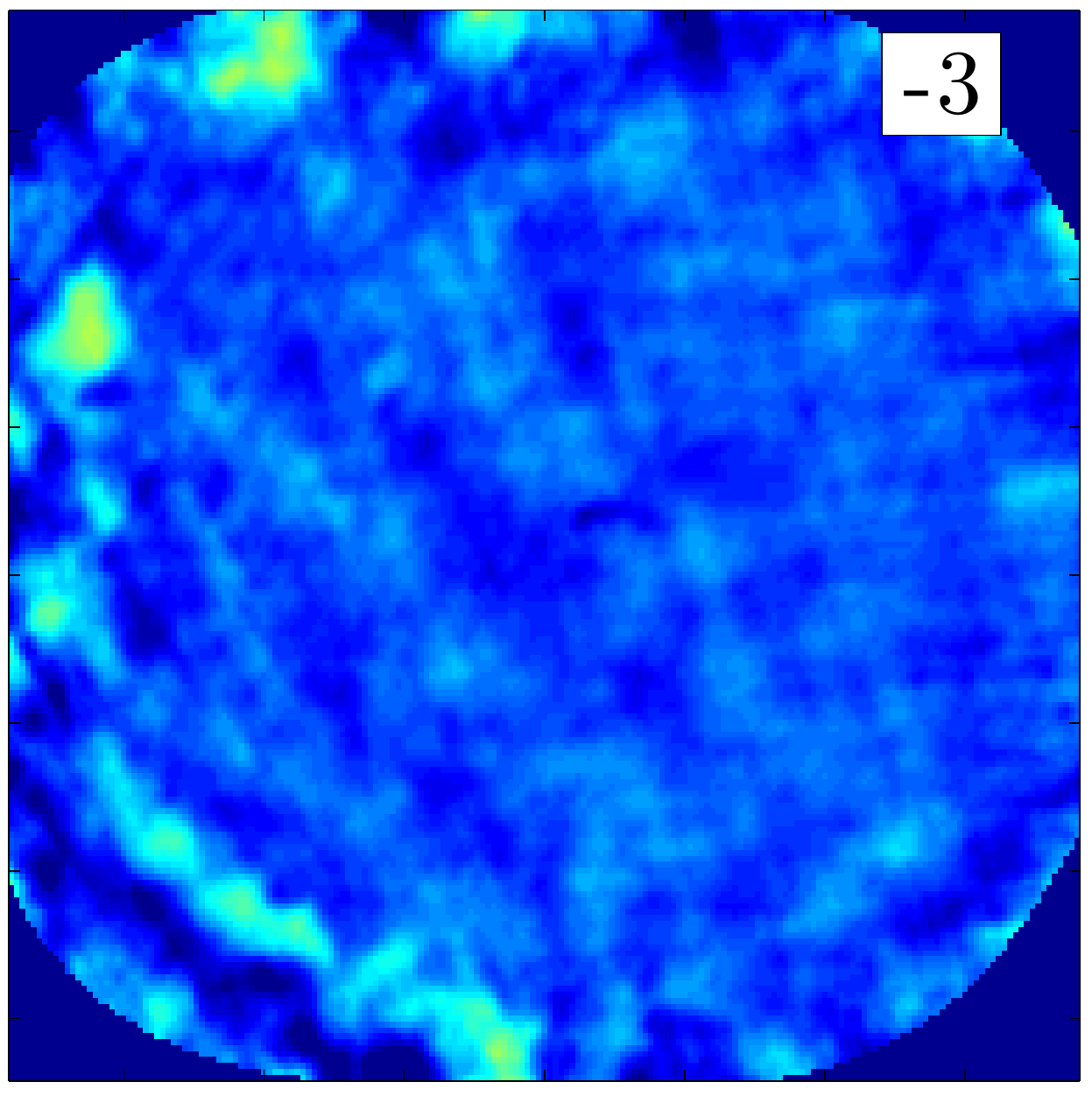}
\includegraphics[height=3.9cm]{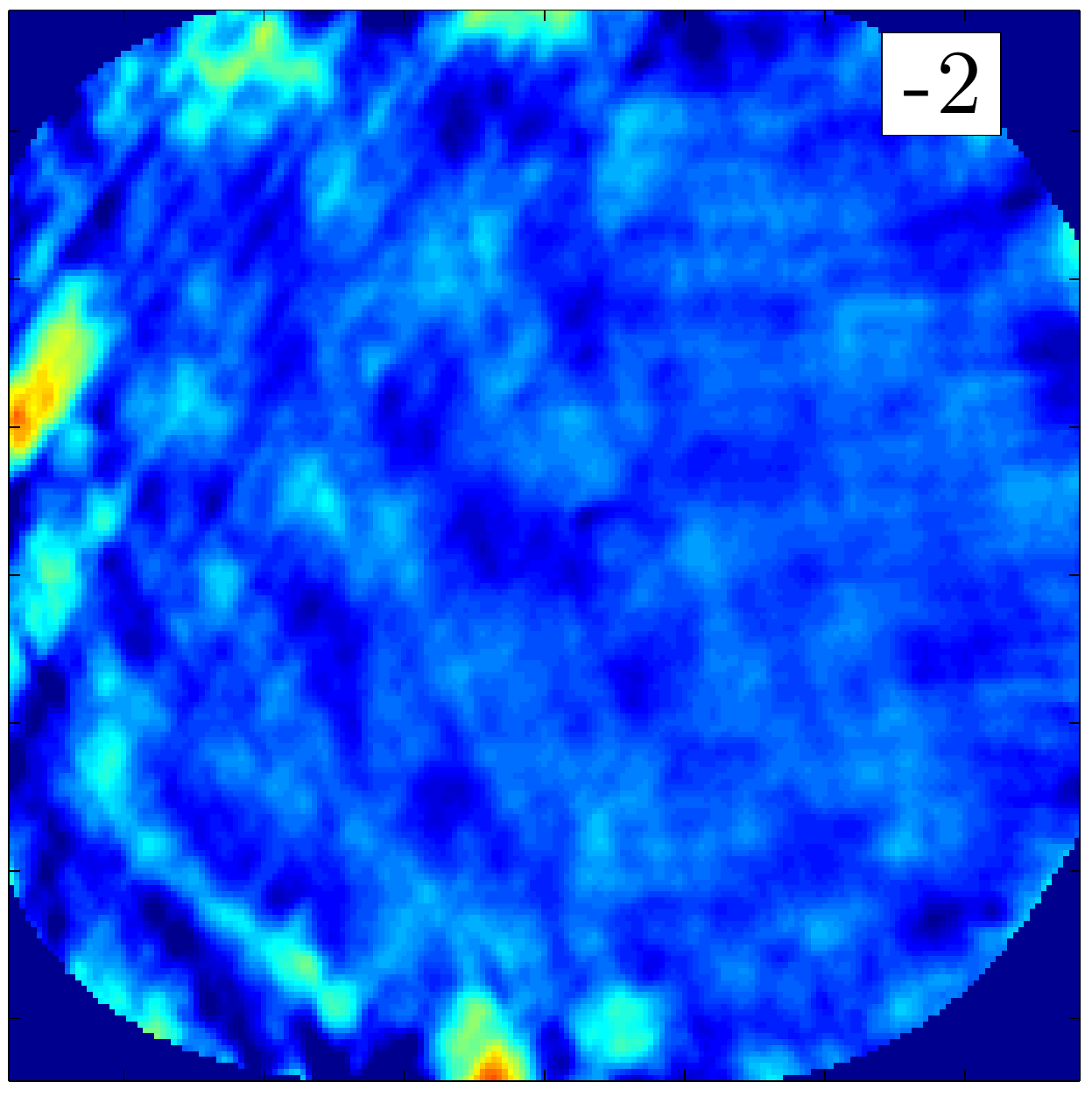}
\includegraphics[width=4.72cm]{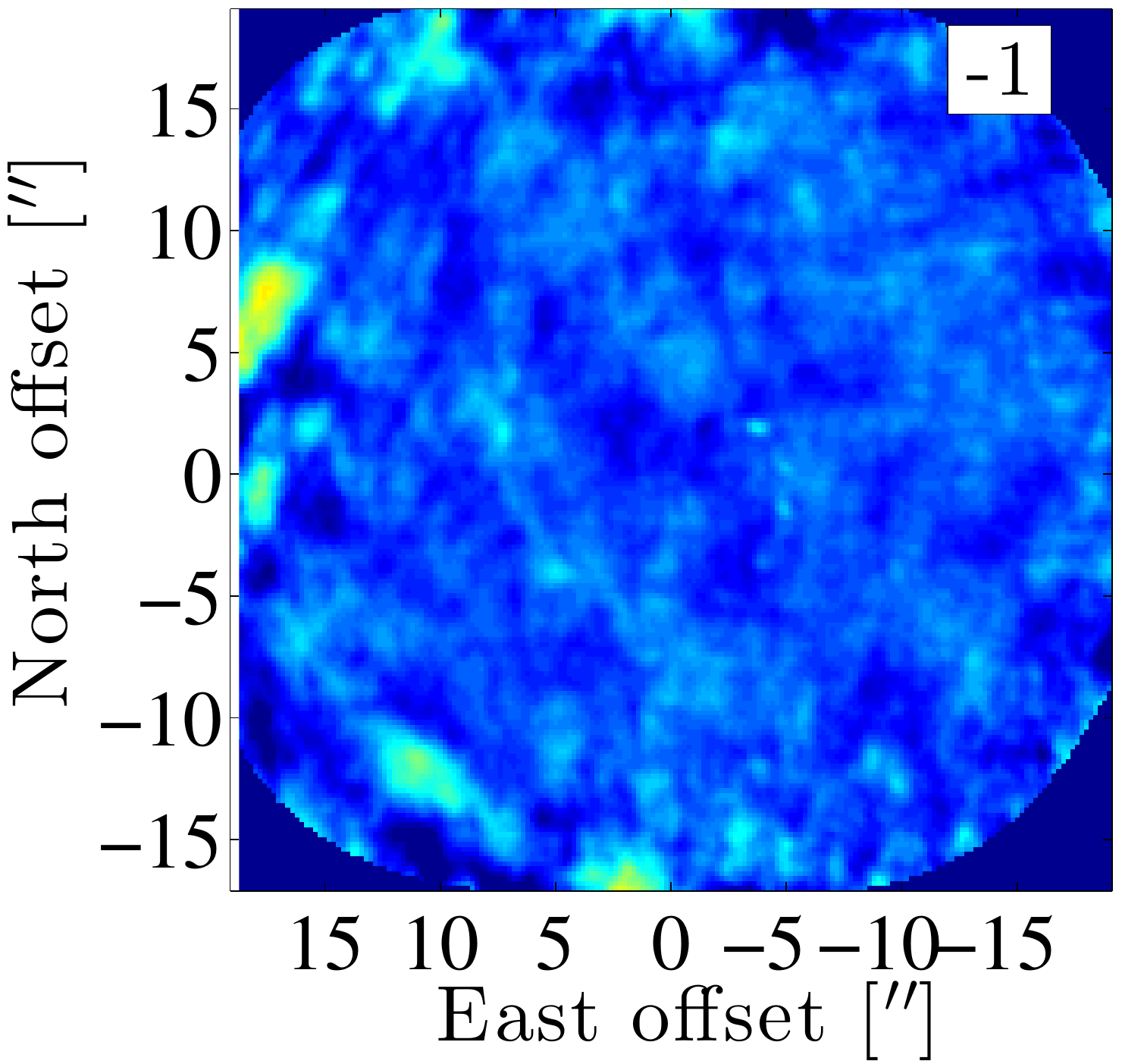}
\includegraphics[width=3.83cm]{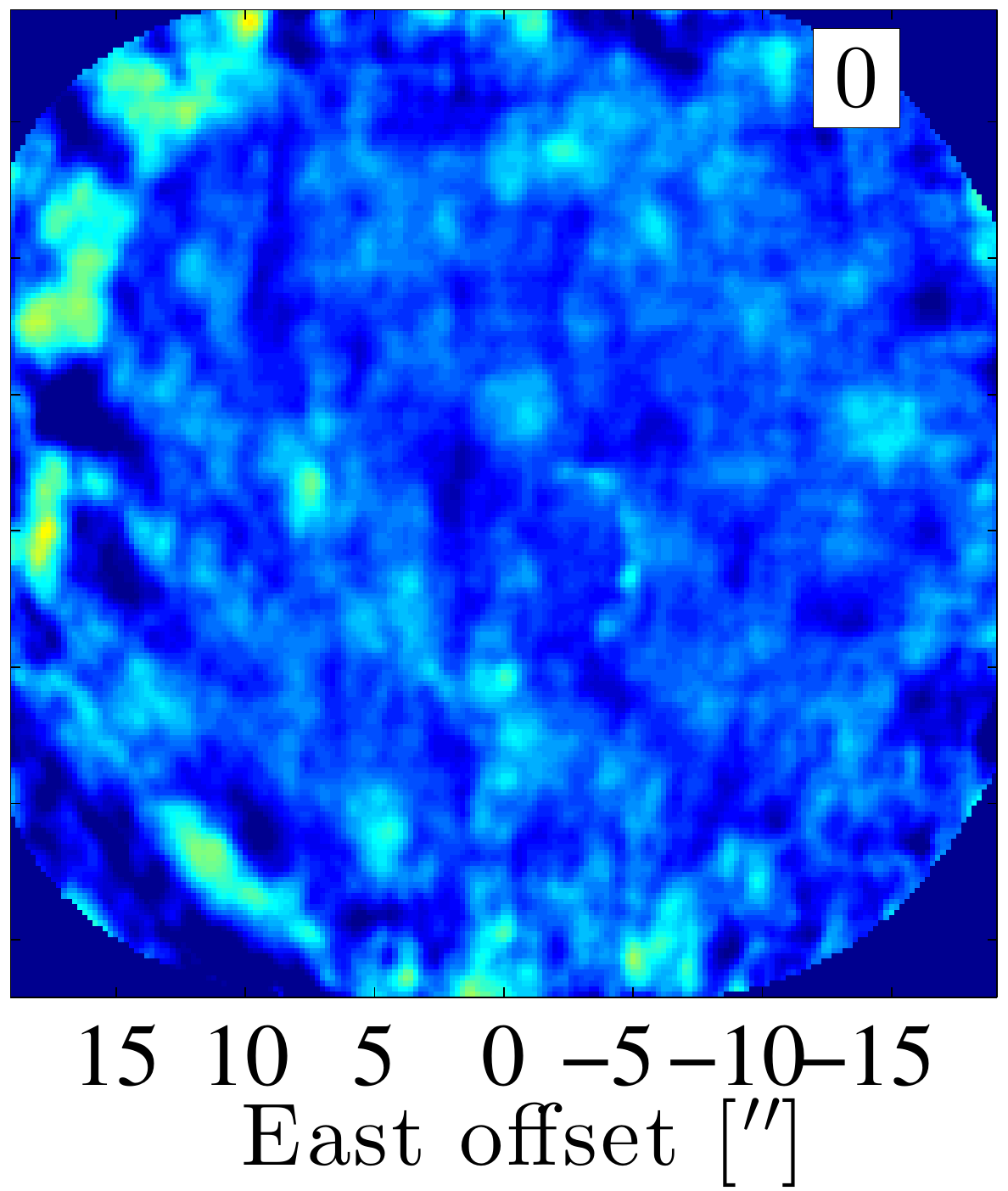}
\caption{Channel maps from the model with $e$=0.2 (see text for explanation). The velocity relative to the line centre of each channel is given in the upper right corner legend. Only the blue-shifted emission is shown since the maps are completely symmetric around the line centre.}
\end{figure*}

\begin{figure*}
\flushleft
\label{mchannelmaps}
\includegraphics[height=3.9cm]{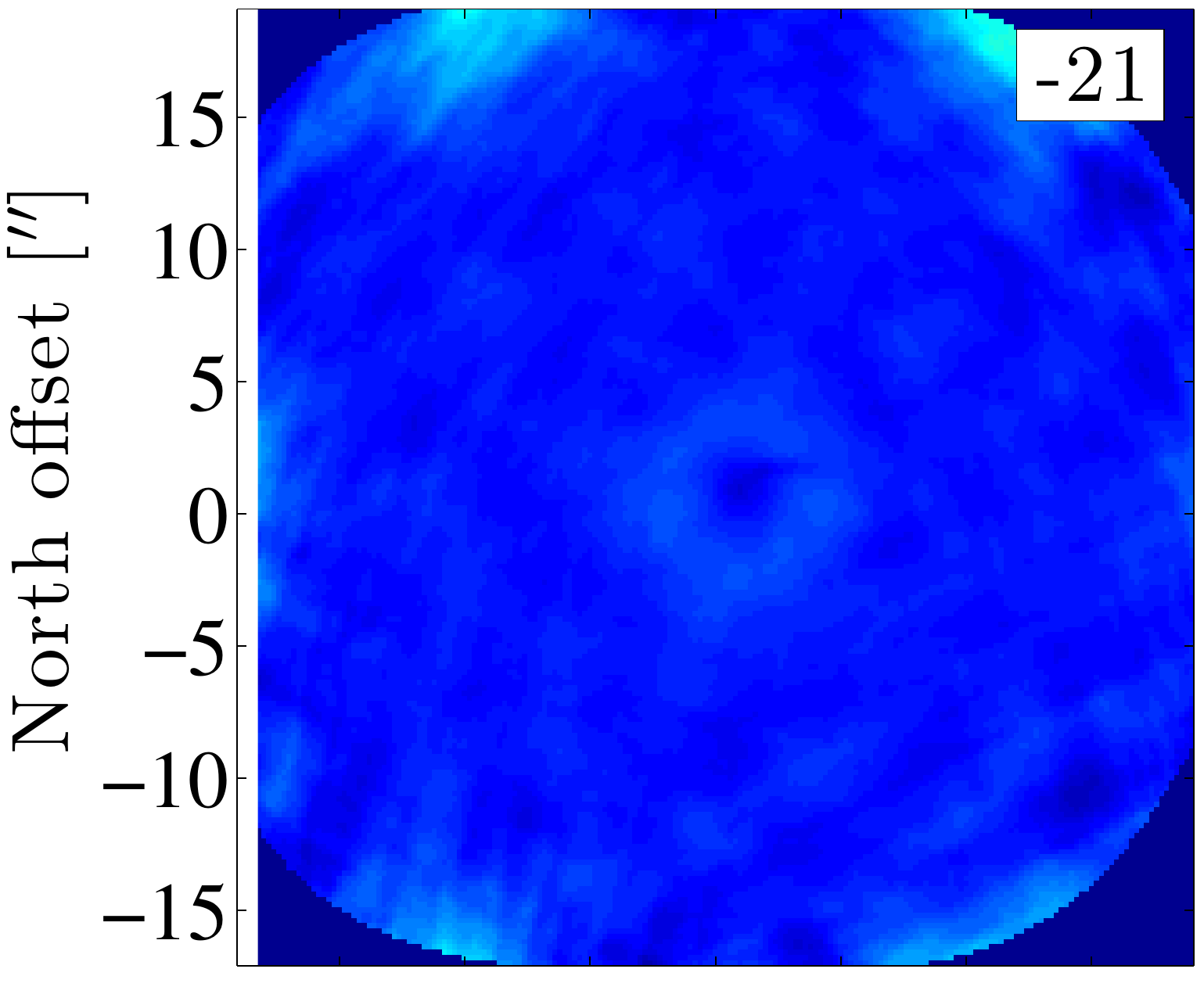}
\includegraphics[height=3.9cm]{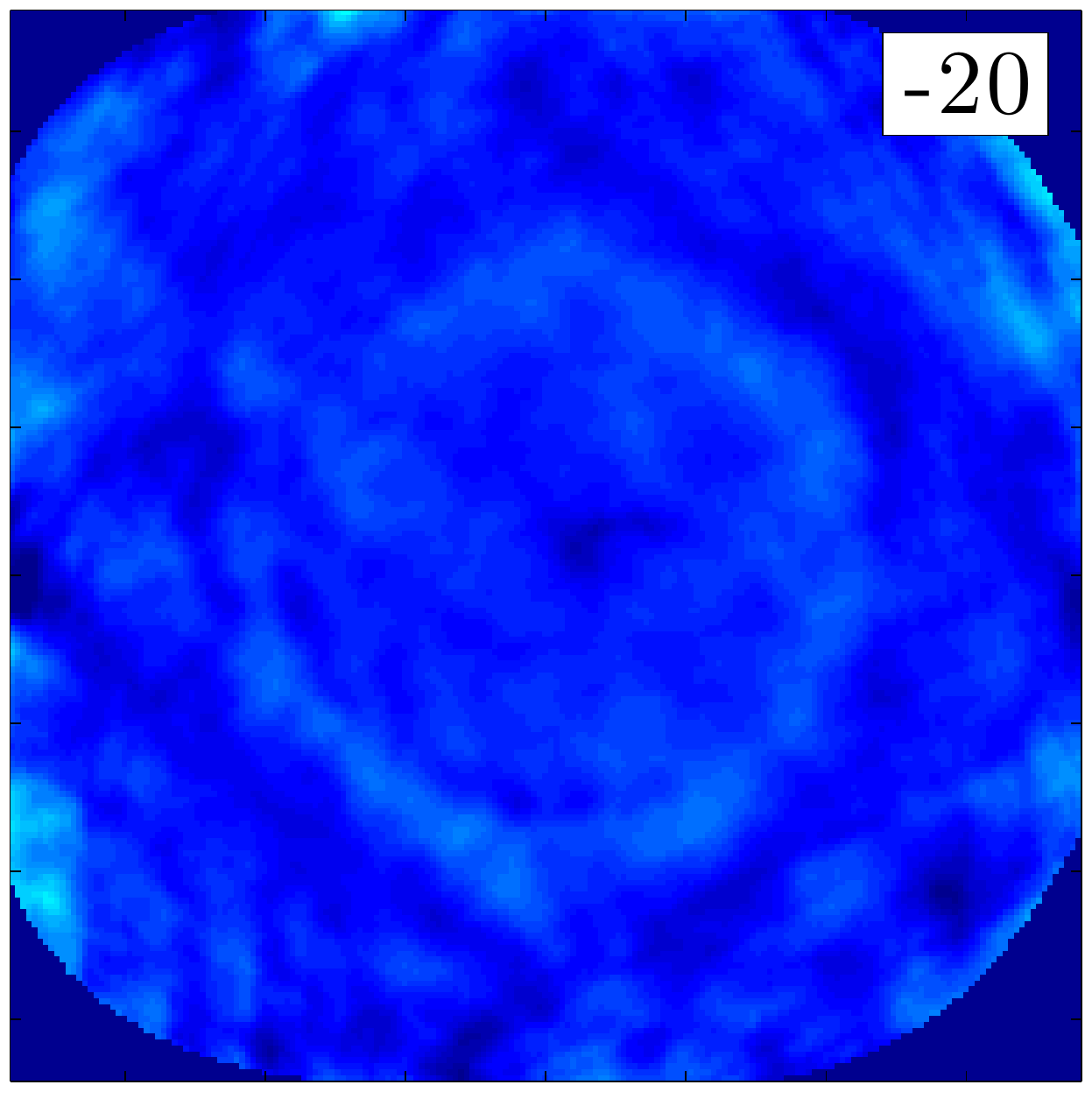}
\includegraphics[height=3.9cm]{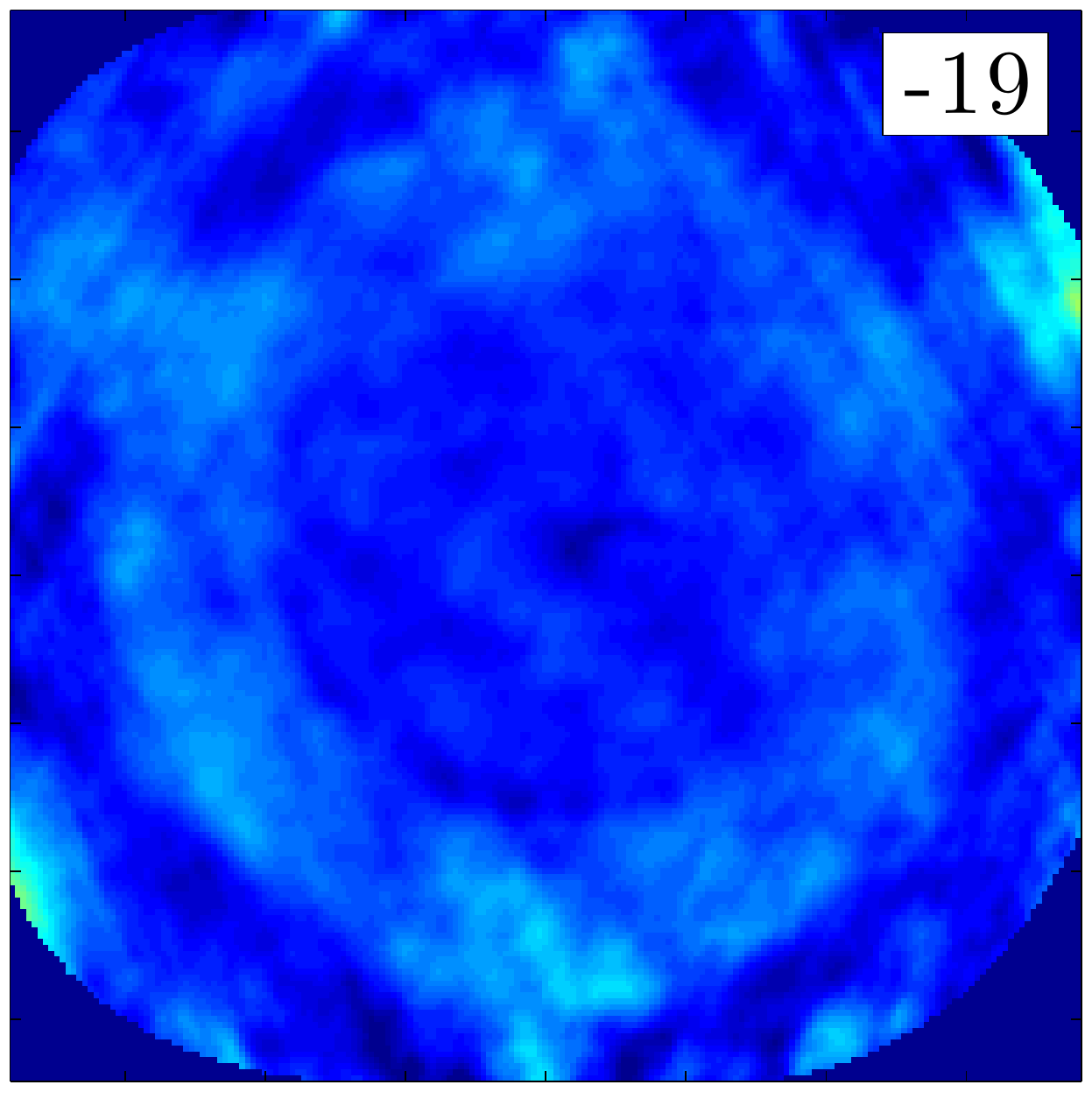}
\includegraphics[height=3.9cm]{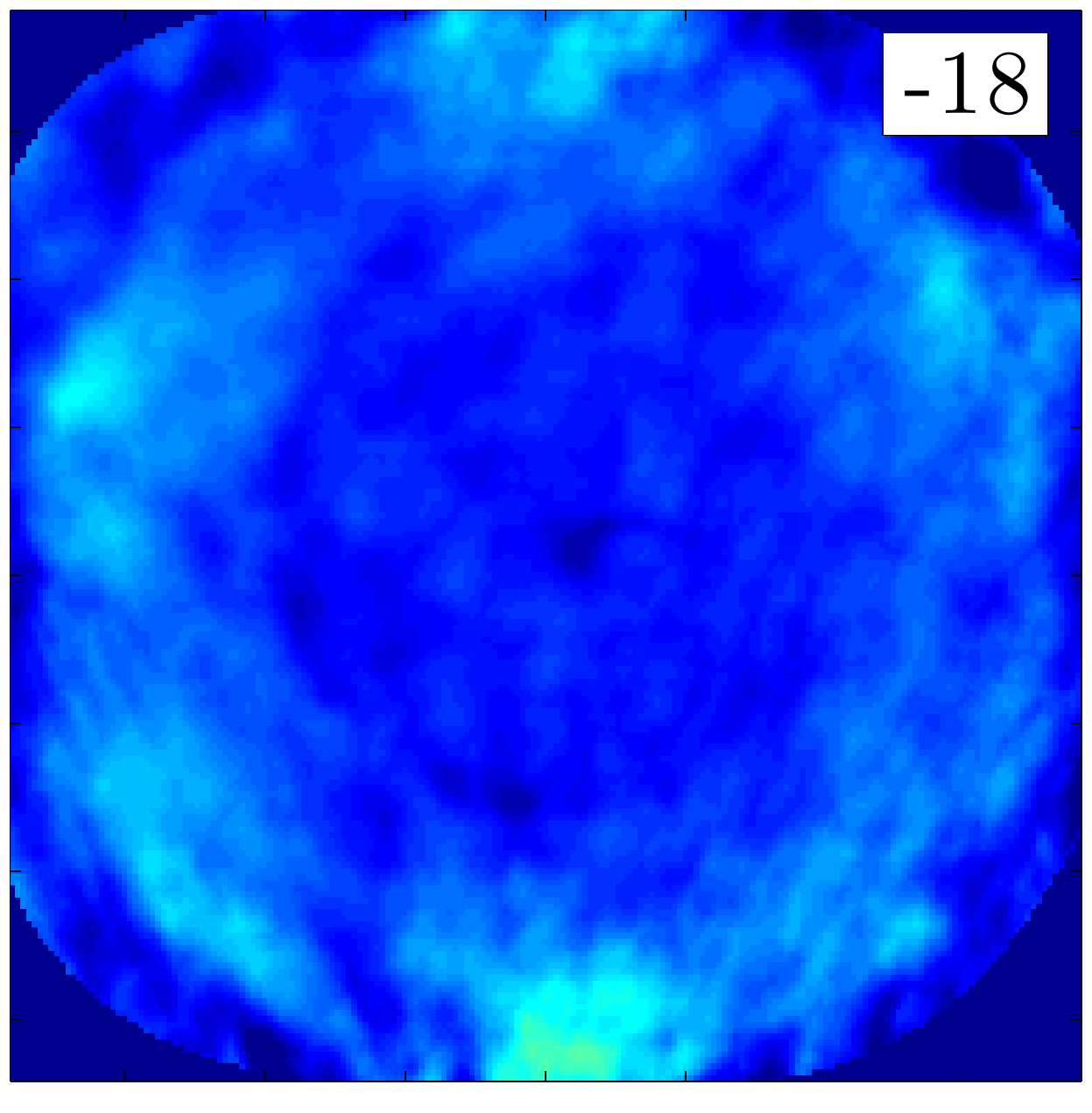}
\includegraphics[height=3.9cm]{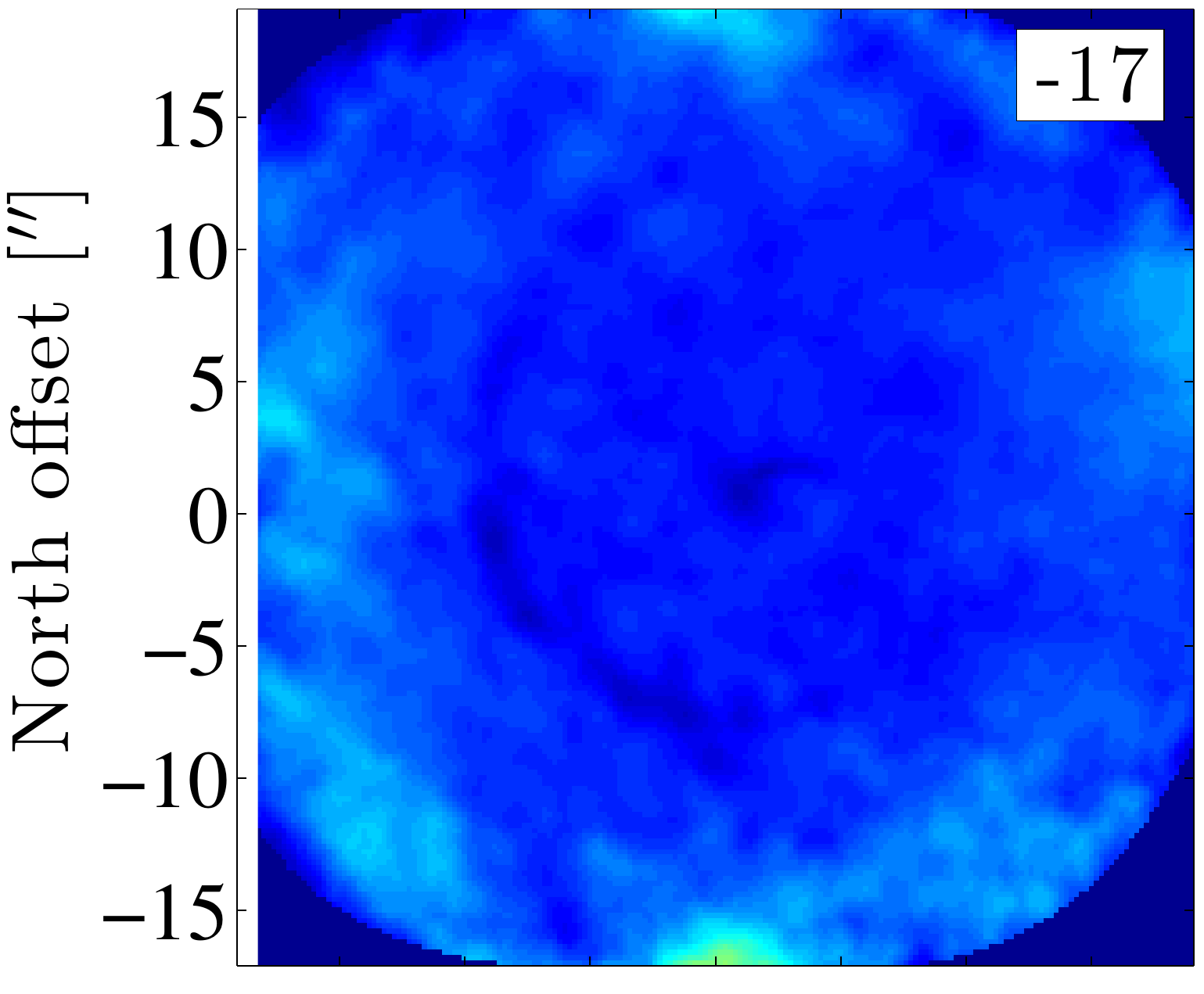}
\includegraphics[height=3.9cm]{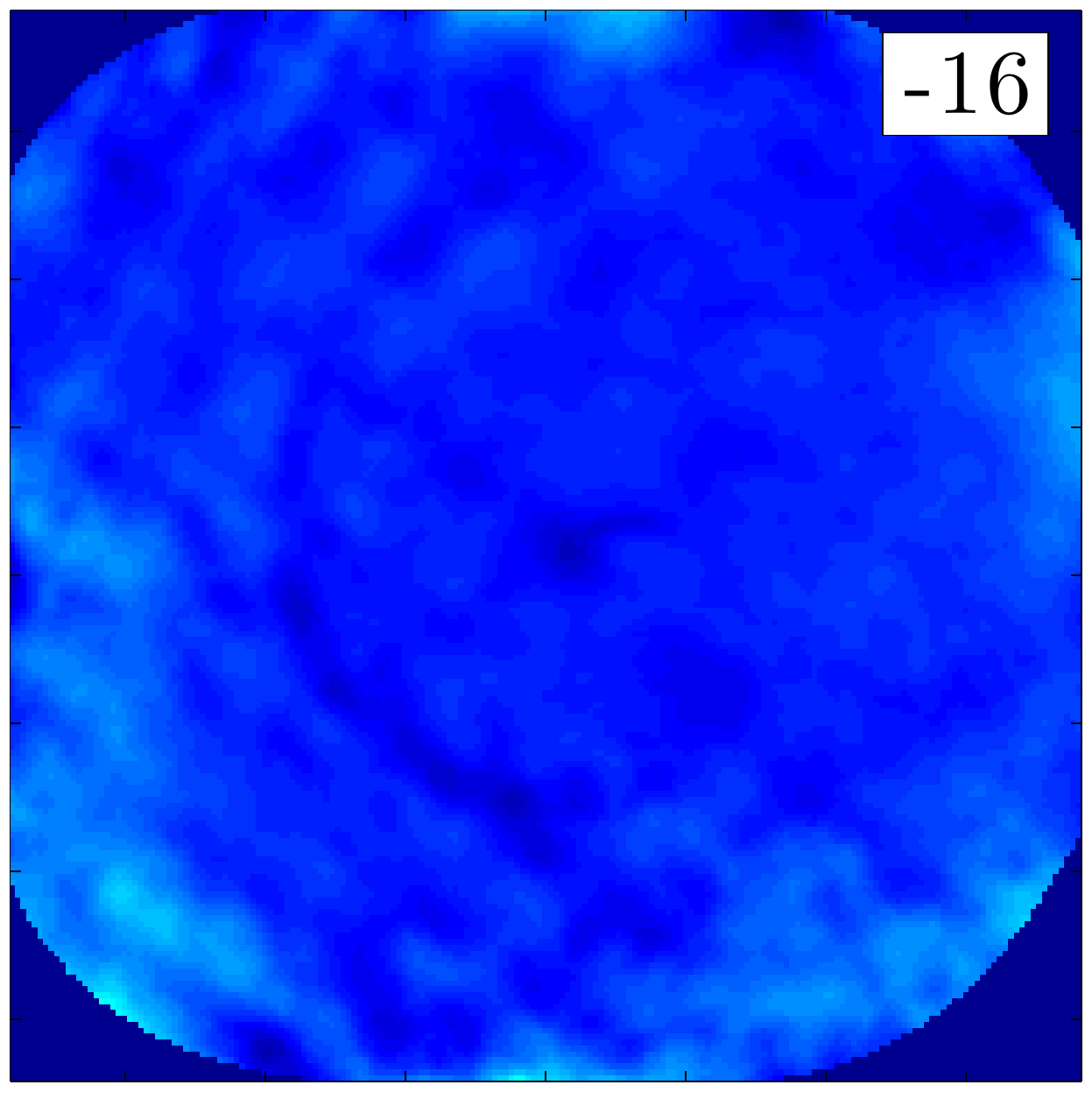}
\includegraphics[height=3.9cm]{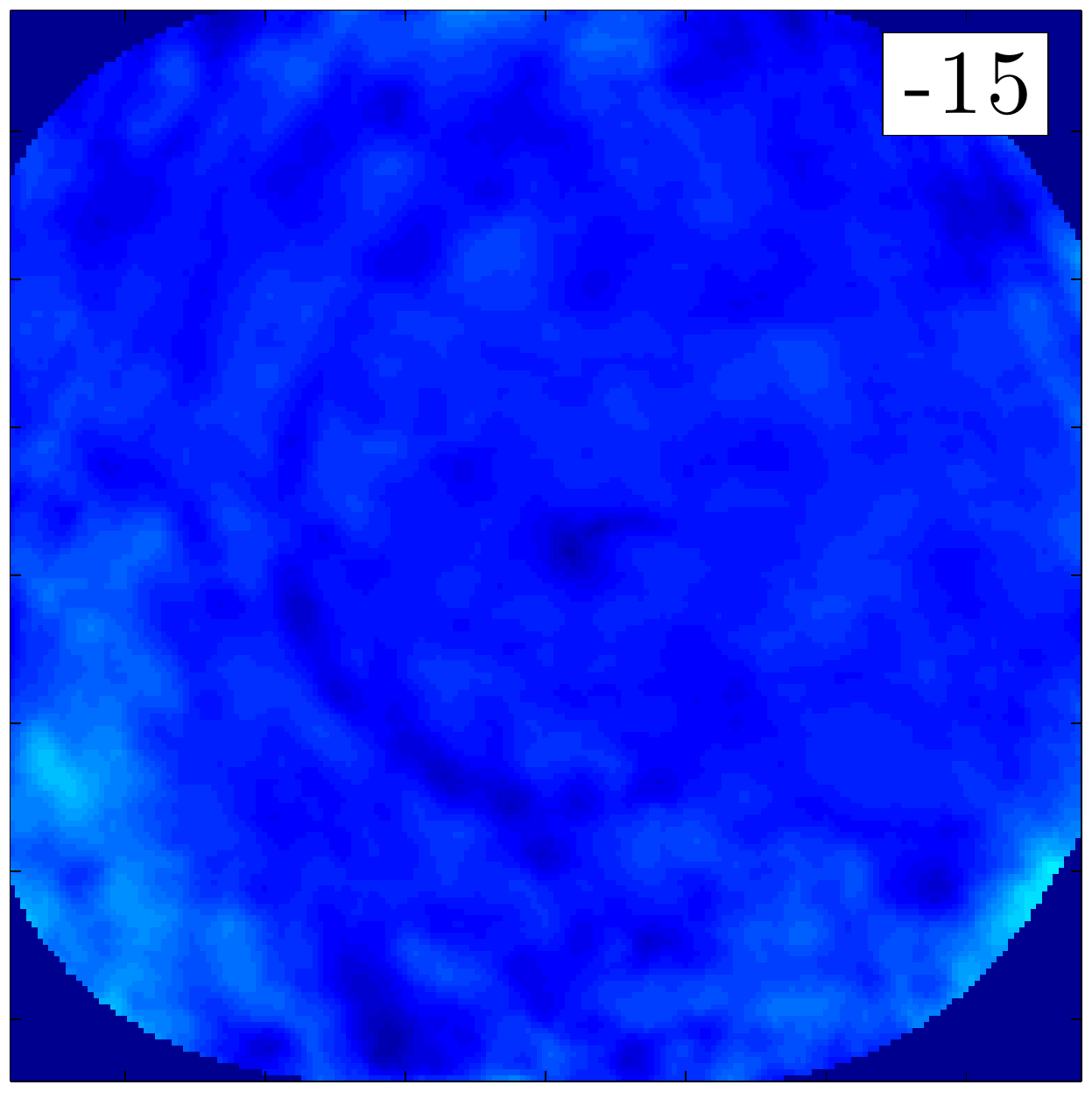}
\includegraphics[height=3.9cm]{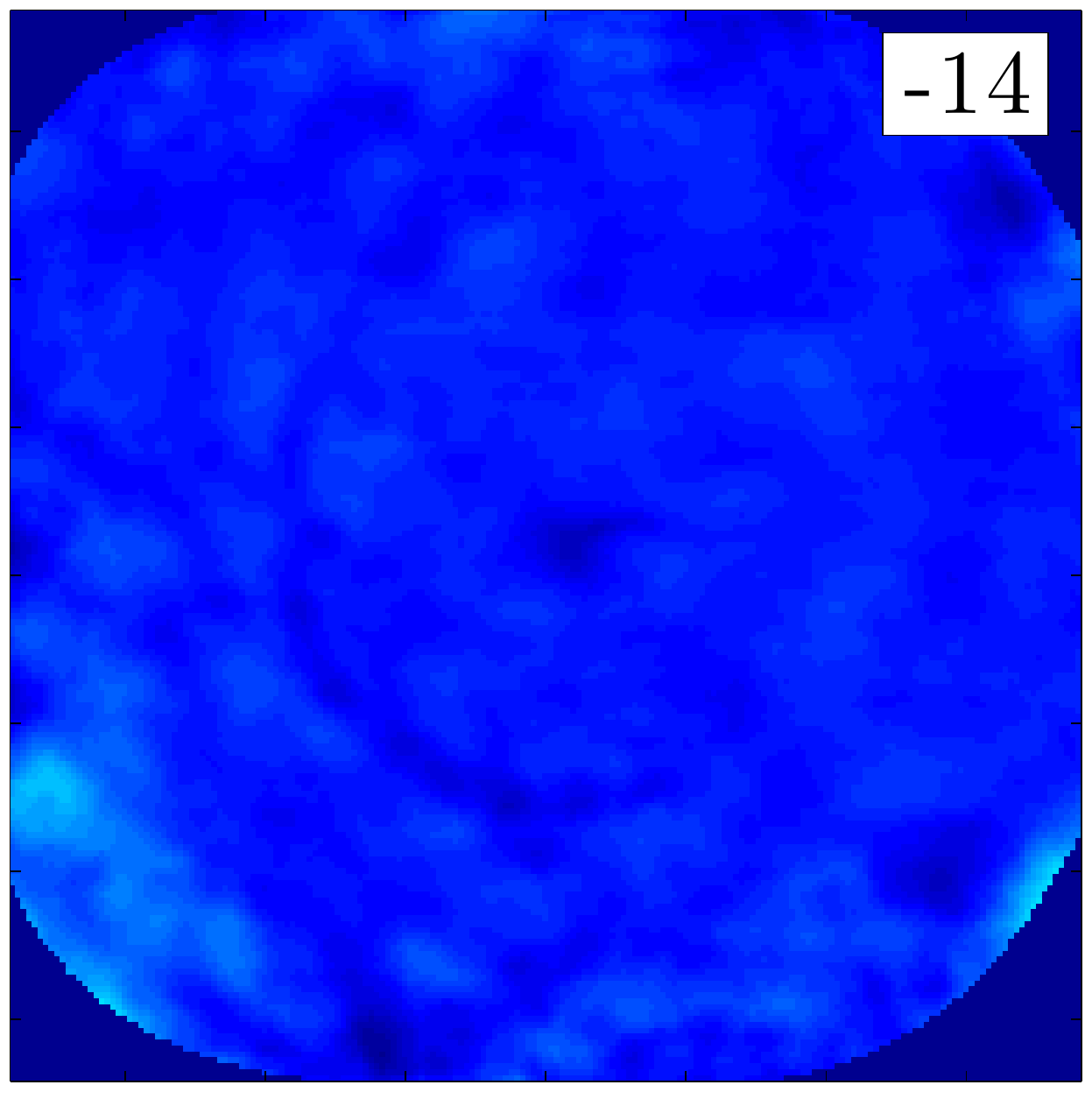}
\includegraphics[height=3.9cm]{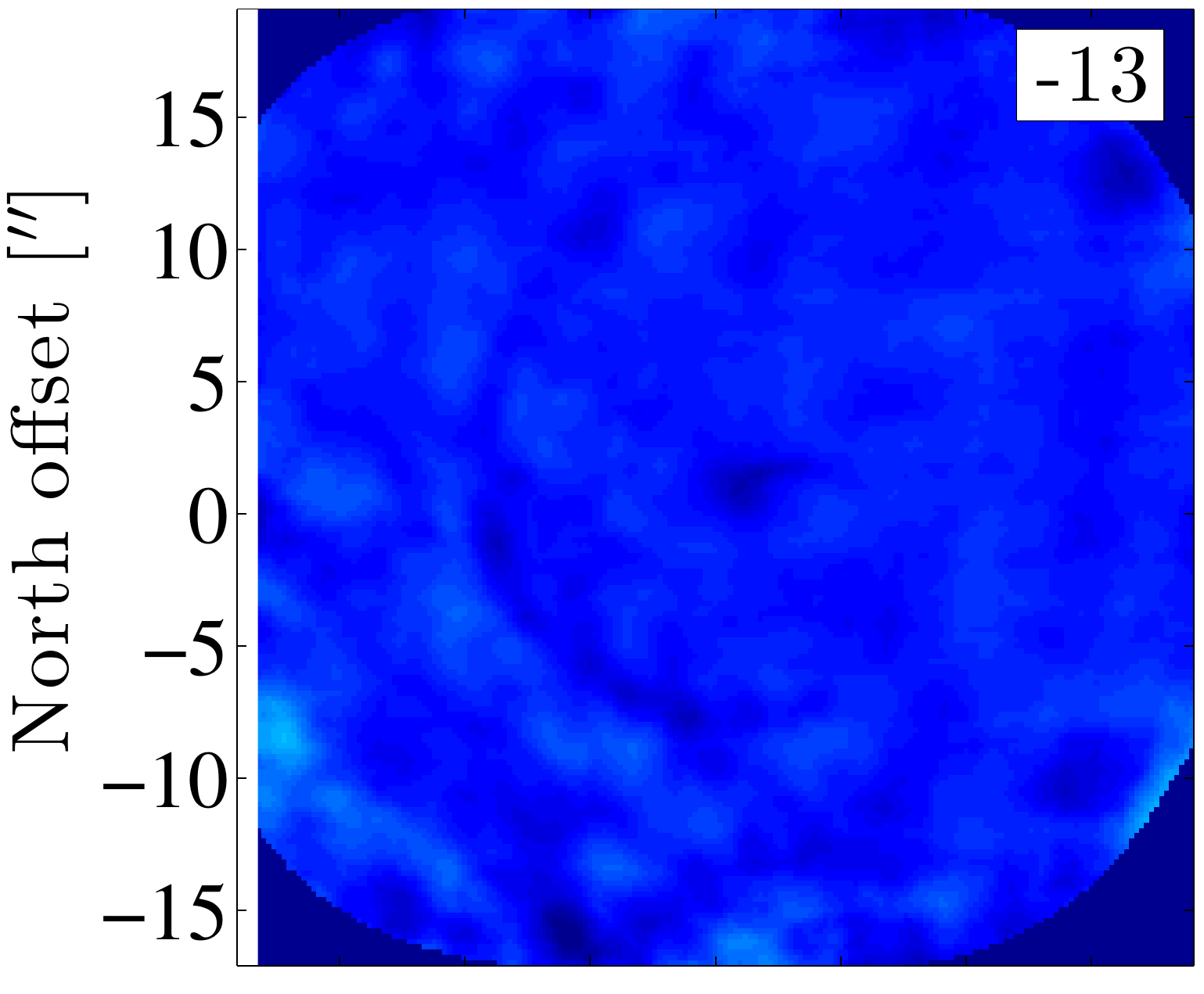}
\includegraphics[height=3.9cm]{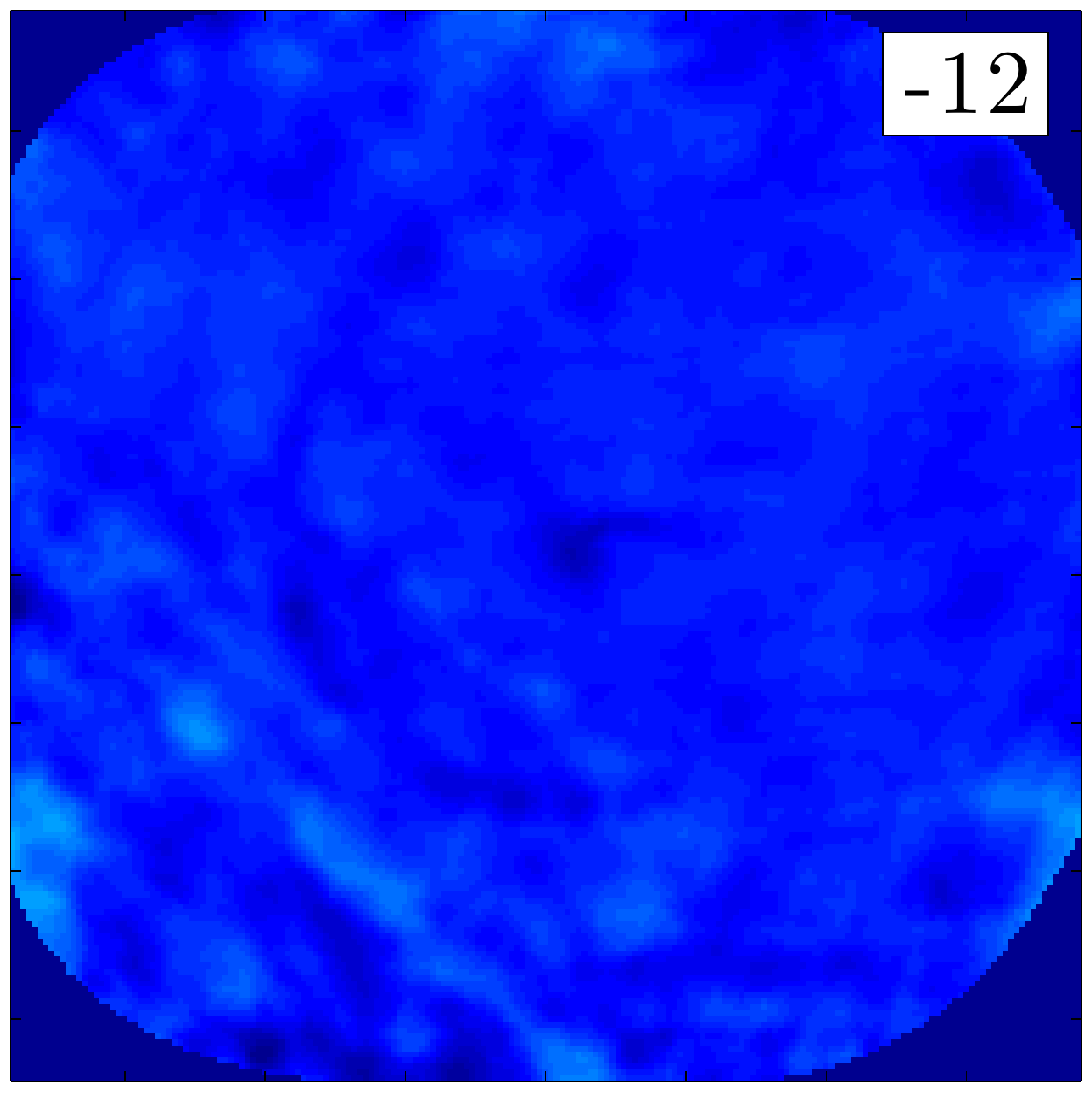}
\includegraphics[height=3.9cm]{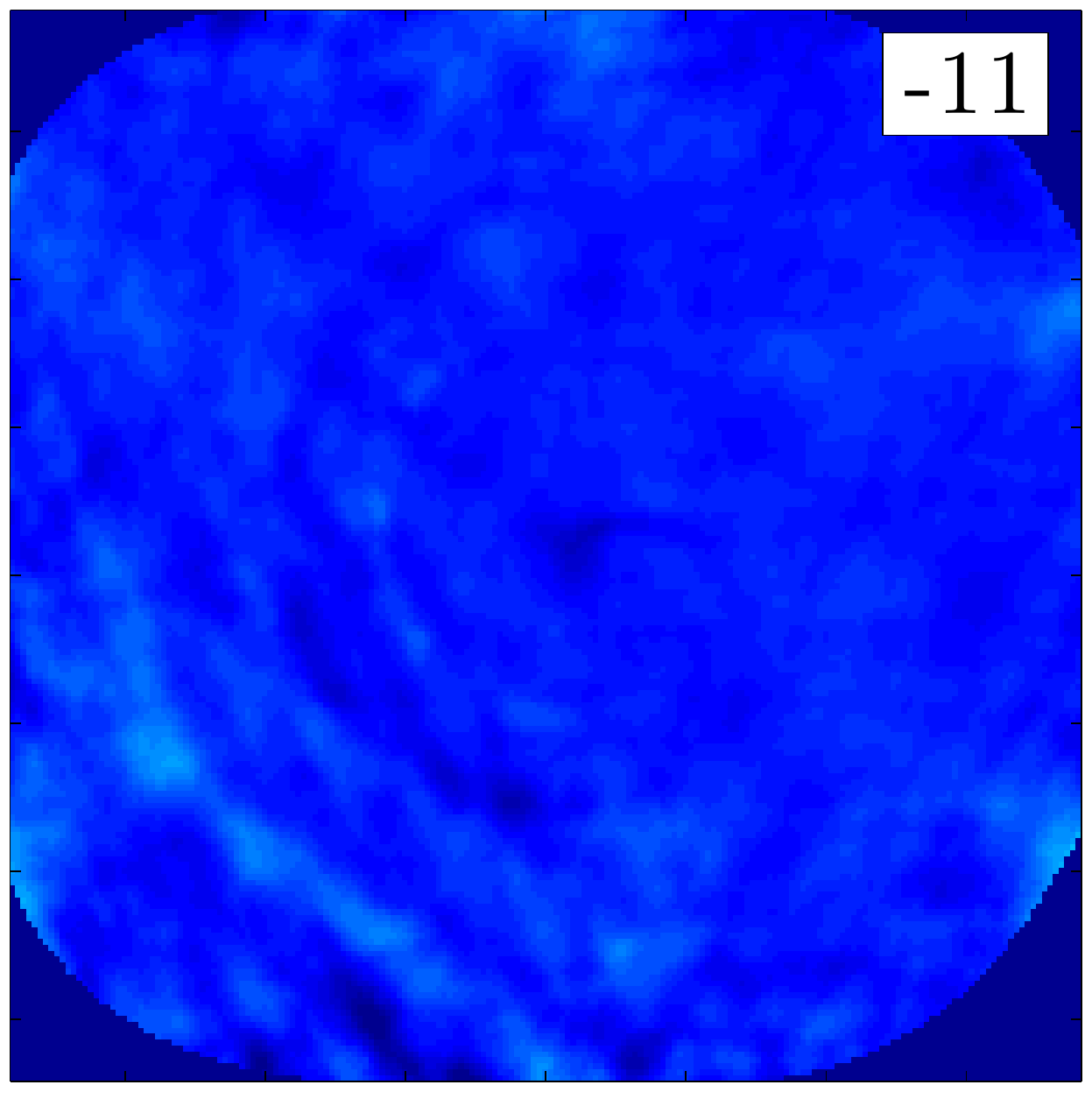}
\includegraphics[height=3.9cm]{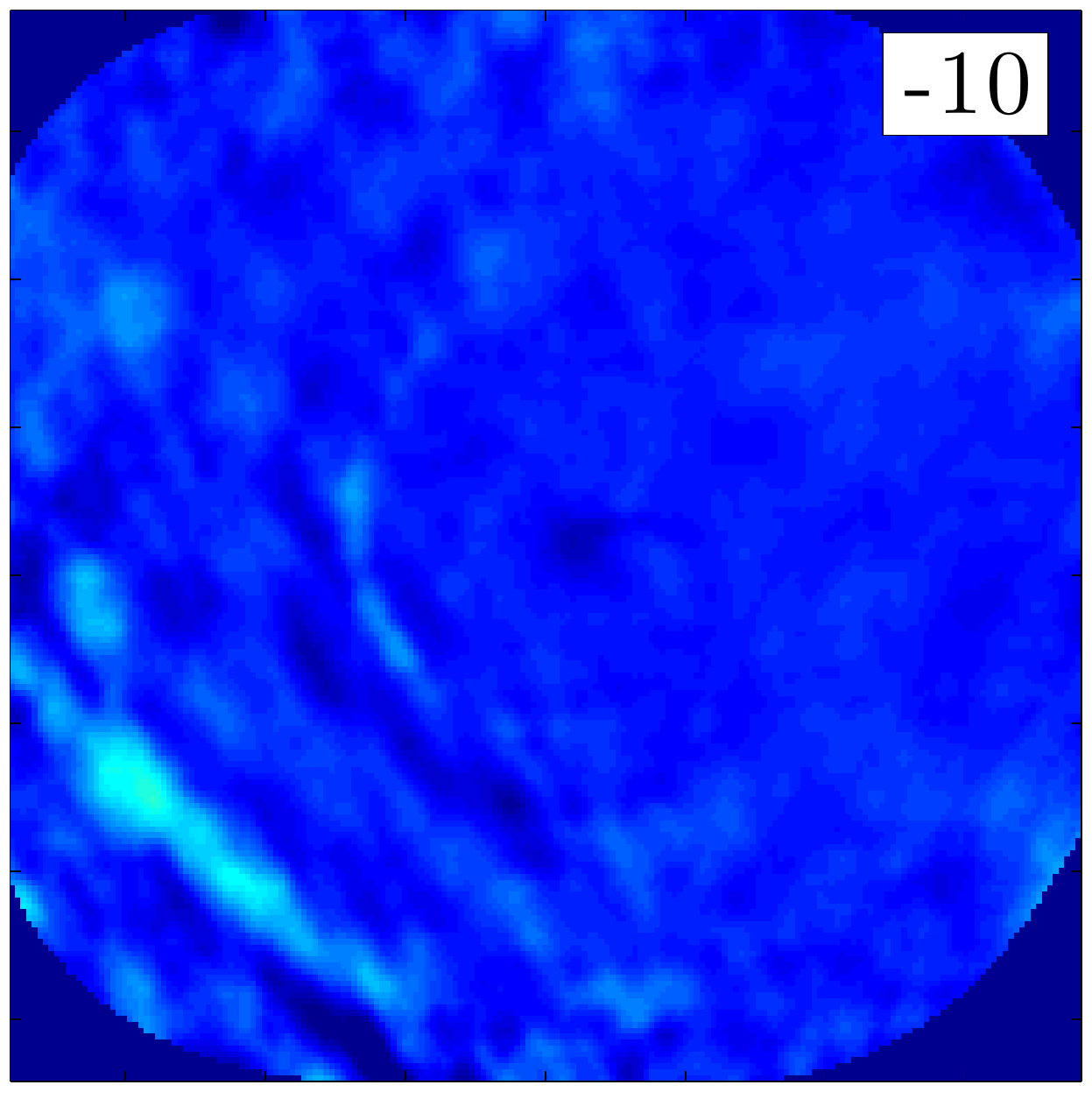}
\includegraphics[height=3.9cm]{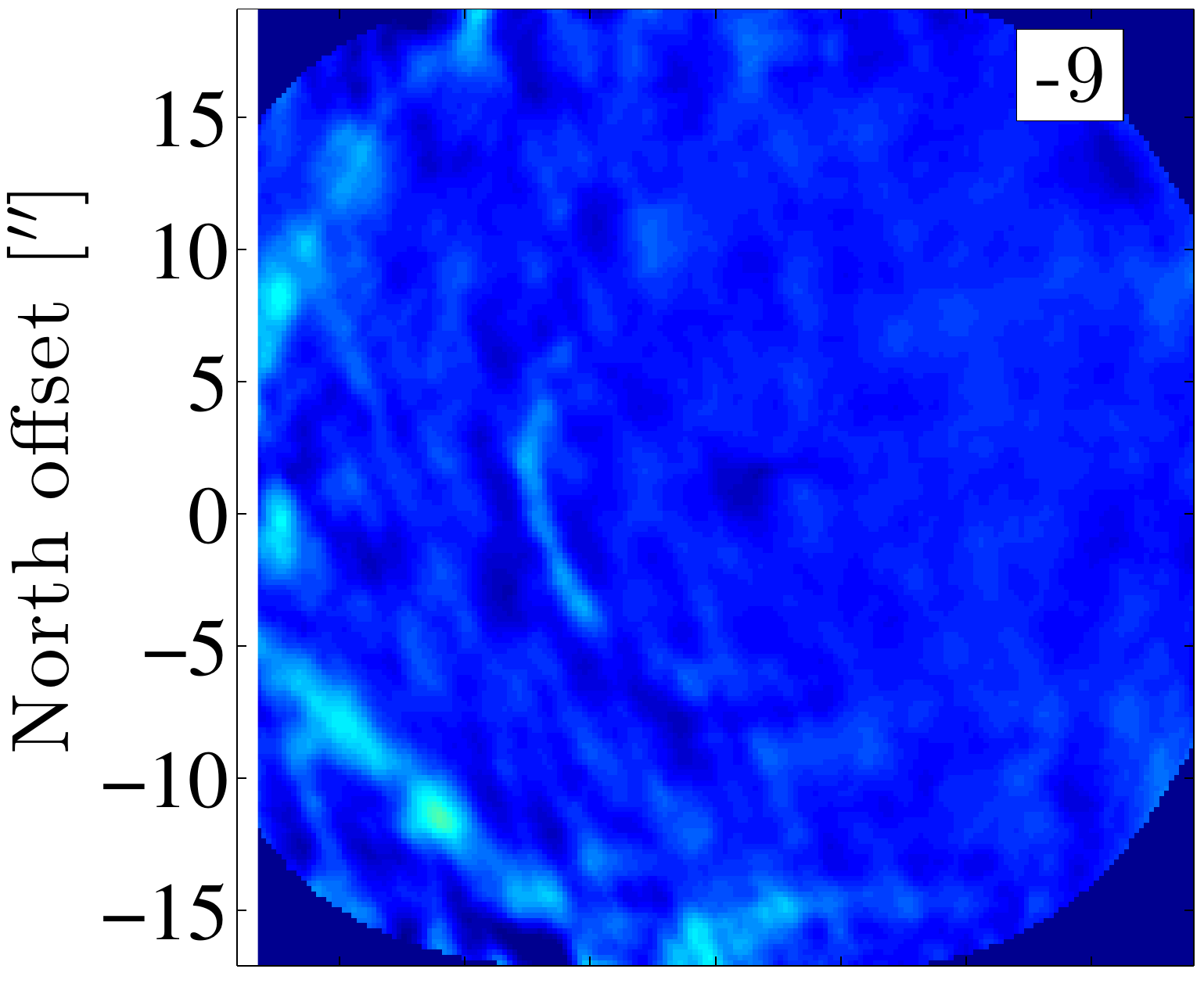}
\includegraphics[height=3.9cm]{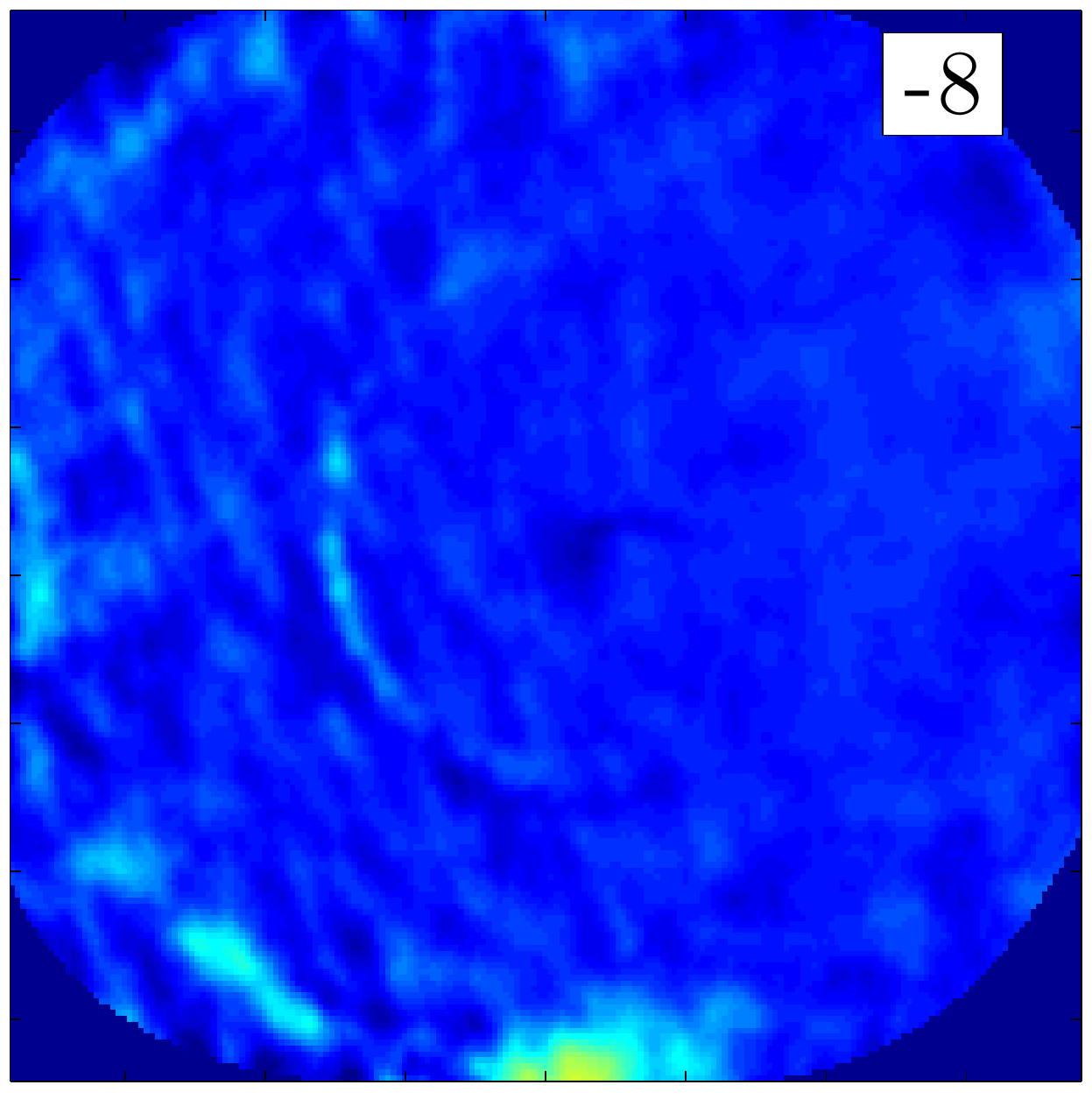}
\includegraphics[height=3.9cm]{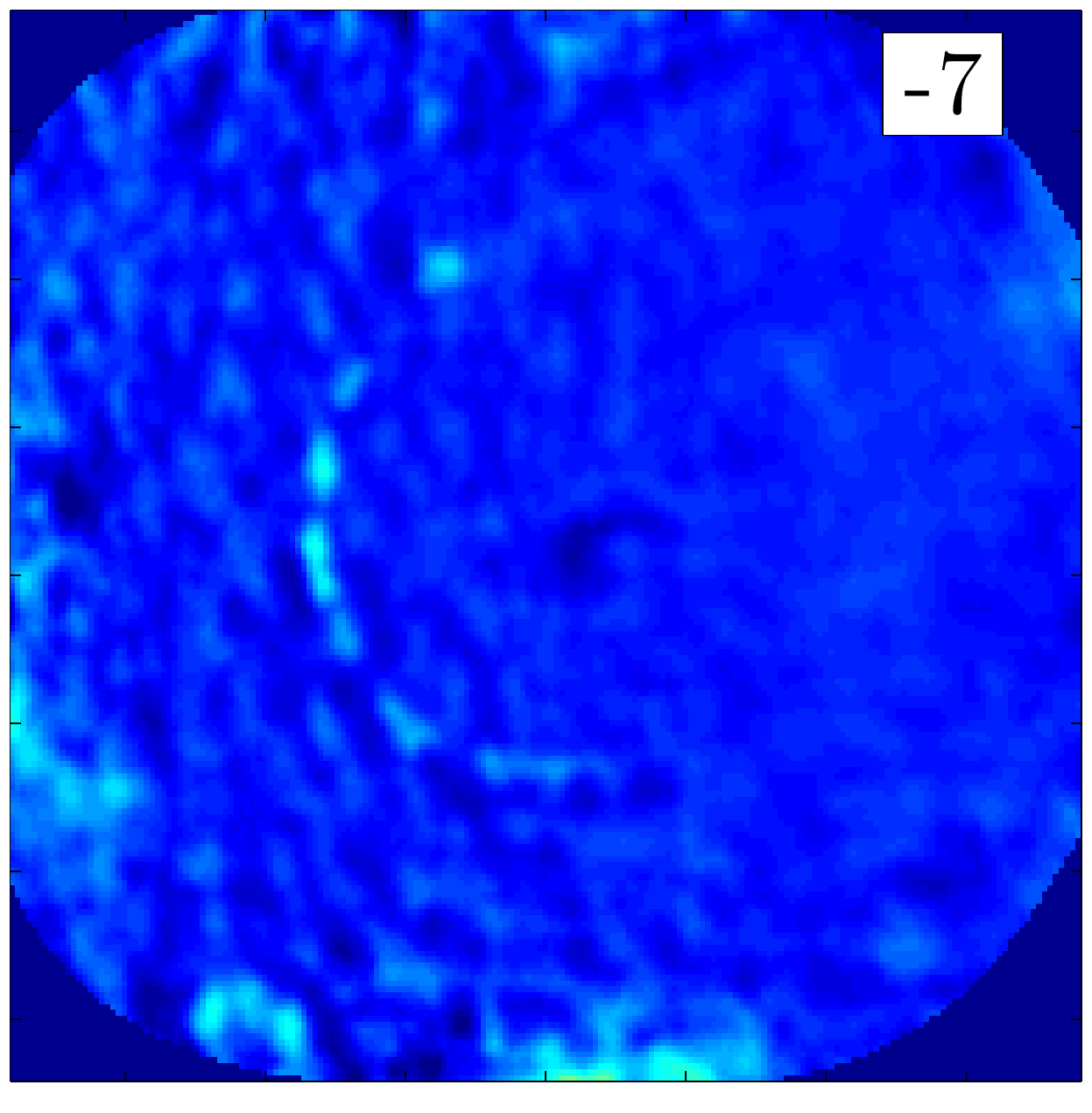}
\includegraphics[height=3.9cm]{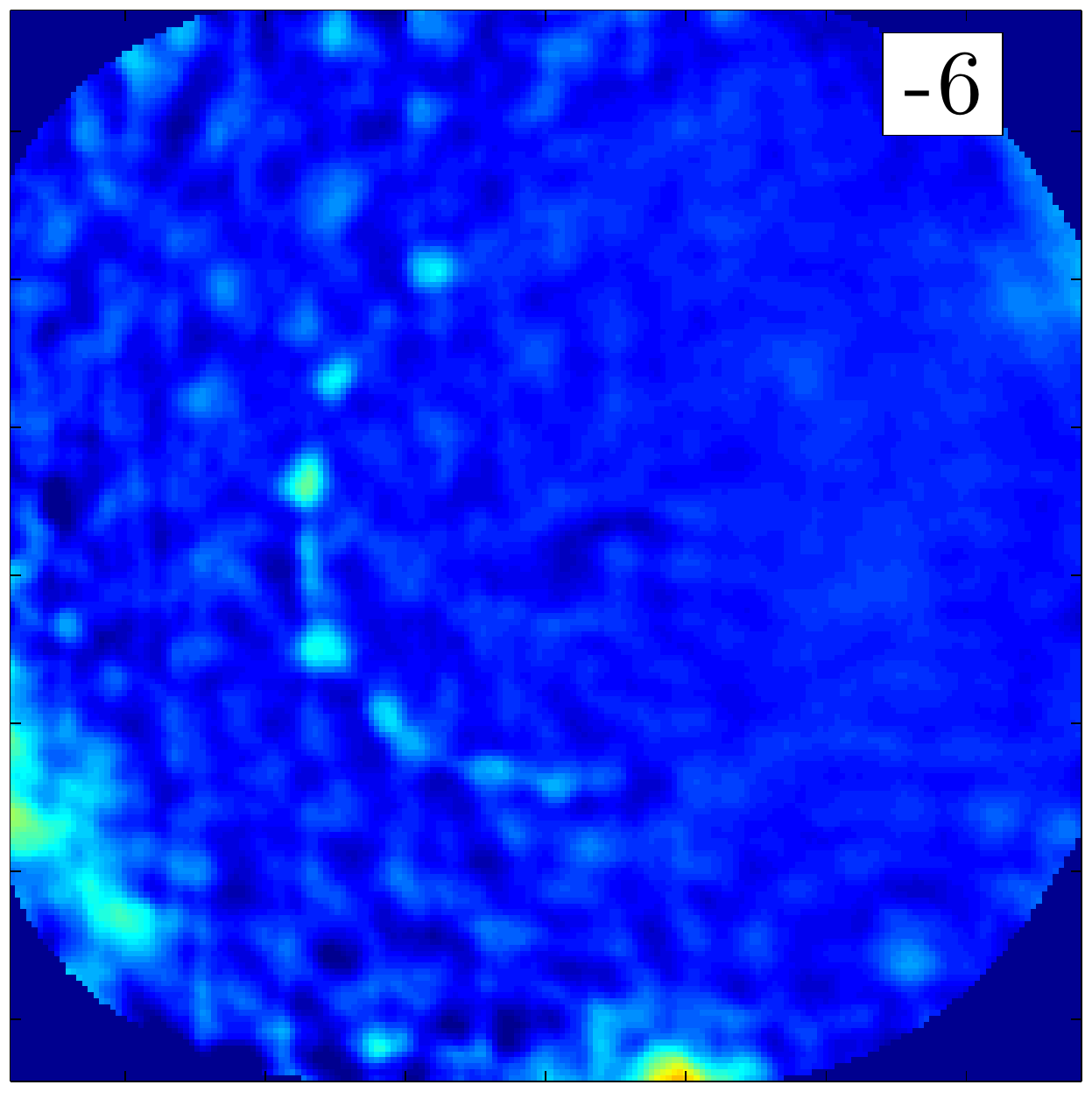}
\includegraphics[height=3.9cm]{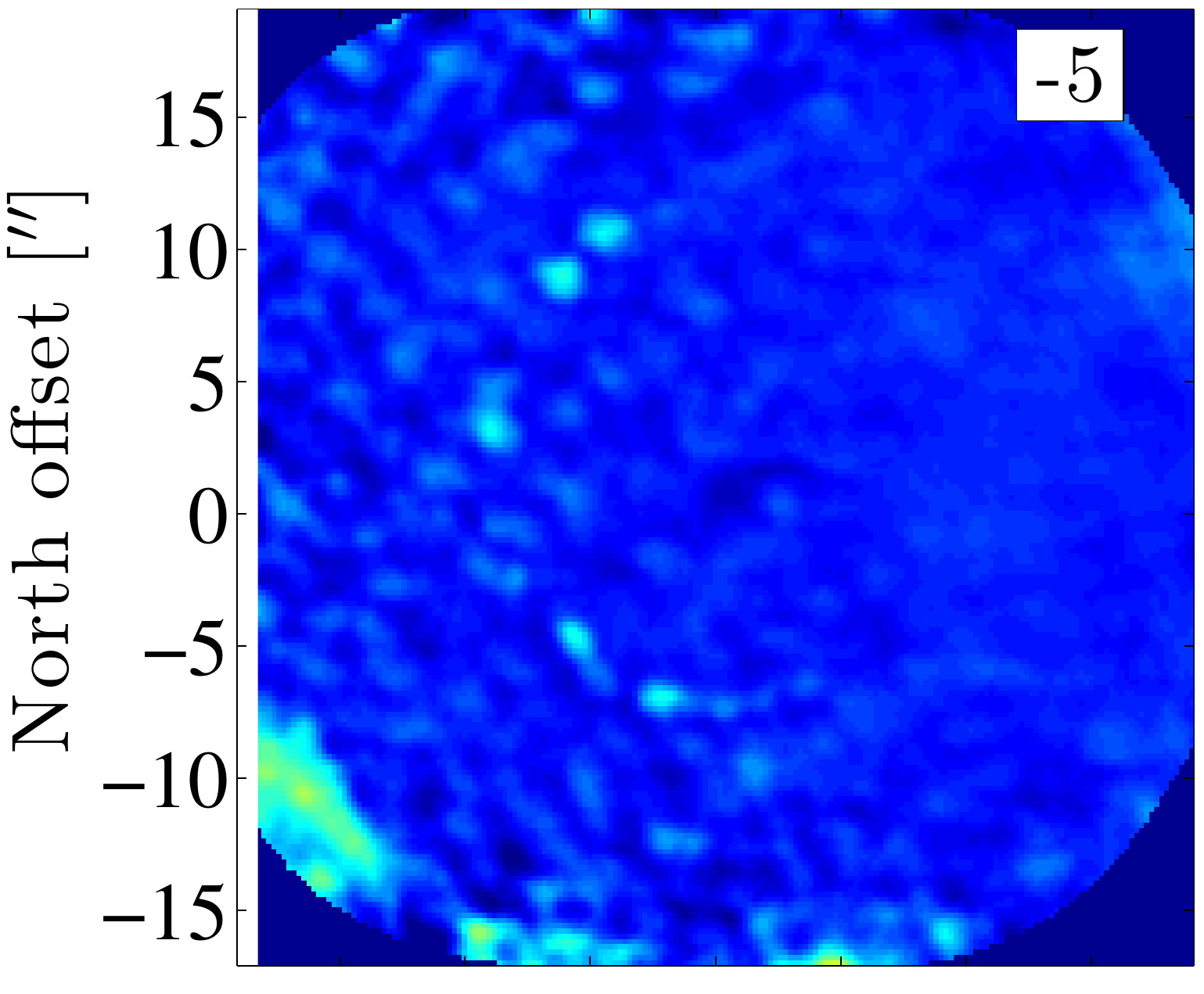}
\includegraphics[height=3.9cm]{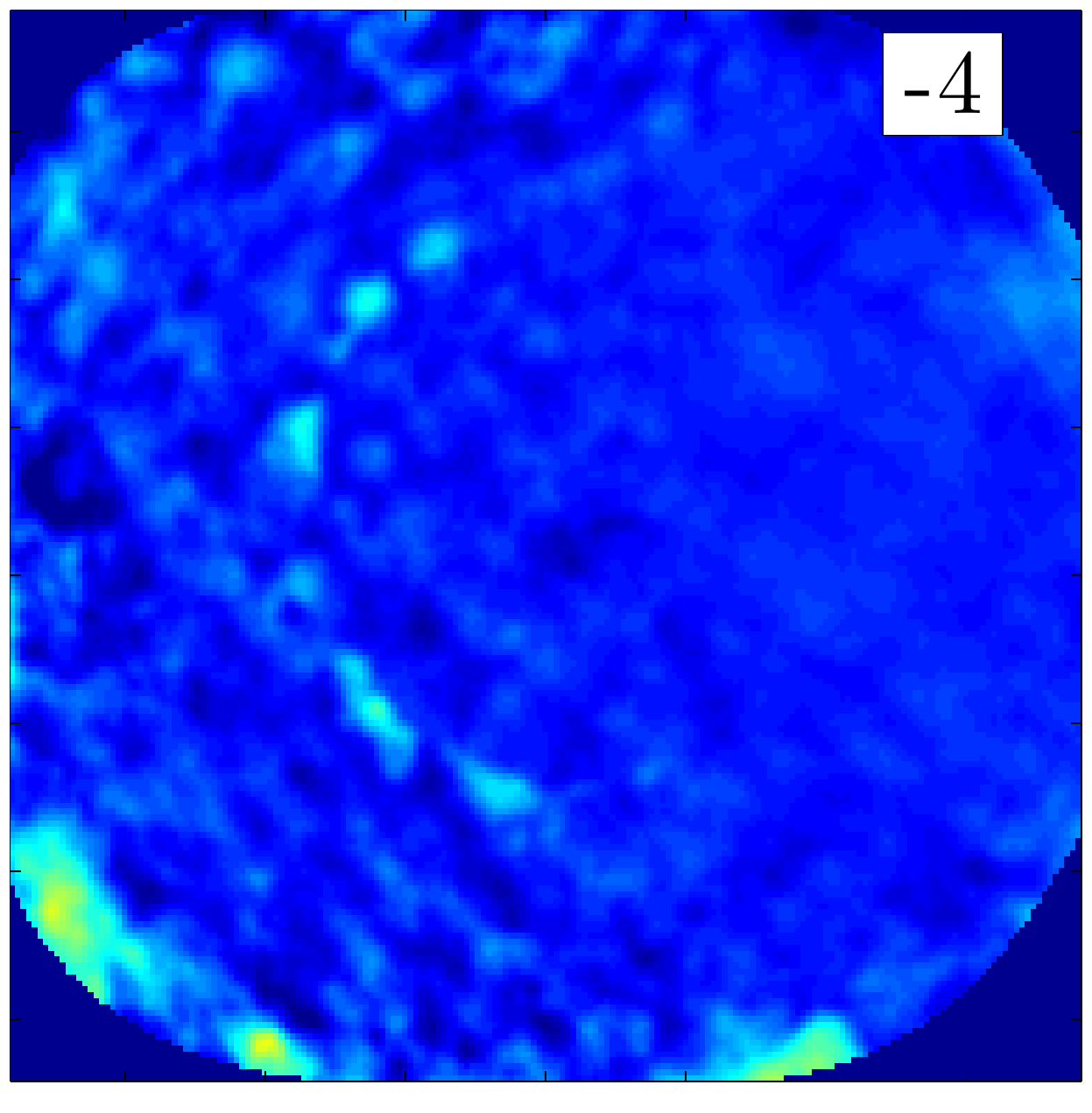}
\includegraphics[height=3.9cm]{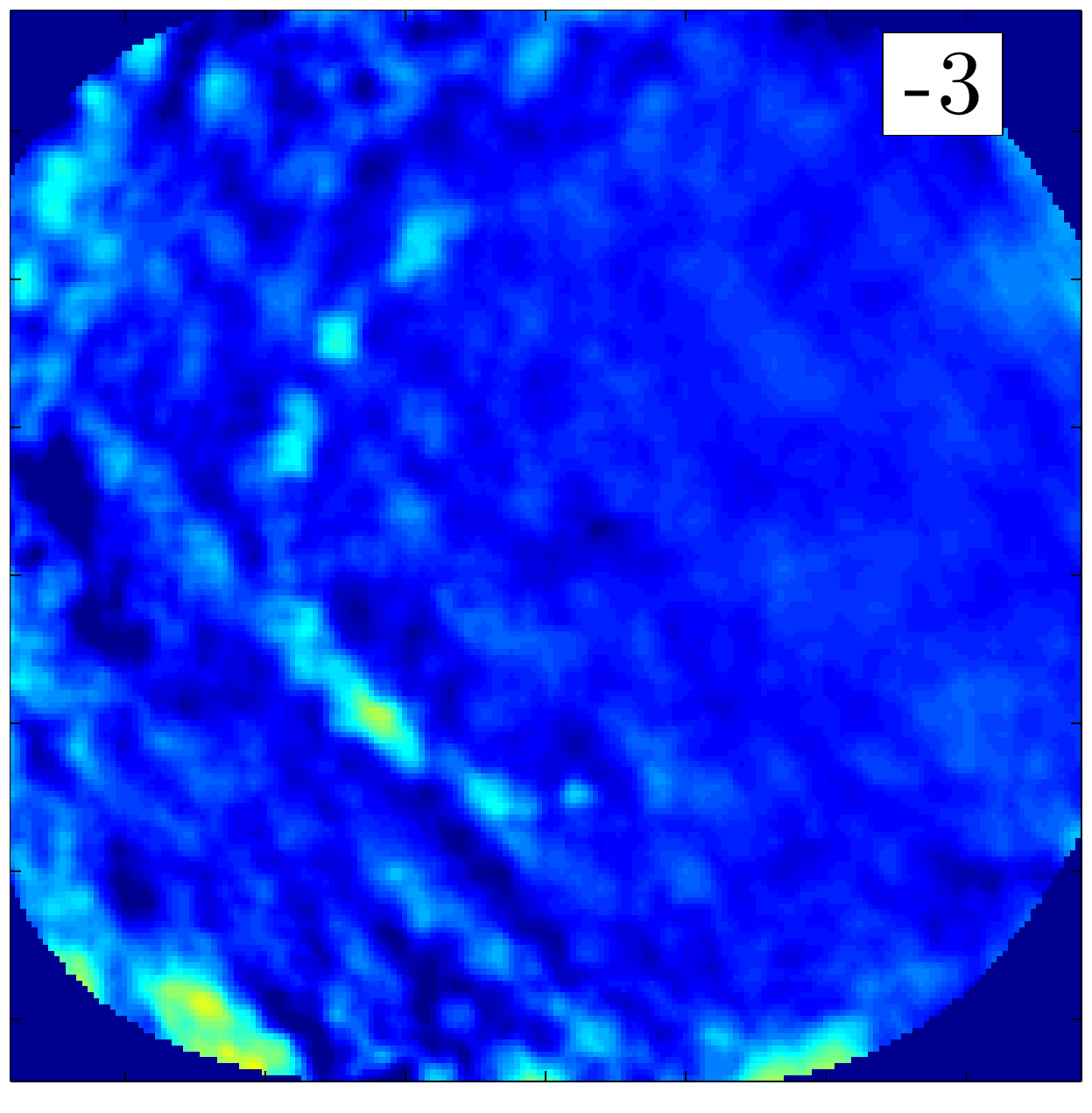}
\includegraphics[height=3.9cm]{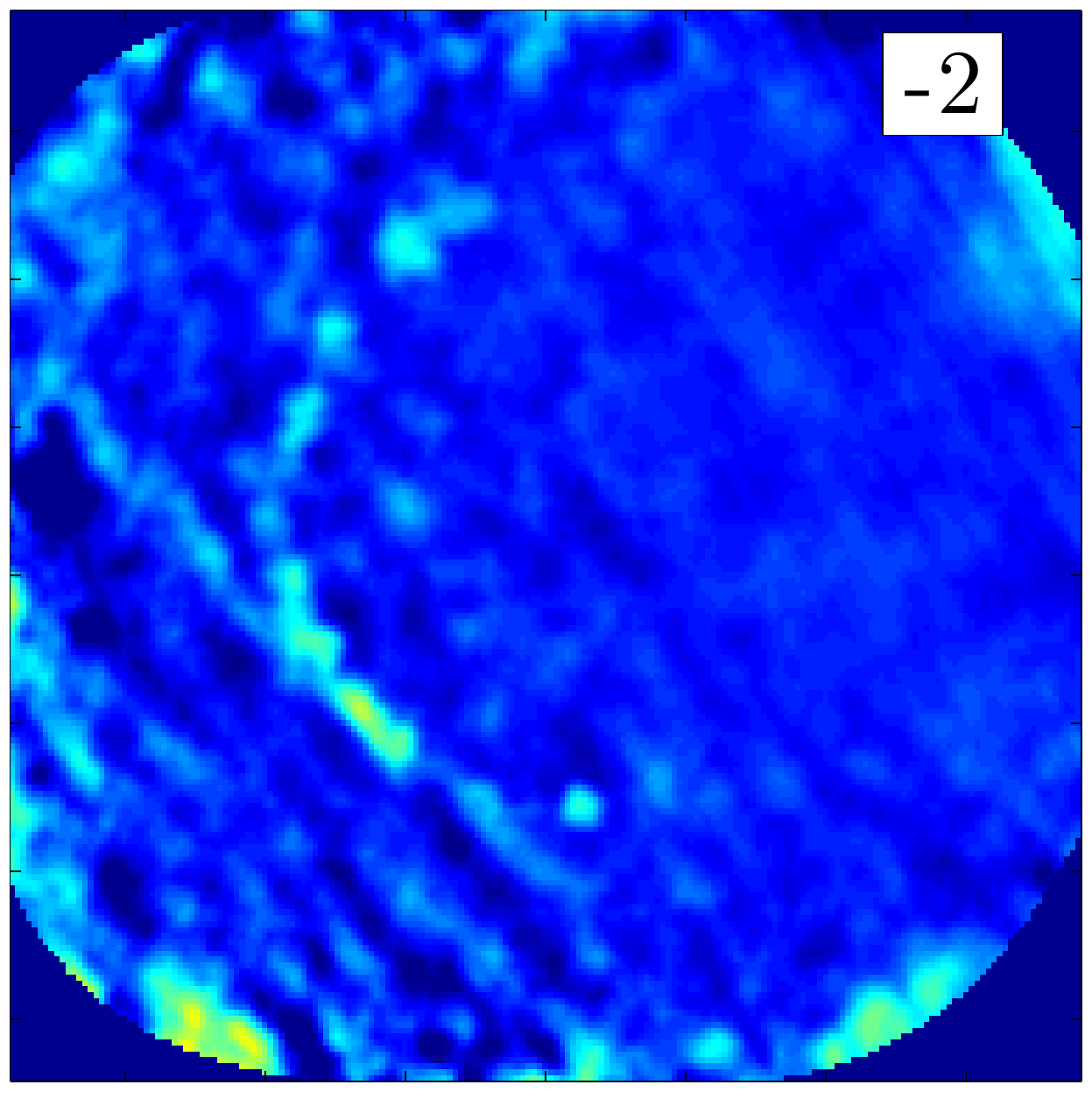}
\includegraphics[width=4.72cm]{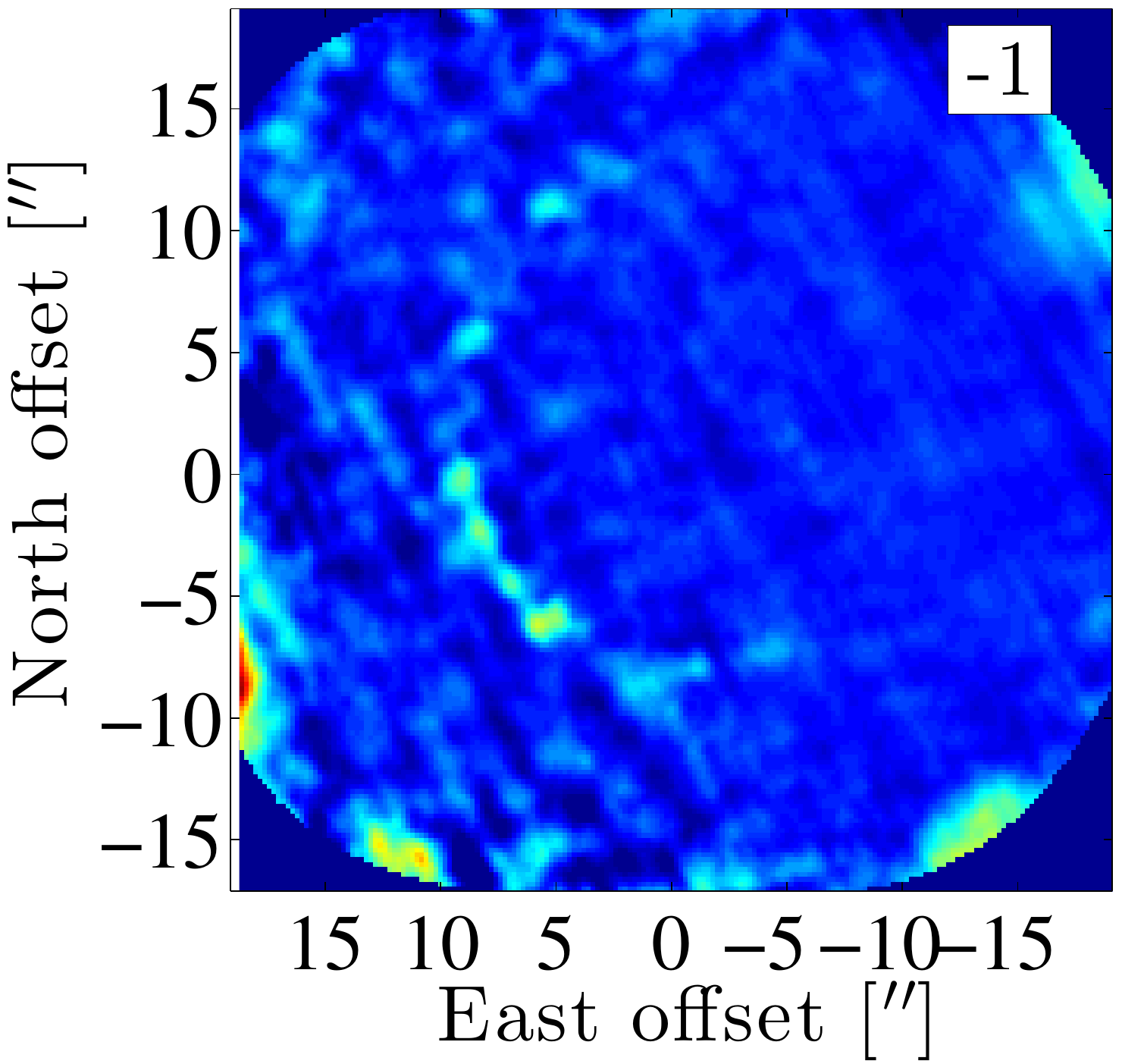}
\includegraphics[width=3.83cm]{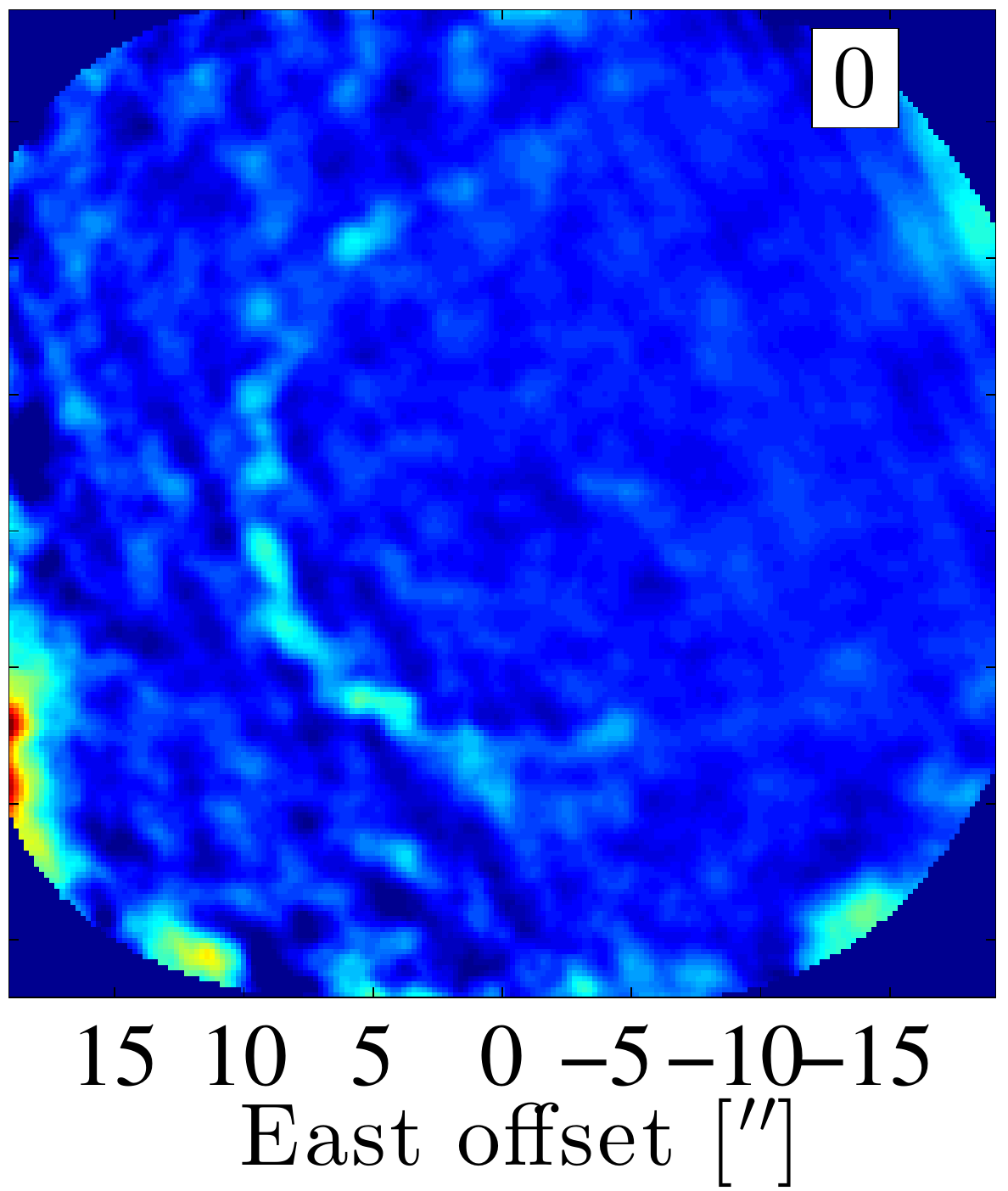}
\caption{ Channel maps from the model with $e$=0.6 (see text for explanation). The velocity relative to the line centre of each channel is given in the upper right corner legend. Only the red-shifted emission is shown since the maps are completely symmetric around the line centre.}
\end{figure*}

\end{document}